\documentclass[12 pt]{article}
\usepackage{amsfonts}
\usepackage{amsmath}
\usepackage{amsthm}
\usepackage{braket}
\usepackage{graphicx}
\usepackage{hyperref}
\usepackage[usenames, dvipsnames]{color}
\usepackage{amssymb}
\usepackage{dsfont}
\usepackage{verbatim}
\usepackage{mathrsfs}
\usepackage{sidecap}
\usepackage[T1]{fontenc}
\usepackage[utf8]{inputenc}
\usepackage{authblk}
\usepackage[a4paper, total={6in, 9in}]{geometry}
\usepackage[nottoc,notlot,notlof]{tocbibind}
\usepackage{float}
\usepackage{eufrak}
\usepackage{leftidx}

\title{\textbf{Fermionic projected entangled-pair states and topological phases}}
\author[a]{Nick Bultinck}
\author[b]{Dominic J. Williamson}
\author[a]{Jutho Haegeman}
\author[a,b]{Frank Verstraete}
\affil[a]{\small{\emph{Department of Physics and Astronomy, Ghent University, Krijgslaan 281
S9, B-9000 Ghent, Belgium}}}
\affil[b]{\small{\emph{Vienna Center for Quantum Technology, University of Vienna, Boltzmanngasse
5, 1090 Vienna, Austria}}}

\newcommand{\otimesg}{\ensuremath{\otimes_{\mathfrak{g}}}}

\begin{document}
  \maketitle

\begin{abstract}
We study fermionic matrix product operator algebras and identify the associated algebraic data. Using this algebraic data we construct fermionic tensor network states in two dimensions that have non-trivial symmetry-protected or intrinsic topological order. The tensor network states allow us to relate physical properties of the topological phases to the underlying algebraic data. We illustrate this by calculating defect properties and modular matrices of supercohomology phases. Our formalism also captures Majorana defects as we show explicitly for a class of $\mathbb{Z}_2$ symmetry-protected and intrinsic topological phases. The tensor networks states presented here are well-suited for numerical applications and hence open up new possibilities for studying interacting fermionic topological phases.
\end{abstract}

\tableofcontents

\section{Introduction} \label{sec:intro}

In recent years there has been substantial progress in the understanding of topological phases in spin systems and their representations via tensor network states. Tensor networks are ideally suited for describing topological phases of matter because, nonlocal, topological features of a system are captured by the symmetries of local tensors. In one-dimensional spin systems Matrix Product States (MPS) were used to classify all Symmetry-Protected Topological (SPT) phases \cite{Pollmann1,Pollmann2,Chen1,Schuch2}. A complete understanding of two-dimensional SPT phases in terms of Projected Entangled-Pair States (PEPS) was developed in \cite{Chen2,Williamson,molnar}. A first systematic study of intrinsic topological order in PEPS was done in Ref. \cite{Schuch}, where the concept of $G$-injectivity was introduced. The concept of $G$-injectivity was soon after generalized to twisted $G$-injectivity \cite{Buerschaper} and to matrix product operator (MPO)-injectivity \cite{Sahinoglu}, the latter  describing the same class of topological phases as those captured by string-net models \cite{stringnet,turaevviro}. A detailed understanding of the anyonic excitations in MPO-injective PEPS and how to construct them was developed in \cite{MPOpaper}.

For topological fermionic systems, the understanding is much less developed. Building on the work of Ref. \cite{Fidkowski2} a complete description of interacting fermionic SPT phases in one dimension using fermionic MPS (fMPS) was given in Refs. \cite{fMPS,Turzillo}. In \cite{Schuch3,Dubail,Wahl}, it was shown that free fermions systems with nonzero thermal Hall conductance can be represented as Gaussian PEPS. The first steps in generalizing MPO-injectivity to fermionic PEPS were reported in Refs. \cite{Williamson2,Wille}, but those formulations did not develop the theory of Majorana defects.

In this work we will focus on topological phases with zero thermal Hall conductance in two dimensions and develop a general formalism for understanding the universal properties of fermionic tensor network states representing these phases of matter. We do this by first studying fermionic Matrix Product Operator (fMPO) algebras. The structural data associated to such algebras, which can be seen as a fermionic version of the fusion categories underlying bosonic topological tensor networks, will allow us to construct the relevant topological PEPS. Similarly to the bosonic case, the crucial property giving rise to the non-trivial topological order is the \emph{pulling through} equation. The advantage of the tensor network language is that many interesting universal physical properties of the topological phases can be calculated in a straightforward way. We illustrate this by calculating the symmetry properties of defects and the modular matrices of symmetry-twisted states on a torus for Gu-Wen or supercohomology phases \cite{supercohomology}. We also show that the formalism presented here goes beyond supercohomology and fermionic string-net phases \cite{Gu,Gu2} and captures systems with Majorana defects \cite{ware,Tarantino},  and our construction is hence related to the state sum constructions of spin topological field theories reported in Ref. \cite{bhardwaj}.

Many equivalent formulations of fermionic tensor networks based on fermionic mode operators, Grassmann variables or swap gates exist in the literature \cite{corboz1,Kraus,barthel2009contraction,Corboz,Gu3}. In this work we use the graded vector space approach presented in Ref. \cite{fMPS}, as it turns out to be the natural framework for generalizing the MPO symmetries of the bosonic case.

\section{Fermionic tensor networks}

In this section we review the fermionic tensor network formalism as introduced in \cite{fMPS}. To define fermionic tensors we will make use of super vector spaces. A super vector space $V$ has a natural direct sum structure

\begin{equation}
V = V^0\oplus V^1\, ,
\end{equation}
where vectors in $V^0$ or in $V^1$ are called homogeneous vectors. A vector in $V^0$ $(V^1)$ is said to have even (odd) parity. We denote the parity of homogeneous basis vectors $|i\rangle$ as

\begin{equation}
|i| = \Big\{ \begin{matrix} 0 &  \text{ if } |i\rangle \in V^0 \\ 1 & \text{ if } |i\rangle \in V^1
\end{matrix}\, .
\end{equation}
The tensor product of two homogeneous vectors $|i\rangle$ and $|j\rangle$ is again a homogeneous vector and has parity $|i|+|j|$ mod 2. This implies that $V$ and the associated operation of taking tensor products is $\mathbb{Z}_2$ graded. We denote the graded tensor product as

\begin{equation}
|i\rangle \otimesg |j\rangle \in V\otimesg V\, .
\end{equation}
For super vector spaces we will always use the following canonical tensor product isomorphism:

\begin{eqnarray}
\mathcal{F}: V\otimesg W \rightarrow W\otimesg V: |i\rangle\otimesg |j\rangle \rightarrow (-1)^{|i||j|}|j\rangle\otimesg |i\rangle\, .
\end{eqnarray}
This isomorphism of course connects the mathematical concept of super vector spaces to physical systems of fermions. The dual vector space $V^*$ inherits the $\mathbb{Z}_2$ grading from $V$ and $\mathcal{F}$ can be extended in the following way:

\begin{eqnarray}
\mathcal{F}: V^*\otimesg W \rightarrow W\otimesg V^*: \langle i|\otimesg |j\rangle \rightarrow (-1)^{|i||j|}|j\rangle\otimesg \langle i|\, ,
\end{eqnarray}
and similarly for the action on $V^*\otimesg W^*$.

Fermionic tensors are defined in the graded tensor product of super vector spaces. We will always restrict to homogeneous tensors, i.e. those tensors that have a well-defined parity. Let us now introduce the contraction map $\mathcal{C}$:

\begin{equation}\label{Crestricted}
\mathcal{C}: V^*\otimesg V: \langle i|\otimesg |j\rangle \rightarrow \langle i|j\rangle = \delta_{i,j}\, .
\end{equation}
The contraction map $\mathcal{C}$ can be generalized to arbitrary tensor contractions in the following way: first we take the graded tensor product of the tensors one wishes to contract, secondly, use $\mathcal{F}$ to bring the bra and ket to be contracted next to each other and last, apply $\mathcal{C}$ as defined in \eqref{Crestricted}. For tensor contraction to be well defined it is crucial that the tensors have a definite parity, as we explain in more detail at the end of this section. Note that following the fermionic contraction rules, we get

\begin{equation}
\mathcal{C}(\ket{i} \otimesg \bra{j}) = (-1)^{|i||j|} \mathcal{C}(\bra{j} \otimesg \ket{i}) = (-1)^{|i|} \delta_{i,j}\, ,
\end{equation}
which results in the fermionic super trace. Vice versa, if we want to write the ordinary trace of an operator as a tensor contraction, we need to insert an additional parity tensor on the contracted index. As an illustration of more general fermionic tensor contraction, let us define the following fermionic tensors (we will not always explicitly denote the graded tensor product symbol $\otimesg$)
\begin{align*}
C &= \sum_{\alpha\beta\gamma}C_{\alpha\beta\gamma}|\alpha)|\beta)(\gamma| \\
D &= \sum_{\lambda\kappa}D_{\lambda\kappa}|\lambda)(\kappa|\, ,
\end{align*}
where we wish to contract the $\beta$ index of $C$ with the $\kappa$ index of $D$. As a first step we take the graded tensor product of $C$ and $D$:
\begin{displaymath}
C\otimes_{\mathfrak{g}} D = \sum_{\alpha\beta\gamma\lambda\kappa}C_{\alpha\beta\gamma}D_{\lambda\kappa}|\alpha)|\beta)(\gamma|\otimes_{\mathfrak{g}}|\lambda)(\kappa|\, .
\end{displaymath}
Next, we bring the $\kappa$ bra next to the $\beta$ ket using fermionic reordering:
\begin{displaymath}
\mathcal{F}(C\otimes_{\mathfrak{g}} D) = \sum_{\alpha\beta\gamma\lambda\kappa}C_{\alpha\beta\gamma}D_{\lambda\kappa}(-1)^{|\kappa|(|\lambda|+|\gamma|+|\beta|)}|\alpha)(\kappa|\,|\beta)(\gamma|\,|\lambda)\, .
\end{displaymath}
If the tensors $C$ and $D$ are even, this is equivalent to
\begin{displaymath}
\mathcal{F}(C\otimes_{\mathfrak{g}} D) = \sum_{\alpha\beta\gamma\lambda\kappa}C_{\alpha\beta\gamma}D_{\lambda\kappa}(-1)^{|\kappa|+|\kappa||\alpha|}|\alpha)(\kappa|\,|\beta)(\gamma|\,|\lambda)\, .
\end{displaymath}
Now we apply the contraction to obtain the final tensor:
\begin{displaymath}
F \equiv \sum_{\alpha\gamma\lambda}\left(\sum_\beta C_{\alpha\beta\gamma}D_{\lambda\beta}(-1)^{|\beta|+|\beta||\alpha|}\right)|\alpha)(\gamma|\,|\lambda)\, .
\end{displaymath}
Note that in the definition of fermionic tensors we have to include an internal ordering of the basis vectors. It therefore only makes sense to compare tensors that have the same internal ordering, but we can easily switch to a different ordering by absorbing minus signs from the fermionic reordering in the tensor components. Tensor identities obtained in this way will of course continue to hold when suitably transformed to a different internal ordering.

With this definition of tensor contraction the diagrammatic notation familiar from bosonic tensor networks still applies to the fermionic case. However, note that the diagrammatic notation does not unambiguously specify the order in which the tensors are put in the tensor product before contracting. This choice is irrelevant as long as all tensors have total even parity, or there is at most one tensor with odd parity, since we can then always swap the order of the tensors before performing contractions. In later sections, we will also need to consider diagrams with two odd tensors, and will be more careful in that case. Another important point is that the order in which the contractions are performed is also irrelevant, on which we further elaborate. Let us thereto highlight some special cases that relate to matrix multiplication and are noteworthy for the following sections. Two-index tensors of the form $\sum_{\alpha,\beta} C_{\alpha,\beta} \ket{\alpha}\bra{\beta}$, $\sum_{\gamma,\delta} D_{\gamma,\delta} \ket{\gamma}\bra{\delta}$ will give rise to ordinary matrix multiplication of the components when contracting index $\beta$ with $\gamma$, resulting in $\sum_{\alpha,\delta} (C D)_{\alpha,\delta} \ket{\alpha}\bra{\delta}$. As expected, we can introduce an identity tensor $\sum_{\beta',\gamma'} \delta_{\beta',\gamma'}\ket{\beta'}\bra{\gamma'}$ in between this contraction (now contracting $\beta$ with $\beta'$ and $\gamma'$ with $\gamma$) without changing the result. If we want to contract index $\beta$ and $\gamma$ of $\sum_{\alpha,\beta} C_{\alpha,\beta} \bra{\alpha}\ket{\beta}$ and $\sum_{\gamma,\delta} D_{\gamma,\delta} \bra{\gamma}\ket{\delta}$, we obtain $\sum_{\alpha,\beta,\delta} C_{\alpha,\beta}D_{\beta,\delta} (-1)^{|\beta|} \bra{\alpha}\ket{\delta} = \sum_{\alpha,\delta} (C P D)_{\alpha,\delta} \bra{\alpha}\ket{\delta}$, with $P$ the parity matrix. The identity tensor for this contraction is $\sum_{\beta',\gamma'} P_{\beta',\gamma'} \bra{\beta'} \ket{\gamma'} = \sum_{\beta',\gamma'} (-1)^{|\beta'|} \delta_{\beta',\gamma'} \bra{\beta'} \ket{\gamma'}\stackrel{\mathcal{F}}{\rightarrow} \sum_{\beta',\gamma'} \delta_{\gamma',\beta'} \ket{\gamma'}\bra{\beta'}$. The identity tensor in this case is thus equivalent to the former identity tensor, but just expressed with a different internal ordering. For the diagrammatic tensor notation to be well-defined, the identity tensor should indeed not depend on the type of contraction, i.e. whether bra is contracted with ket or vice versa depends on which tensor is taken first and which second, and this is not specified by the diagrammatic notation. From the above observations it follows that once every individual tensor is specified (with internal ordering) every diagram with contracted indices can be unambiguously translated in a fermionic tensor contraction. We will use the diagrammatic notation extensively in the remainder of this manuscript. 

As a final point about fermionic tensor contraction, we consider multi-index tensors which can be interpreted as matrices with compound indices. Contracting index $\beta$ with $\gamma$, as well as $\beta'$ with $\gamma'$, in the two tensors $\sum_{\alpha,\alpha',\beta,\beta'} C_{(\alpha,\alpha'),(\beta,\beta')} \ket{\alpha}\ket{\alpha'} \bra{\beta}\bra{\beta'}$ and $\sum_{\gamma,\gamma',\delta,\delta'} D_{(\gamma,\gamma'),(\delta,\delta')} \ket{\gamma}\ket{\gamma'} \bra{\delta}\bra{\delta'}$ gives rise to $\sum_{\alpha,\alpha',\delta,\delta'} (CD)_{(\alpha,\alpha'),(\delta,\delta')} \ket{\alpha} \ket{\alpha'} \bra{\delta} \bra{\delta'}$. Note that in order to obtain simple matrix multiplication, the order of the indices in the tensor components and the order of the indices in the fermionic basis vectors are chosen differently.

\section{fMPO algebras}\label{sec:superalgebras}

Similar to the bosonic case \cite{MPOpaper}, we start with a finite number of irreducible fMPOs which arise as the virtual symmetries of the topologically ordered PEPS and which constitute a $C^*$ algebra. Specifically, we consider $N$ irreducible fMPOs of length $L$ $\{O^L_a|a=1\dots N\}$ that are closed under multiplication and Hermitian conjugation for every $L$:

\begin{eqnarray}
O^L_aO^L_b & = & \sum_{c = 1}^N N_{ab}^c O_c^L\\
\left(O^L_a\right)^\dagger & \equiv & O^L_{a^*}\, ,
\end{eqnarray}
with $N_{ab}^c \in \mathbb{N}$ and $O^L_{a^*} \in \{O^L_a|a=1\dots N\}$. The reason for these requirements is that we want to be able to construct a Hermitian projector $P^L = \sum_{a=1}^N w_a O^L_a$ from the irreducible fMPOs, which then determines the virtual support space of a PEPS tensor.

The fMPOs are constructed from even fermionic tensors 
\begin{equation}
\mathsf{B}[a] = \sum_{i,j,\alpha,\beta}\left(B_a^{ij}\right)_{\alpha,\beta}|\alpha)|i\rangle\langle j|(\beta|\, \;\text{ with } |i|+|j|+|\alpha|+|\beta| = 0 \text{ mod } 2
\end{equation}
and the parity tensor $\mathsf{P} = \sum_\alpha (-1)^{|\alpha|}|\alpha)(\alpha|$ as:
\begin{eqnarray}\label{eq:mpo}
O^L_a & \equiv & \mathcal{C}(\mathsf{P}\otimesg\mathsf{B[a]}\otimesg\mathsf{B[a]}\otimesg\dots\otimesg\mathsf{B[a]}) \nonumber\\
 & = & \sum_{\{i\}\{j\}} \text{tr}\left(B_a^{i_1j_1}B_a^{i_2j_2}\dots B_a^{i_Lj_L} \right)|i_1\rangle\langle j_1|\otimesg|i_2\rangle\langle j_2|\otimesg\dots\otimesg |i_L\rangle\langle j_L|\, .
\end{eqnarray}
The reason for inserting the extra parity matrix arises from the PEPS construction explained in the following section, which indeed ensures that such a parity tensor is inserted in every closed virtual loop. Physically, this parity tensor encodes anti-periodic boundary conditions. Note that the parity matrix gets canceled by the super trace generated by the fermionic contraction rules, such that the final expression in terms of the tensor components is identical to that of the bosonic MPO algebras with periodic boundary conditions, and enables us to recycle many of the results. However, unlike in the bosonic case, there are two types of irreducible fMPOs. In Ref.~\cite{fMPS}, it was shown that irreducibility for a fMPO implies that the matrices $B^{ij}$ span a simple $\mathbb{Z}_2$ graded matrix algebra over $\mathbb{C}$, which come in two different types: the even and odd type \cite{Wall}. An even simple $\mathbb{Z}_2$ graded algebra is simple as an ungraded algebra implying that its center consists of multiples of the identity. An odd simple $\mathbb{Z}_2$ graded algebra is not simple as an ungraded algebra and its graded center consists of multiples of the identity and multiples of $Y$, where $Y$ is an odd matrix satisfying $Y^2 \propto \mathds{1}$. Without loss of generality we adopt the convention that $Y^2 = -\mathds{1}$. The type of irreducible fMPO will be denoted by $\epsilon_a \in \{0,1\}$, where $\epsilon_a = 0$ implies that $O^L_a$ is of even type while $\epsilon_a=1$ implies $O^L_a$ is of odd type, which we will also refer to as Majorana type. For simplicity, we take $\epsilon_a$ to be a $\mathbb{Z}_2$ grading of the fMPO algebra. Another consequence of the anti-periodic boundary conditions is that both types of irreducible fMPOs have a total fermion parity that is even, whereas fMPOs with periodic boundary conditions have a total fermion parity that matches the value $\epsilon$ of the underlying algebra.

\subsection{Fusion tensors} 
Multiplying two fMPOs $O^L_a$ and $O^L_b$ gives rise to a new fMPO with a tensor that can be written as
$$\sum_{\alpha,\alpha',i,k,\beta,\beta'} (B_{ab}^{ik})_{(\alpha,\alpha'),(\beta,\beta')} |\alpha) |\alpha') |i\rangle \langle k | (\beta'|(\beta|$$
where the ordering was chosen such that the fMPO coefficients reduce to a matrix product of the matrices $B_{ab}^{ik}$, which are given by
$$(B_{ab}^{ik})_{(\alpha,\alpha'),(\beta,\beta')} = (-1)^{|\alpha'|(|\alpha|+|\beta|)}\sum_j (B_a^{ij})_{\alpha,\beta} (B_b^{jk})_{\alpha',\beta'}$$
Similar to the bosonic case, the fact that $O^L_aO^L_b =  \sum_{c = 1}^N N_{ab}^c O_c^L$ for every $L$ implies the existence of a gauge transformation $X_{ab}$ that simultaneously brings the matrices $B_{ab}^{ik}$ into a canonical form (block upper triangular), where the diagonal blocks correspond to $B_c^{ik}$ appearing $N_{ab}^c$ times \cite{Cirac}.

From the columns of the the gauge transform $X_{ab}$ and the rows of its inverse $X_{ab}^{-1}$, we can build fermionic splitting and fusion tensors $\mathsf{X^{c}_{ab,\mu}}$ and $\mathsf{X^{c+}_{ab,\mu}}$ ($\mu = 1,\ldots,N_{ab}^c$), such that

\begin{equation}
\mathcal{C}(\mathsf{X^{c+}_{ab,\mu}}\otimesg\mathsf{B[a]}\otimesg\mathsf{B[b]}\otimesg \mathsf{X_{ab,\mu}^c}) = \mathsf{B[c]}\,.\label{eq:reduction} 
\end{equation}
We introduce the following graphical notation for the tensors $\mathsf{B[a]}$, $\mathsf{X_{ab,\mu}^c}$ and $\mathsf{X_{ab,\mu}^{c+}}$

\begin{equation}\label{eq:graphical}
\includegraphics[width=0.8\textwidth]{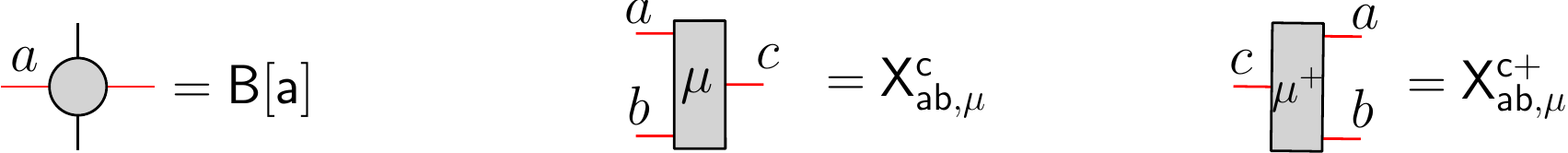}\, ,
\end{equation}
where the red (horizontal) indices represent the internal fMPO indices and the black (vertical) indices represent the external fMPO indices. We can then denote the contraction in equation \eqref{eq:reduction} graphically as

\begin{equation}\label{eq:reduction2}
\includegraphics[width=0.29\textwidth]{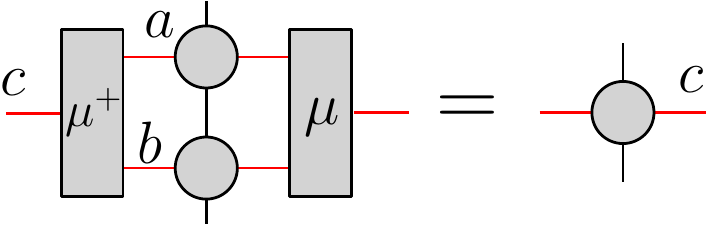}\, .
\end{equation}
Note that although the fMPO tensors $\mathsf{B[a]}$ have even parity, the fusion tensors have a well defined parity that can be either even or odd. This parity depends on the degeneracy label $\mu$ and adds a $\mathbb{Z}_2$ grading denoted as $|\mu|$ to the degeneracy space. 

The fusion tensors satisfy following properties:

\begin{equation} \label{eq:orthogonality}
\includegraphics[width=0.42\textwidth]{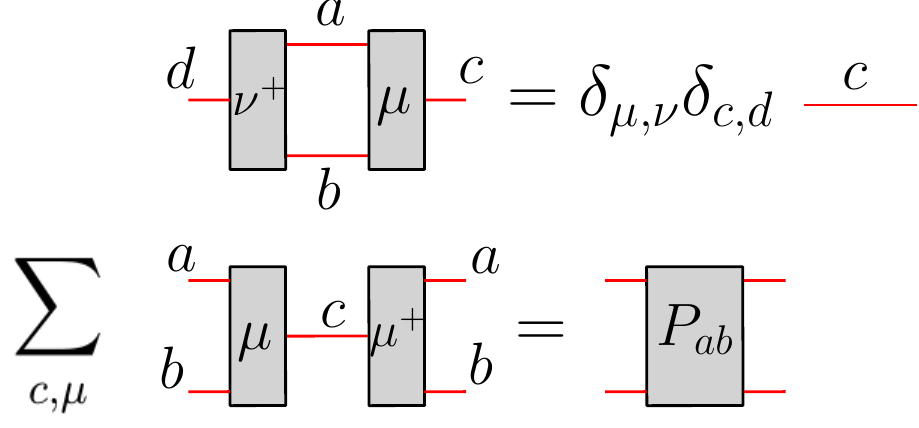}\, ,
\end{equation}
where $P_{ab}$ is the projector onto the support of the internal indices of the fMPO tensor $\mathcal{C}(\mathsf{B[a]}\otimesg\mathsf{B[b]})$. For our purposes we are interested in fMPOs that satisfy a slightly stronger condition than equation \eqref{eq:reduction2}. Namely, we assume that the following \emph{zipper condition} holds:

\begin{equation}\label{eq:zipper}
\includegraphics[width=0.4\textwidth]{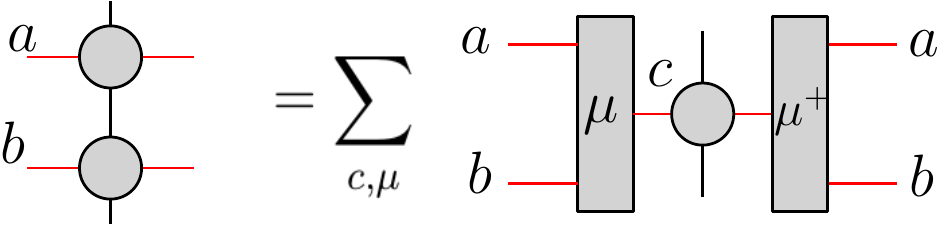}\,.
\end{equation}

Up to this point, the properties of fMPO super algebras are very similar to those of bosonic MPO algebras. We will now discuss the implications of the presence of $\epsilon_a = 1$ irreducible fMPOs. Because the graded center of the matrices $B[a]^{ij}$ for $\epsilon_a =1$ contains the odd matrix $Y$, it is clear that we can contract $\mathsf{Y}$ onto any index of a fusion tensor corresponding to an irreducible fMPO with $\epsilon =1$ to get another fusion tensor that also satisfies the defining equations \eqref{eq:reduction2} and \eqref{eq:zipper}. Because $\mathsf{Y}$ is odd this changes the parity of the fusion tensor $\mathsf{X_{ab,\mu}^c}$. Let us start with the situation $\epsilon_a=\epsilon_b=1$ and consider the matrix

\begin{equation}\label{eq:center1}
\includegraphics[width=0.16\textwidth]{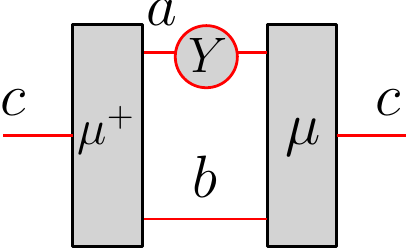}\, ,
\end{equation}
where without loss of generality we take $\mathsf{X_{ab,\mu}^c}$ and $\mathsf{X_{ab,\mu}^{c+}}$ to have even parity. Eq. \eqref{eq:center1} represents an odd matrix that commutes with the matrices $B[c]^{ij}$ because of Eq. \eqref{eq:zipper}. But $\epsilon_c =0$ so the center of the matrix algebra $B[c]^{ij}$ consists only of multiples of the identity. For this reason, the matrix in Eq. \eqref{eq:center1} is zero when $\epsilon_a = \epsilon_b = 1$. Similar reasoning shows that also the odd matrix

\begin{equation}\label{eq:center2}
\includegraphics[width=0.16\textwidth]{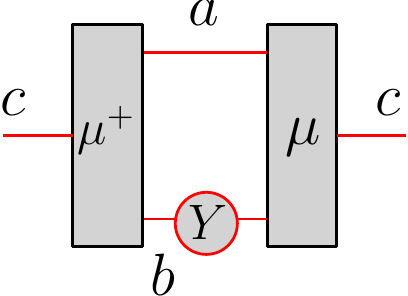}
\end{equation}
is zero when $\epsilon_a=\epsilon_b = 1$. On the other hand, the matrix

\begin{equation}\label{eq:center3}
\includegraphics[width=0.17\textwidth]{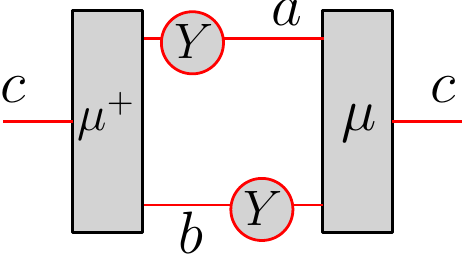}
\end{equation}
is an even matrix commuting with all matrices $B[c]^{ij}$, which implies that it is a multiple of the identity. Since $(\mathsf{Y}\otimesg \mathsf{Y})^2 = -\mathds{1}\otimesg\mathds{1}$ we thus find that the matrix in Eq.~\eqref{eq:center3} equals $\pm i\mathds{1}$. Combining all the properties just derived we can conclude that $N_{ab}^c$ is a multiple of two when $\epsilon_a = \epsilon_b = 1$. The index $\mu$ labeling the fusion tensors $\mathsf{X}_{ab,\mu}^c$ has a natural tensor product structure $\mu = (\hat{\mu},|\mu|)$, where $\hat{\mu} \in \{1,\dots,N_{ab}^c/2\}$ and $|\mu|$ also denotes the parity of the fusion tensor $\mathsf{X}_{ab,(\hat{\mu},|\mu|)}^c$. We will adopt following graphical notation for the fusion tensors and the property derived from matrix \eqref{eq:center3}:

\begin{equation}\label{eq:Ymove1}
\includegraphics[width=0.53\textwidth]{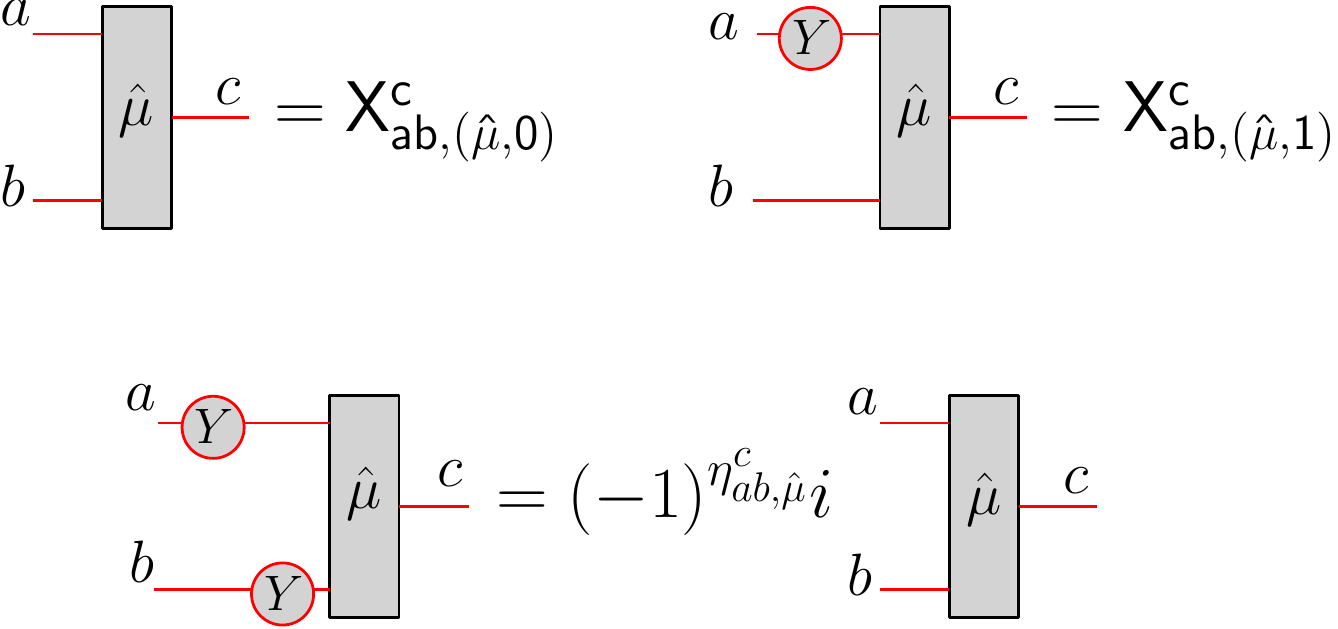}\, ,
\end{equation}
where $\eta^c_{ab,\hat{\mu}} \in \{0,1\}$ are discrete quantities that are part of the algebraic structure defining the fMPO super algebra. 

Let us revisit the matrix in Eq.~\eqref{eq:center1} when $\epsilon_a = 1$ and $\epsilon_b = 0$. Now $\epsilon_c = 1$ so the fact that this odd matrix commutes with all $B[c]^{ij}$ implies that it is a multiple of $\mathsf{Y}$. Since $(\mathsf{Y}\otimesg \mathds{1})^2 = -\mathds{1}\otimesg\mathds{1}$ this implies that

\begin{equation}\label{eq:Ymove2}
\includegraphics[width=0.4\textwidth]{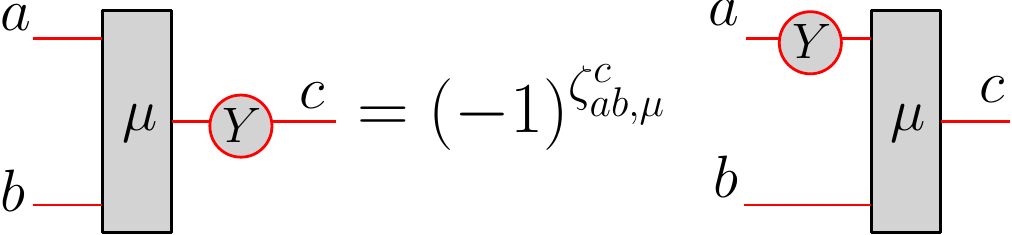}\, .
\end{equation}
Similar reasoning for the matrix in Eq.~\eqref{eq:center2} when $\epsilon_a = 0$ and $\epsilon_b = 1$ shows that

\begin{equation}\label{eq:Ymove3}
\includegraphics[width=0.4\textwidth]{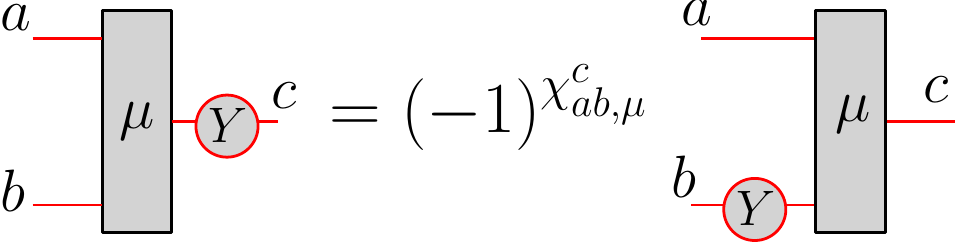}\, .
\end{equation}
So when $\epsilon_c = 1$ there is no further restriction on $N_{ab}^c$ and the parity of the fusion tensor for each $\mu$ is completely arbitrary. We will keep the graphical notation introduced in Eq.~\eqref{eq:graphical} for the even parity fusion tensor $\mathsf{X_{ab,\mu}^c}$ and use the left hand sides of Eq.~\eqref{eq:Ymove2} and \eqref{eq:Ymove3} as a graphical notation for the odd fusion tensors $\mathsf{X_{ab,\mu}^c}$. In Appendix~\ref{app:fusionfMPO} we give a more detailed derivation of the fusion tensors and their properties. 

\subsection{$F$ move and pentagon equation}\label{sec:fmovepentagon}

Associativity of the product of three fMPOs $O^L_aO^L_bO^L_c$ clearly implies that $\sum_d N_{ab}^dN_{dc}^e = \sum_f N_{bc}^f N_{af}^e$. Associativity also allows one to derive an important property of the fusion tensors. The fMPO tensor of $O^L_aO^L_bO^L_c$, $\mathcal{C}(\mathsf{B[a]}\otimesg\mathsf{B[b]}\otimesg\mathsf{B[c]})$, can be written as a sum in two different ways by either applying equation \eqref{eq:zipper} first to $\mathcal{C}(\mathsf{B[a]}\otimesg\mathsf{B[b]})$ or first to $\mathcal{C}(\mathsf{B[b]}\otimes\mathsf{B[c]})$. Let us first consider the case where $\epsilon \equiv 0$. Equality of the two sums in this case implies that the fusion tensors satisfy \footnote{The proof is similar to the bosonic case \cite{MPOpaper}.}

\begin{equation}\label{eq:Fmove}
\includegraphics[width=0.5\textwidth]{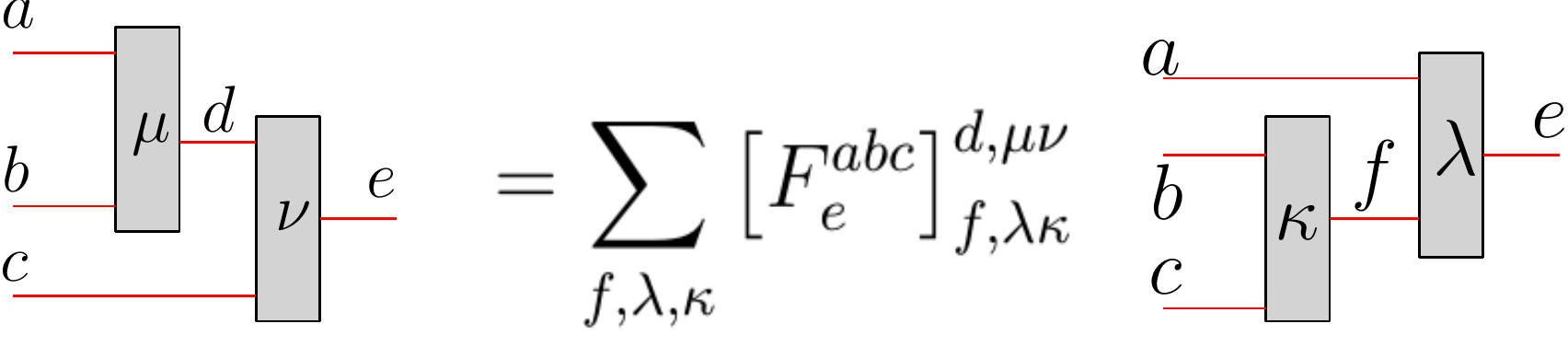}\, ,
\end{equation}
where $\left[F^{abc}_e \right]^{d,\mu\nu}_{f,\lambda\kappa}$ is an invertible even matrix. We will often refer to this identity as an $F$-move and to the matrices $\left[F^{abc}_e \right]^{d,\mu\nu}_{f,\lambda\kappa}$ as the $F$-symbols.

\begin{figure}
  \centering
    \includegraphics[width=0.5\textwidth]{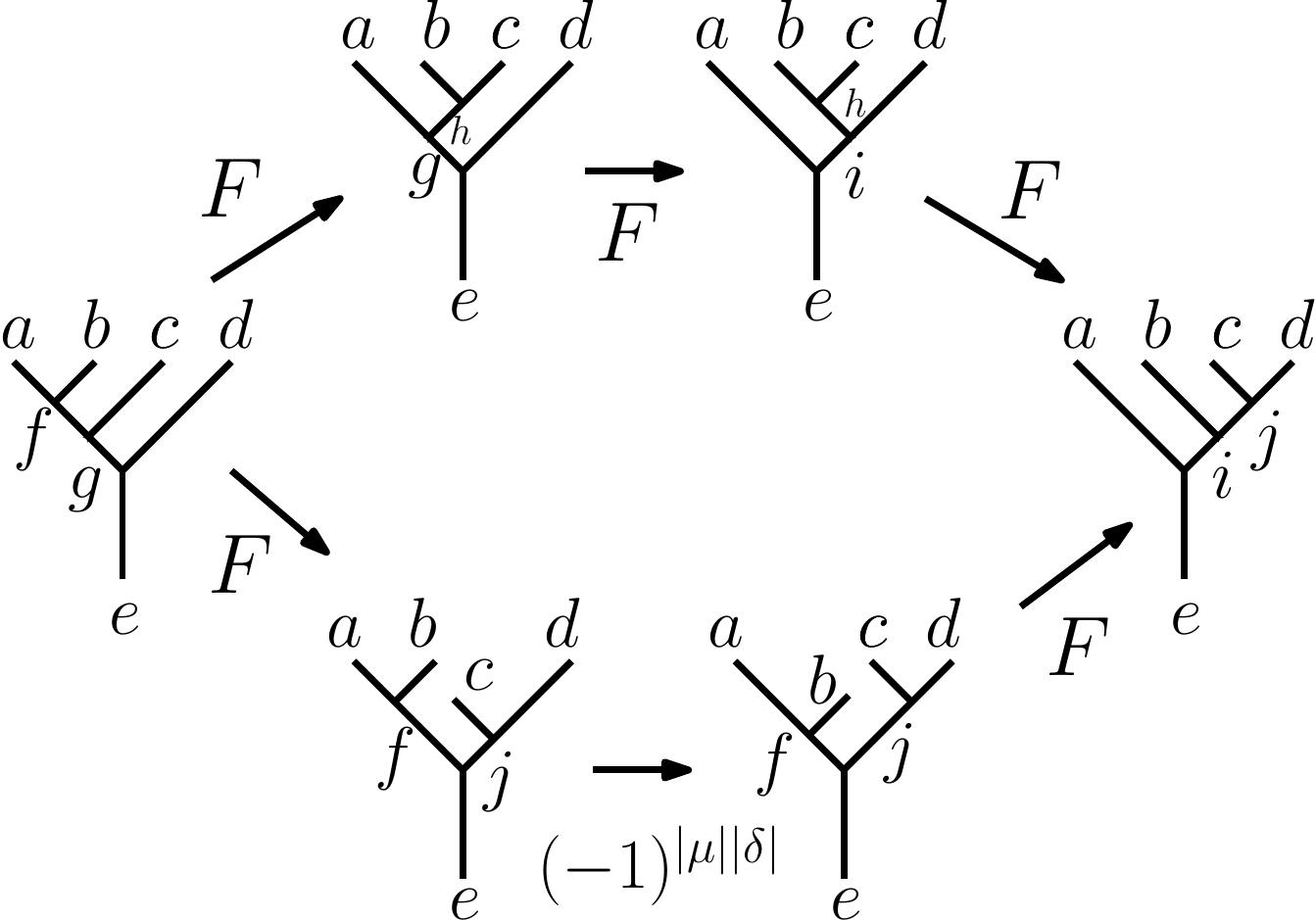}
  \caption{Schematic representation of the two paths giving rise to the super pentagon equation. The upper path consists of three $F$-moves and is similar to the bosonic case. In the lower path there are two $F$-moves and one fermionic reordering of the fusion tensors, leading to a potential minus sign depending on their parity.}\label{fig:pentagon}
\end{figure}

As is familiar from bosonic fusion categories, the $F$-symbols have to satisfy a consistency equation called the (super) pentagon equation. This consistency condition arises from equating the two different paths one can follow to get from $\mathcal{C}((\mathsf{X^f_{ab,\mu}}\otimesg\mathds{1}\otimesg\mathds{1})\otimesg(\mathsf{X^g_{fc,\nu}}\otimesg\mathds{1})\otimesg\mathsf{X^e_{gd,\rho}})$ to $\mathcal{C}((\mathds{1}\otimesg\mathds{1}\otimesg\mathsf{X^j_{cd,\delta}})\otimesg(\mathds{1}\otimesg\mathsf{X^i_{bj,\gamma}})\otimesg\mathsf{X^e_{ai,\omega}})$ using $F$-moves. These different paths are shown in figure \ref{fig:pentagon}. Written down explicitly, the super pentagon equation is

\begin{equation}\label{eq:pentagon}
\sum_{h,\sigma,\lambda,\kappa}[F^{abc}_g]^{f,\mu\nu}_{h,\sigma\lambda}[F^{ahd}_e]^{g,\sigma\rho}_{i,\omega\kappa}[F^{bcd}_i]^{h,\lambda\kappa}_{j,\gamma\delta} = \sum_\sigma [F^{fcd}_e]^{g,\nu\rho}_{j,\sigma\delta}[F^{abj}_e]^{f,\mu\sigma}_{i,\omega\gamma}(-1)^{|\mu| |\delta|}\, ,
\end{equation}
where $|\mu|$ ($|\delta|$) denotes the parity of fusion tensor $\mathsf{X^f_{ab,\mu}}$ ($\mathsf{X^j_{cd,\delta}}$). We see that for $\epsilon \equiv 0$, the only difference between the fermionic pentagon equation and the standard, bosonic pentagon equation is the minus sign depending on $|\mu|$ and $|\delta|$. This sign arises from the reordering of two fusion tensors so that a subsequent $F$-move can be applied. This step is also shown in figure \ref{fig:pentagon}. For $\epsilon \equiv 0$ the super pentagon equation was previously derived in the construction of fermionic string-net models \cite{Gu,Gu2}.

Let us now also take fMPOs with $\epsilon =1$ into account. As in equation \eqref{eq:Fmove}, we want to relate $\mathcal{C}(\mathsf{X}_{ab,\mu}^{d} \otimesg \mathsf{X}_{dc,\nu}^e)$ and $ \mathcal{C}(\mathsf{X}_{bc,\kappa}^{f} \otimesg \mathsf{X}_{af,\lambda}^e)$, which both reduce $\mathcal{C}(\mathsf{B[a]}\otimesg\mathsf{B[b]}\otimesg\mathsf{B[c]})$ to a direct sum of $\mathsf{B[e]}$. Since $\mathsf{B[e]}$ has a non-trivial center $\{\mathds{1}_e,\mathsf{Y}_e\}$ when $\epsilon_e=1$ we find

\begin{equation*}
\mathcal{C}(\mathsf{X}_{ab,\mu}^{d} \otimesg \mathsf{X}_{dc,\nu}^e)	= \begin{cases}
\sum_{f,\lambda,\kappa}  \mathcal{C}(\mathsf{X}_{bc,\kappa}^{f} \otimesg \mathsf{X}_{af,\lambda}^e\otimesg ([F^{abc}_e]^{d,\mu\nu}_{f,\lambda\kappa} \mathds{1}_e)),&\epsilon_e=0\\
\sum_{f,\lambda,\kappa}  \mathcal{C}(\mathsf{X}_{bc,\kappa}^{f} \otimesg \mathsf{X}_{af,\lambda}^e \otimesg ([F^{abc}_e]^{d,\mu\nu}_{f,\lambda\kappa} \mathds{1}_e + [G^{abc}_e]^{d,\mu\nu}_{f,\lambda\kappa}\mathsf{Y}_e)),&\epsilon_e=1
\end{cases}
\end{equation*}
From parity consideration, it follows for $\epsilon_e=0$ that $[F^{abc}_e]^{d,\mu\nu}_{f,\lambda\kappa}=0$ if $|\mu|+|\nu|+|\kappa|+|\lambda|\mod 2 \neq 0$. For $\epsilon_e=1$, we have $[F^{abc}_e]^{d,\mu\nu}_{f,\lambda\kappa}=0$ if $|\mu|+|\nu|+|\kappa|+|\lambda|\mod 2 = |\mu|+|\kappa|\mod 2 \neq 0$ and $[G^{abc}_e]^{d,\mu\nu}_{f,\lambda\kappa}=0$ if $|\mu|+|\nu|+|\kappa|+|\lambda|\mod 2 = |\mu|+|\kappa|\mod 2 \neq 1$. If the fusion tensors are isometric, such that $\mathsf{X}^{c+}_{ab,\mu} = \mathsf{X}^{c\dagger}_{ab,\mu}$, we find that
\begin{equation}
	\begin{cases}
		\sum_{f,\lambda\kappa} [\bar{F}_e^{abc}]^{d',\mu'\nu'}_{f,\lambda\kappa} [F_e^{abc}]^{d,\mu\nu}_{f,\lambda\kappa} = \delta_{d,d'}\delta_{\mu,\mu'}\delta_{\nu,\nu'},&\epsilon_e = 0,\\
		\sum_{f,\lambda\kappa}[\bar{F}_e^{abc}]^{d',\mu'\nu'}_{f,\lambda\kappa} [F_e^{abc}]^{d,\mu\nu}_{f,\lambda\kappa} + [\bar{G}_e^{abc}]^{d',\mu'\nu'}_{f,\lambda\kappa} [G_e^{abc}]^{d,\mu\nu}_{f,\lambda\kappa} = \delta_{d,d'}\delta_{\mu,\mu'}\delta_{\nu,\nu'},&\epsilon_e = 1,\\
		\sum_{f,\lambda\kappa} [\bar{F}_e^{abc}]^{d',\mu'\nu'}_{f,\lambda\kappa} [G_e^{abc}]^{d,\mu\nu}_{f,\lambda\kappa} - [\bar{G}_e^{abc}]^{d',\mu'\nu'}_{f,\lambda\kappa} [F_e^{abc}]^{d,\mu\nu}_{f,\lambda\kappa} = 0,&\epsilon_e = 1.
	\end{cases}
\end{equation}
This means that, for $\epsilon_e=0$, $F^{abc}_{e}$ is itself a unitary matrix (note that it's square as $\sum_{d} N_{ab}^dN_{dc}^e= \sum_{f} N_{bc}^f N_{af}^e$), while for $\epsilon_e=1$, the matrix $F^{abc}_e \otimes \mathds{1} + G^{abc}_e\otimes y$ is unitary and symplectic.

Having the $F$-move interact with the virtual fMPO indices is inconvenient in order to derive the super pentagon equation and to construct an explict fPEPS tensor satisfying the pulling through equation in the following section. Indeed, the latter requires that we have scalar coefficient $[F_{e}^{abc}]^{d,\mu\nu}_{f,\lambda\kappa}$ rather than a matrix. We can therefore switch to a different convention for the fusion tensors, where we redefine 
$\frac{1}{\sqrt{2}}\mathsf{X}_{ab,\mu}^{c} \to \tilde{\mathsf{X}}_{ab,(\hat{\mu},0)}^{c}$ and $\frac{1}{\sqrt{2}}\mathcal{C}(\mathsf{X}_{ab,\mu}^{c} \otimesg \mathsf{Y}_c) \to \tilde{\mathsf{X}}_{ab,(\hat{\mu},1)}^{c}$ when $\epsilon_c=1$, while $\tilde{\mathsf{X}}_{ab,(\hat{\mu},|\mu|)}^{c} = \mathsf{X}_{ab,(\hat{\mu},|\mu|)}^{c}$ when $\epsilon_a=\epsilon_b=1$ and $\tilde{\mathsf{X}}_{ab,\mu}^{c} = \mathsf{X}_{ab,\mu}^{c}$ when $\epsilon_a=\epsilon_b=0$. In all cases, $|\mu|$ denotes the parity of the fusion tensor $\tilde{\mathsf{X}}_{ab,\mu}^{c}$. The factors $\frac{1}{\sqrt{2}}$ are introduced such that
\begin{equation}
\sum_{c,\mu} \mathcal{C}(\tilde{\mathsf{X}}_{ab,\mu}^{c} \otimesg \tilde{\mathsf{X}}_{ab,\mu}^{c+}) = \sum_c \sum_{\hat{\mu}=1}^{N_{ab}^{c}}\sum_{|\mu|=0,1}\mathcal{C}(\tilde{\mathsf{X}}_{ab,(\hat{\mu},|\mu|)}^{c} \otimesg \tilde{\mathsf{X}}_{ab,(\hat{\mu},|\mu|)}^{c+})
\end{equation}
still defines a properly normalized projector onto the support subspace of the tensor $\mathsf{B}_{ab}$, while
\begin{equation}
\mathcal{C}(\tilde{\mathsf{X}}_{ab,\mu}^{c+} \otimesg \tilde{\mathsf{X}}_{ab,\nu}^{d} ) = \delta_{c,d} \begin{cases} \delta_{\mu,\nu} \mathds{1}_c, &\epsilon_c=0\\
 \frac{1}{2} (\delta_{\mu,\nu} \mathds{1}_{c} - Y_{\mu,\nu} \mathsf{Y}_c),&\epsilon_c=1	
 \end{cases}
\end{equation}
with $Y_{\mu,\nu} = \delta_{\hat{\mu},\hat{\nu}}y_{|\mu|,|\nu|} = \delta_{\hat{\mu},\hat{\nu}}( |\nu| - |\mu|)$. The latter expression for the case $\epsilon_c$ is reminiscent of the pseudo-inverse of a Majorana fMPS. 

The fusion tensors $\tilde{\mathsf{X}}_{ab,\mu}^c$ have the degeneracy structure $\mu=(\hat{\mu},|\mu|)$ as soon as either $\epsilon_a$, $\epsilon_b$ or $\epsilon_c$ is nonzero. Contraction with $\mathsf{Y}_c$ switches between $(\hat{\mu},0)$ and $(\hat{\mu},1)$ if $\epsilon_c=1$, i.e.
\begin{align}
\mathcal{C}(\tilde{\mathsf{X}}_{ab,\mu}^{c}\otimesg \mathsf{Y}_c) &= \sum_{\nu} Y_{\nu,\mu} \tilde{\mathsf{X}}_{ab,\nu}^{c} \quad \text{if $\epsilon_c=1$}\, .
\end{align}
For the case with $\epsilon_a = \epsilon_b=1$ we have:
\begin{align}
\mathcal{C}(\mathsf{Y}_a \otimesg \tilde{\mathsf{X}}_{ab,\mu}^{c}) &= \sum_{\nu} (M_{ab}^c)_{\nu,\mu} \tilde{\mathsf{X}}_{ab,\nu}^{c} \\
\mathcal{C}(\mathsf{Y}_b \otimesg \tilde{\mathsf{X}}_{ab,\mu}^{c}) &= \sum_{\nu} (L_{ab}^c)_{\nu,\mu} \tilde{\mathsf{X}}_{ab,\nu}^{c}\, ,
\end{align}
where $M_{ab}^{c}$ and $L_{ab}^{c}$ are odd (i.e.\ they are nonzero only for $|\mu|\neq |\nu|$). From the results of the previous section it follows that $\left(M_{ab}^c\right)_{\mu\nu} = \delta_{\hat{\mu},\hat{\nu}}y_{|\mu|,|\nu|}$ and $L_{ab}^c = (-1)^{\eta_{ab}^c+1}i\delta_{\hat{\mu},\hat{\nu}} x_{|\mu|,|\nu|}$, with $x_{|\mu||\nu|} = 1 - \delta_{|\mu||\nu|}$.

When $\epsilon_c=0$ we have $\mu=1,\ldots,N_{ab}^c$ whereas if $\epsilon_c=1$, we have $\hat{\mu}=1,\ldots,N_{ab}^c$ and thus $\mu=1,\ldots,2N_{ab}^c$. But here, $N_{ab}^c$ only represents the number of times $O_c$ originates from multiplying $O_a$ and $O_b$ if these fMPOs are built from the fermionic tensors $B_a$, $B_b$ and $B_c$ without normalization factor. Since we take $\epsilon$ to act as a $\mathbb{Z}_2$ grading, we can define all Majorana fMPOs to have an additional global factor $1/\sqrt{2}$, so that in the case $\epsilon_a=\epsilon_b=1$ we would also have $\mu=1,\ldots,2N_{ab}^c$, i.e. $\hat{\mu}=1,\ldots,N_{ab}^c$, if we fix $N_{ab}^{c}$ in the relation $O_a O_b = \sum_{c} N_{ab}^{c} O_c$.

The advantage of working with an overcomplete basis of fusion tensors is that we can now write the $F$-move as an even transformation acting purely on the degeneracy spaces and not on the virtual indices of fMPOs, exactly as in the bosonic case, i.e.\ we can write
\begin{equation}
\mathcal{C}(\tilde{\mathsf{X}}_{ab,\mu}^{d} \otimesg \tilde{\mathsf{X}}_{dc,\nu}^e)	= 
\sum_{f,\lambda,\kappa} [\tilde{F}^{abc}_e]^{d,\mu\nu}_{f,\lambda\kappa} \mathcal{C}(\tilde{\mathsf{X}}_{bc,\kappa}^{f} \otimesg \tilde{\mathsf{X}}_{af,\lambda}^e).
\end{equation}
Let us explain this in more detail by providing an explicit recipe for going from the $F$-symbols to the $\tilde{F}$-symbols. 
\begin{enumerate}
\item Step 1: We first write
\begin{equation}
\mathcal{C}(\mathsf{X}_{ab,\mu}^{d} \otimesg \tilde{\mathsf{X}}_{dc,\nu}^e)	= 
\sum_{f,\lambda,\kappa} [(f_1)^{abc}_e]^{d,\mu\nu}_{f,\lambda\kappa} \mathcal{C}(\mathsf{X}_{bc,\kappa}^{f} \otimesg \tilde{\mathsf{X}}_{af,\lambda}^e)\, ,
\end{equation}
where $[(f_1)^{abc}_e]^{d,\mu\nu}_{f,\lambda\kappa}=[F^{abc}_e]^{d,\mu\nu}_{f,\lambda\kappa}$ if $\epsilon_e=0$, and 
\begin{align}
	[(f_1)^{abc}_e]^{d,\mu(\hat{\nu},0)}_{f,(\hat{\lambda},0)\kappa} &= [F^{abc}_e]^{d,\mu\hat{\nu}}_{f,\hat{\lambda}\kappa}&[(f_1)^{abc}_e]^{d,\mu(\hat{\nu},1)}_{f,(\hat{\lambda},0)\kappa} &= [G^{abc}_e]^{d,\mu\hat{\nu}}_{f,\hat{\lambda}\kappa}\\
	[(f_1)^{abc}_e]^{d,\mu(\hat{\nu},0)}_{f,(\hat{\lambda},1)\kappa} &= -[G^{abc}_e]^{d,\mu\hat{\nu}}_{f,\hat{\lambda}\kappa}&[(f_1)^{abc}_e]^{d,\mu(\hat{\nu},1)}_{f,(\hat{\lambda},1)\kappa} &= [F^{abc}_e]^{d,\mu\hat{\nu}}_{f,\hat{\lambda}\kappa}
\end{align}
if $\epsilon_e=1$. From the properties of $F$ and $G$, we can check that $(f_1)^{abc}_e$ is still a unitary matrix, and is even, i.e.\ its elements $[(f_1)^{abc}_e]^{d,\mu\nu}_{f,\lambda\kappa}$ vanish if $|\mu|+|\nu|+|\kappa|+|\lambda| \mod 2 \neq 0$. Furthermore, in the isometric case, $(f_1)^{abc}_e$ is unitary, i.e.
\begin{equation}
\sum_{f,\lambda\kappa} [(\bar{f}_1)_e^{abc}]^{d',\mu'\nu'}_{f,\lambda\kappa} [(f_1)_e^{abc}]^{d,\mu\nu}_{f,\lambda\kappa} = \delta_{d',d}\delta_{\mu',\mu}\delta_{\nu',\nu}\, ,
\end{equation}
from which also follows
\begin{equation}
\sum_{d,\mu,\nu} [(f_1)_e^{abc}]^{d,\mu\nu}_{f,\lambda\kappa} [(\bar{f}_1)_e^{abc}]^{d,\mu\nu}_{f',\lambda'\kappa'}  = \delta_{f,f'}\delta_{\kappa,\kappa'}\delta_{\lambda,\lambda'}\, .
\end{equation}

\item Step 2: 
\begin{equation}
\mathcal{C}(\tilde{\mathsf{X}}_{ab,\mu}^{d} \otimesg \tilde{\mathsf{X}}_{dc,\nu}^e)	= 
\sum_{f,\lambda,\kappa} [(f_2)^{abc}_e]^{d,\mu\nu}_{f,\lambda\kappa} \mathcal{C}(\mathsf{X}_{bc,\kappa}^{f} \otimesg \tilde{\mathsf{X}}_{af,\lambda}^e)
\end{equation}
with $[(f_2)^{abc}_e]^{d,\mu\nu}_{f,\lambda\kappa}=[(f_1)^{abc}_e]^{d,\mu\nu}_{f,\lambda\kappa}$ if $\epsilon_d=0$. If $\epsilon_d=1$, we obtain
\begin{align}
	[(f_2)^{abc}_e]^{d,(\hat{\mu},0)\nu}_{f,\lambda\kappa} &= \frac{1}{\sqrt{2}}[(f_1)^{abc}_e]^{d,\hat{\mu}\nu}_{f,\lambda\kappa},\\
	[(f_2)^{abc}_e]^{d,(\hat{\mu},1)\nu}_{f,\lambda\kappa} &= \frac{1}{\sqrt{2}}\sum_{\nu'} (M_{dc}^{e})_{\nu',\nu} [(f_1)^{abc}_e]^{d,\hat{\mu}\nu'}_{f,\lambda\kappa}.
\end{align}
Note that $(f_2)^{abc}_e$ is still even, because $M_{dc}^{e}$ is odd. Furthermore, in the isometric case, we obtain
\begin{equation}
\sum_{f,\lambda\kappa} [(\bar{f}_2)_e^{abc}]^{d',\mu'\nu'}_{f,\lambda\kappa} [(f_2)_e^{abc}]^{d,\mu\nu}_{f,\lambda\kappa} = 
\delta_{d',d}\begin{cases}
\delta_{\mu',\mu}\delta_{\nu',\nu},&\epsilon_d = 0\\
\left[\delta_{\mu',\mu} \delta_{\nu',\nu} + Y_{\mu',\mu} (M_{dc}^{e})_{\nu',\nu} \right]/2,&\epsilon_d = 1\\
\end{cases}
\end{equation}
and
\begin{equation}
\sum_{d,\mu,\nu} [(f_2)_e^{abc}]^{d,\mu\nu}_{f,\lambda\kappa}[(\bar{f}_2)_e^{abc}]^{d,\mu\nu}_{f',\lambda'\kappa'}  = \delta_{f,f'}\delta_{\kappa,\kappa'}\delta_{\lambda,\lambda'}.
\end{equation}
Note that if there is a $d$ with $\epsilon_d=1$ present, the matrix $(f_2)^{abc}_e$ has more columns than rows and can therefore no longer be unitary. However, the above expression shows that it is still isometric and defines a projector upon premultiplication with its hermitian conjugate.

\item Step 3:
\begin{equation}
\mathcal{C}(\tilde{\mathsf{X}}_{ab,\mu}^{d} \otimesg \tilde{\mathsf{X}}_{dc,\nu}^e)	= 
\sum_{f,\lambda,\kappa} [\tilde{F}^{abc}_e]^{d,\mu\nu}_{f,\lambda\kappa} \mathcal{C}(\tilde{\mathsf{X}}_{bc,\kappa}^{f} \otimesg \tilde{\mathsf{X}}_{af,\lambda}^e)
\end{equation}
with $[\tilde{F}^{abc}_e]^{d,\mu\nu}_{f,\lambda\kappa}=[(f_2)^{abc}_e]^{d,\mu\nu}_{f,\lambda\kappa}$ if $\epsilon_f=0$. If $\epsilon_f=1$, we obtain the required relation by the following substitution:
\begin{align*}
\mathcal{C}(\mathsf{X}_{bc,\kappa}^{f} \otimesg \tilde{\mathsf{X}}_{af,\lambda}^e)	&= \mathcal{C}\left(\mathsf{X}_{bc,\kappa}^{f} \otimesg \frac{1}{2}(\mathds{1}_f - \mathsf{Y}_f\otimesg \mathsf{Y}_f)\otimesg \tilde{\mathsf{X}}_{af,\lambda}^e\right)\\
	&= \frac{1}{\sqrt{2}} \mathcal{C}\left(\tilde{\mathsf{X}}_{bc,(\hat{\kappa},0)}^{f} \otimesg \tilde{\mathsf{X}}_{af,\lambda}^e\right)-\frac{1}{\sqrt{2}} \mathcal{C}\left(\tilde{\mathsf{X}}_{bc,(\kappa,1)}^{f} \otimesg (\mathsf{Y}_f\otimesg \tilde{\mathsf{X}}_{af,\lambda}^e)\right)\\
	&= \frac{1}{\sqrt{2}} \mathcal{C}\left(\tilde{\mathsf{X}}_{bc,(\hat{\kappa},0)}^{f} \otimesg \tilde{\mathsf{X}}_{af,\lambda}^e\right)-\frac{1}{\sqrt{2}} \sum_{\lambda'} (L_{af}^e)_{\lambda',\lambda} \mathcal{C}\left(\tilde{\mathsf{X}}_{bc,(\kappa,1)}^{f}\otimesg \tilde{\mathsf{X}}_{af,\lambda'}^e)\right)
\end{align*}
So we get for the final $\tilde{F}$-symbols
\begin{align}
	[\tilde{F}^{abc}_e]^{d,\mu\nu}_{f,\lambda(\hat{\kappa},0)} &= \frac{1}{\sqrt{2}}[(f_2)^{abc}_e]^{d,\mu\nu}_{f,\lambda\hat{\kappa}},\\
	[\tilde{F}^{abc}_e]^{d,\mu\nu}_{f,\lambda(\hat{\kappa},1)} &= -\frac{1}{\sqrt{2}}\sum_{\lambda'} (L_{af}^{e})_{\lambda,\lambda'} [(f_2)^{abc}_e]^{d,\mu\nu}_{f,\lambda'\hat{\kappa}}.
\end{align}
The resulting $\tilde{F}^{abc}_e$ is even (because $L_{af}^{e}$ is odd), not necessarily square and in the isometric case satisfies
\begin{equation}\label{eq:isometric1}
\sum_{f,\lambda\kappa} [\bar{\tilde{F}}_e^{abc}]^{d',\mu'\nu'}_{f,\lambda\kappa} [\tilde{F}_e^{abc}]^{d,\mu\nu}_{f,\lambda\kappa} = 
\delta_{d',d}\begin{cases}
\delta_{\mu',\mu}\delta_{\nu',\nu'},&\epsilon_d = 0,\\
\left[\delta_{\mu',\mu} \delta_{\nu',\nu} + Y_{\mu',\mu} (M_{dc}^{e})_{\nu',\nu} \right]/2,&\epsilon_d = 1,\\
\end{cases}
\end{equation}
and
\begin{equation}\label{eq:isometric2}
\sum_{d,\mu,\nu} [\tilde{F}_e^{abc}]^{d,\mu\nu}_{f,\lambda\kappa}[\bar{\tilde{F}}_e^{abc}]^{d,\mu\nu}_{f',\lambda'\kappa'}  = \delta_{f,f'}
\begin{cases}
\delta_{\kappa,\kappa'}\delta_{\lambda,\lambda'},&\epsilon_f = 0,\\
\left[\delta_{\kappa,\kappa'} \delta_{\lambda',\lambda'} + Y_{\kappa,\kappa'} (L_{af}^{e})_{\lambda,\lambda'} \right]/2,&\epsilon_f = 1.\\
\end{cases}
\end{equation}
\end{enumerate}
Fusing the product of four MPOs using these fusion tensors in two different ways gives rise to the super pentagon equation for $\tilde{F}$.

\subsection{Frobenius-Schur indicator}

As a final point on fMPO super algebras, we want to consider the irreducible fMPOs for which $a^*=a$, i.e. the irreducible fMPOs satisfying $\left(O^L_a\right)^\dagger = O^L_a$. It was shown in Ref. ~\cite{MPOpaper} that in the bosonic case one can associate an invariant $\varkappa_a \in \{-1,1\}$ to such MPOs, which coincides with the Frobenius-Schur indicator from fusion categories. In the fermionic case, this invariant has a natural generalization. A crucial observation to obtain the correct generalization is that Hermitian conjugation involves a reordering of the basis vectors for operators that act on the graded tensor product of super vector spaces. Hermitian conjugation is most naturally defined in the following basis, where contraction coincides with matrix multiplication of the components:
\begin{equation}
\left(\sum_{i_1,i_2,j_1,j_2}M_{i_1,i_2,j_1,j_2}|i_1\rangle | i_2\rangle \langle j_2| \langle j_1|\right)^\dagger = \sum_{i_1,i_2,j_1,j_2}\bar{M}_{i_1,i_2,j_1,j_2}|j_1\rangle | j_2\rangle \langle i_2| \langle i_1|\, .
\end{equation}
However, the natural basis in which fMPOs are expressed is of the form $|i_1\rangle\langle j_1|\otimesg |i_2\rangle\langle j_2|$, on which Hermitian conjugation then acts as 

 \begin{equation}
\left(|i_1\rangle\langle j_1|\otimesg |i_2\rangle\langle j_2|\right)^\dagger = (-1)^{(|i_1| + |j_1|)(|i_2|+|j_2|)} |j_1\rangle\langle i_1|\otimesg|j_2\rangle\langle i_2|\,.
\end{equation} 
So Hermitian conjugation does not only result in complex conjugation for the components but also produces additional signs. For this reason it might not be clear at first sight that $\left(O_a^L\right)^\dagger$ is actually also an fMPO. However, the minus sign produced by Hermitian conjugation is the same as the minus sign one gets from reordering of fermion modes under reflection symmetry, and we know this sign can be absorbed in the fMPO tensors by redefining them as $B^{i,j} \rightarrow P^{|i|+|j|}B^{i,j}$ (or equivalently as  $B^{i,j}P^{|i|+|j|}$) \cite{fMPS}, where $P$ is the matrix containing the components of $\mathsf{P}$ as defined earlier. One can check this by explicitly evaluating the redefined fMPO components:

\begin{align}
&\text{tr}(P^{|i_1|+|j_1|}B^{i_1,j_1}\dots P^{|i_{N-1}|+|j_{N-1}|}B^{i_{N-1},j_{N-1}}P^{|i_{N}|+|j_{N}|}B^{i_N,j_N}) \nonumber \\
&= (-1)^{(|i_{N}|+|j_{N}|)(|i_1|+\dots |j_{N-1}|)} \nonumber \\
& \hspace{1.5 cm} \text{tr}(P^{|i_1|+|j_1|+|i_{N}|+|j_{N}|}B^{i_1,j_1}\dots P^{|i_{N_1}|+|j_{N-1}|}B^{i_{N-1},j_{N-1}}B^{i_N,j_N}) \nonumber  \\
&= (-1)^{(|i_{N}|+|j_{N}|)(|i_1|+\dots |j_{N-1}|) + (|i_{N-1}|+|j_{N-1}|)(|i_1|+\dots |j_{N-2}| )} \nonumber \\ 
& \hspace{1.5 cm}\text{tr}(P^{|i_1|+|j_1|+|i_{N}|+|j_{N}| + |i_{N_1}|+|j_{N-1}|}B^{i_1,j_1}\dots B^{i_{N-1},j_{N-1}}B^{i_N,j_N})  \nonumber \\
&= \dots \nonumber \\
&= (-1)^{(|i_{N}|+|j_{N}|)(|i_1|+\dots |j_{N-1}|) + (|i_{N-1}|+|j_{N-1}|)(|i_1|+\dots |j_{N-2}| ) +\dots + (|i_2|+|j_2|)(|i_1|+|j_1|)} \nonumber\\
& \hspace{1.5 cm}\text{tr}(P^{\sum_{\alpha=1}^N (|i_\alpha|+|j_\alpha|)}B^{i_1,j_1}\dots B^{i_{N-1},j_{N-1}}B^{i_N,j_N})\, .
\end{align}
Since we work with anti-periodic boundary conditions all irreducible fMPOs are even so $\sum_{\alpha=1}^N (|i_\alpha| +|j_\alpha|)= 0$ mod 2, which indeed shows that $P^{|i|+|j|}B^{i,j}$ produces the original fMPO with the desired minus sign.

The property $\left(O_a^L\right)^\dagger = O_a^\dagger$ now implies that the matrices of tensor components $B^{i,j}$ satisfy \cite{Cirac}

\begin{equation}
P^{|i|+|j|}\bar{B}_a^{j,i} = Z_a^{-1} B_a^{i,j}Z_a\, ,
\end{equation}
where $Z_a$ is an invertible matrix with parity $\mu_a$. Iterating this relation twice we find

\begin{equation}\label{eq:Za}
(-1)^{\mu_a(|i|+|j|)}B^{i,j}_a = \left(\bar{Z}^{-1}_aZ^{-1}_a \right)B^{i,j}_a\left(Z_a\bar{Z}_a \right)\, .
\end{equation}
If $\epsilon_a = 0$ the center of the algebra spanned by $B^{i,j}_a$ consists only of multiples of the identity. Therefore, if $\mu_a = 0$, we can conclude from \eqref{eq:Za} that $Z_a\bar{Z}_a = \alpha \mathds{1}$ and thus $\bar{Z}_aZ_a = \bar{\alpha}\mathds{1}$, where without loss of generality we can take $\alpha$ to be a phase by rescaling $Z_a$. Combining these two equations gives $\alpha^2 = 1$ and thus  $Z_a\bar{Z}_a = (-1)^{\rho_a} \mathds{1}$, where $\rho_a\in \{0,1\}$. If $\mu_a = 1$, we similarly find that $Z_a\bar{Z}_a = (-1)^{\rho_a}i P$. For $\epsilon_a = 1$ the center of the algebra spanned by $B_a^{i,j}$ contains the odd matrix $Y$, so that both $Z_a$ and $YZ_a$ are valid gauge transformations satisfying \eqref{eq:Za}. This implies that the parity of $Z_a$ is ambiguous and we can take it to be even. In this case we find similarly to the situation with $\epsilon_a =0$ that $Z_a\bar{Z}_a = (-1)^{\hat{\rho}_a}\mathds{1}$. By defining $Z^1_a \equiv YZ_a$ one can obtain another invariant by $Z_a^1\bar{Z}_a^1 = (-1)^{\tilde{\rho}_a}iP$. One can check that these two invariants are independent. The invariant obtained from the odd gauge transformation $Z_aY$, however, is not independent. So in total we have found eight different possibilities. For $\epsilon_a = 0$ we have four possibilities labeled by $\mu_a$ and $\rho_a$. When $\epsilon_a =1$ we also find four possibilies, labeled by $\hat{\rho}_a$ and $\tilde{\rho}_a$. Using similar techniques as for fermionic matrix product states with time reversal symmetry or reflection symmetry one can show that these eight possibilies form a $\mathbb{Z}_8$ group where the group structure corresponds to taking the graded tensor product of fMPOs  \cite{fMPS}. So if we take the invariant $\epsilon_a$ as part of the definition of the Frobenius-Schur indicator we see that it is isomorphic to $\mathbb{Z}_8$ in the fermionic case, while it is only isomorphic to $\mathbb{Z}_2$ in the bosonic case.

\section{Fixed-point PEPS construction}\label{sec:fixedpoint}

In the previous section we extracted the structural data associated to a fMPO super algebra. In this section we will apply a bootstrap method to construct fermionic PEPS and associated fMPOs from this algebraic data. The fMPOs constructed in this way form explicit representations of the fMPO super algebras described in the previous section, and we can construct such a representation for each consistent set of structural data. Imposing two extra conditions on the $\tilde{F}$-symbols ensures that the PEPS and fMPOs satisfy the pulling through identities, which endow the PEPS with non-trivial topological properties. The topological phases described by the tensor networks constructed in this section coincide with the phases captured by fermionic string-nets \cite{Gu,Gu2}  when $\epsilon \equiv 0$.

\subsection{PEPS tensors}\label{sec:pepstensors}

For simplicity we will restrict our construction to the honeycomb lattice. To specify fermionic tensors one does not only have to specify the coefficients, but also in what ordering of the basis vectors these coefficients are defined. For the fermionic PEPS tensors on the A-sublattice we will choose the following internal ordering:

\begin{equation}\label{eq:tensorA}
\includegraphics[width=0.28\textwidth]{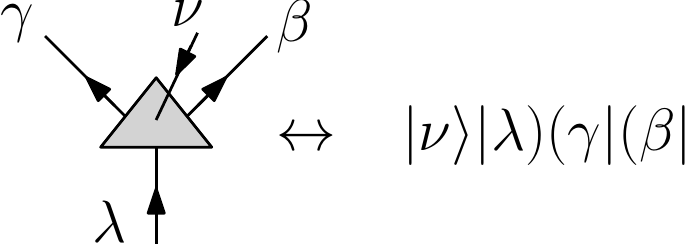}\, ,
\end{equation}
where $\nu$ is the physical index and $\lambda,\gamma,\beta$ are the virtual ones. Note that the arrows in the graphical notation denote which indices correspond to bra's, and which to kets. In the basis just specified, the tensor components are

\begin{equation}\label{eq:tensorAC}
\includegraphics[width=0.77\textwidth]{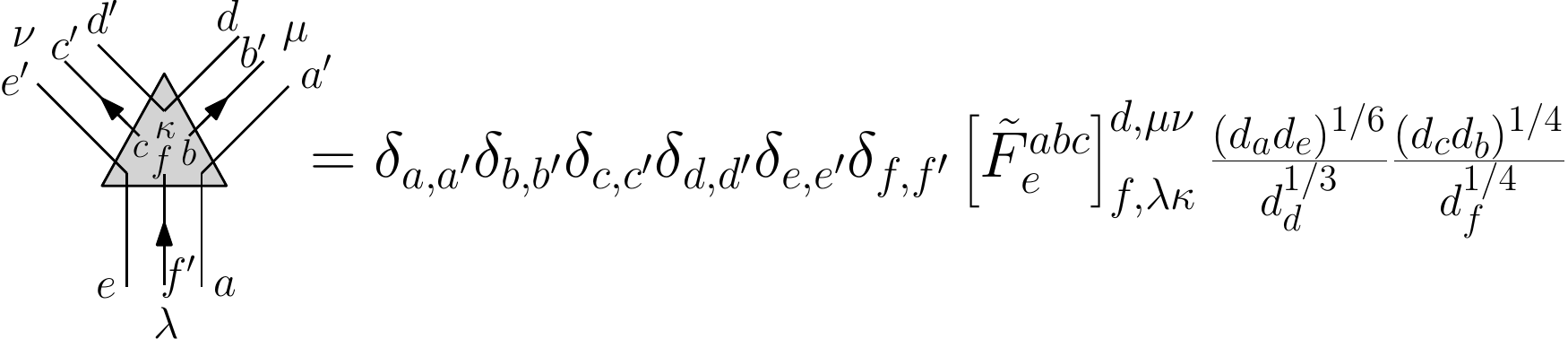}\, .
\end{equation}
This graphical notation requires some explanation. Each index is specified by four labels: three labels are denoted by Latin letters and one label is denoted by a Greek letter, which is also exactly the data that specified a fusion tensor $\mathsf{X^c_{ab,\mu}}$ in the previous section. Each external line in the graphical notation carries a label denoted by a Latin letter. The tensor components are zero when lines that are connected in the body of the tensor carry a different label. This is taken into account by the delta tensors in equation \eqref{eq:tensorAC}. However, in the remainder of this paper these delta conditions will be implicit in our definition of fixed-point tensor components and should be clear from the graphical notation. The physical index is labeled by the three labels carried by the lines that end in the body of the tensor (in the figure these are labels $b,c$ and $f$) and a corresponding Greek label ($\kappa$ in the figure). The possibly non-zero tensor components are given by the $\tilde{F}$-symbols of the previous section, where each of the four tensor indices maps to a fusion tensor that defines the $\tilde{F}$-symbol. The parity of the index also equals the parity of the corresponding fusion tensor. 

The tensors on the B-sublattice are defined with following internal ordering:

\begin{equation}\label{eq:tensorB}
\includegraphics[width=0.32\textwidth]{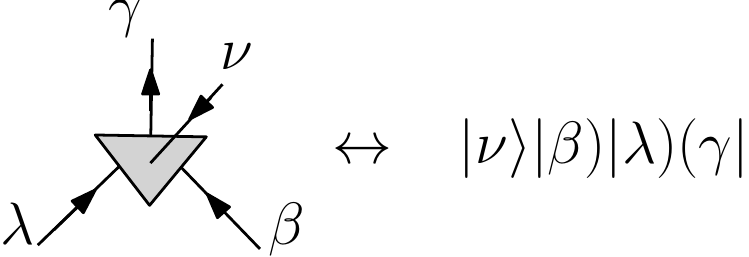}\, .
\end{equation}
And in this basis, the tensor coefficients are analogously represented as

\begin{equation}\label{eq:tensorBC}
\includegraphics[width=0.48\textwidth]{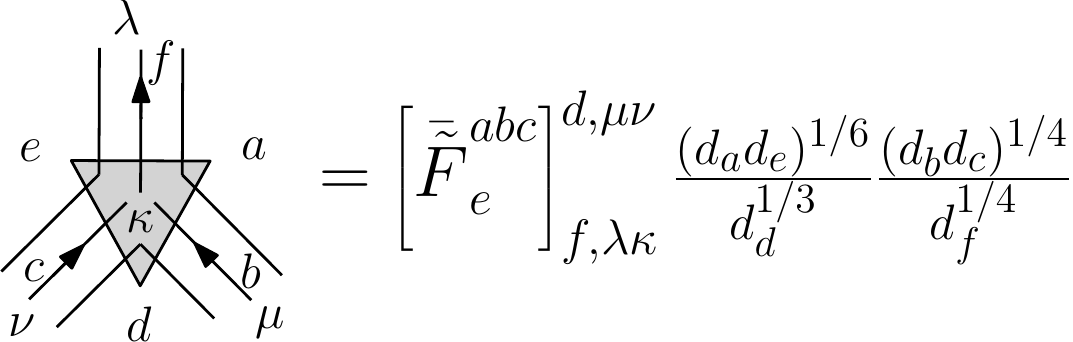}\, ,
\end{equation}
where the bar denotes complex conjugation. All PEPS tensor components are given in terms of $\tilde{F}$ symbols. When $\epsilon\equiv 0$ the $\tilde{F}$-symbols are equivalent to the standard $F$-symbols and the fermionic PEPS is very closely related to the bosonic string-net PEPS. However, when taking Majorana fMPOs into account, the $\tilde{F}$ symbols are a particular choice of associators, and their explicit construction is given in section \ref{sec:fmovepentagon}. 

Let us also comment on the choice of arrows in our definition of the PEPS tensors. Reversing the arrows interchanges bra's with kets and for fermionic PEPS this has a non-trivial effect for the simple reason that $\mathcal{C}((\alpha|\otimesg |\beta)) = \delta_{\alpha,\beta}$ while $\mathcal{C}(|\beta)\otimesg (\alpha|)=(-1)^{|\alpha|}\delta_{\alpha,\beta}$. From this we see that reversing the arrow on a link is equivalent to inserting a parity matrix $P = \mathds{1}_0\oplus -\mathds{1}_1$ on the corresponding virtual index in the contracted network, where $\mathds{1}_0$ ($\mathds{1}_1$) is the identity on the parity even (odd) subspace. So if we would flip all the arrows surrounding a vertex, the three resulting parity matrices on the neighbouring virtual indices can be intertwined to a parity matrix on the physical index since the PEPS tensors are even. This shows that to every fermionic PEPS we can actually associate an entire family of PEPS, that are related to the original one by on-site parity actions, by flipping the arrows surrounding vertices. For this reason, the choice of arrows is very reminiscent of a lattice spin structure.

\subsection{Fermionic pulling through}\label{sec:fpulling}

We will define two types of tensors to construct fMPOs on the virtual level of the fermionic PEPS. The first, right-handed type, defined with the internal ordering

\begin{equation}\label{eq:RH1}
\includegraphics[width=0.3\textwidth]{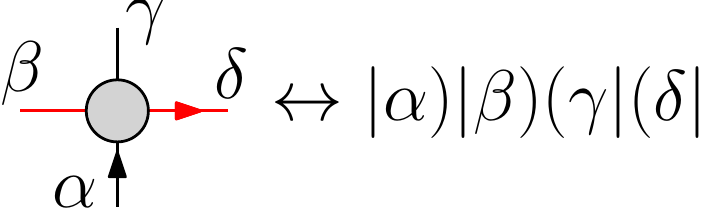}\, ,
\end{equation}
has components which are again determined by the $\tilde{F}$-symbols in the following way:

\begin{equation}\label{eq:RH2}
\includegraphics[width=0.4\textwidth]{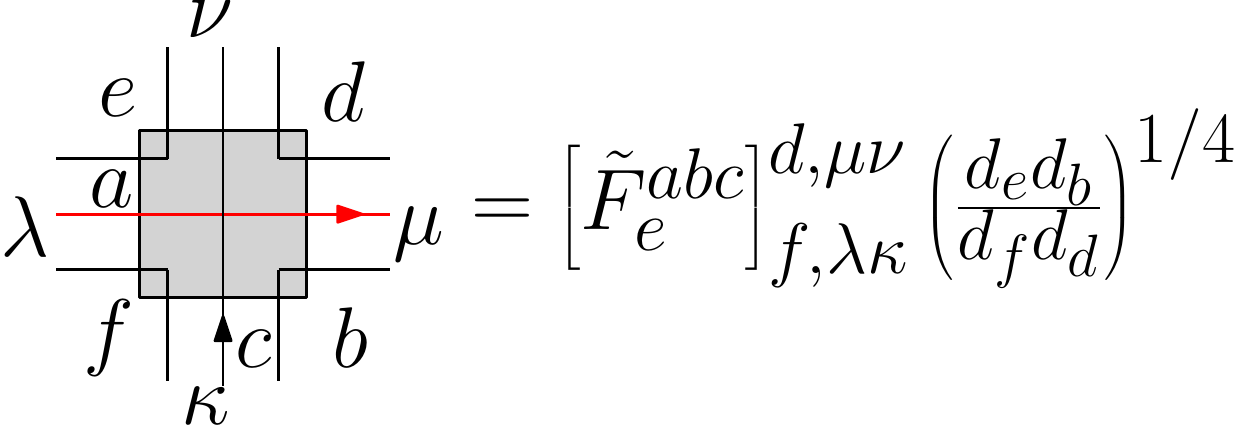}\, .
\end{equation}
In appendix \ref{app:fMPO} we show that the fMPOs constructed from tensors \eqref{eq:RH1}, \eqref{eq:RH2} form an explicit representation of the fMPO algebra whose $\tilde{F}$-symbols we took to define the tensor components. To place the fMPO on the virtual level of the fermionic PEPS we will introduce an additional convention. The closed fMPO should be interpreted as a polygon, i.e. as a closed collection of straight lines and angles between them. On every angle we place a diagonal matrix that inserts some weights, depending on the labels carried by the outer lines. The rule to add the weights is the following: to each label $a$ we associate a positive number $d_a$ (the choice of $d_a$ is not arbitrary as we will see further on) and the weights are then given by $d_a^{\frac{1}{2}(1-\frac{\alpha}{\pi})}$, where $\alpha$ is the inner (outer) angle in radians for the inner (outer) line. For example, when the fMPO contains an angle of $\frac{2\pi}{3}$ the weights are:

\begin{equation}
\includegraphics[width=0.35\textwidth]{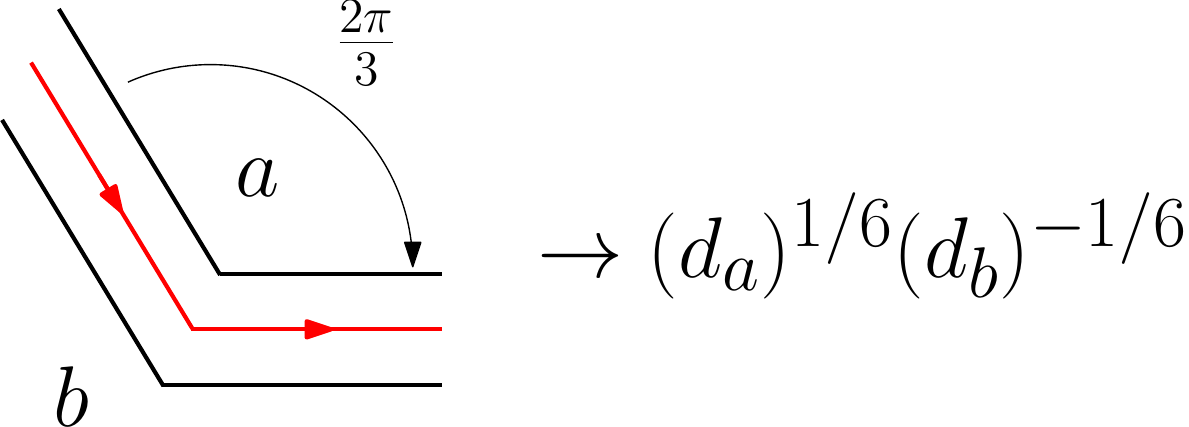}\, .
\end{equation}
For notational simplicity this convention will always be implicit in our graphical notation from now on. 

The reason to define the right-handed fMPO tensors as in (\ref{eq:RH1},\ref{eq:RH2}) is that the pentagon equation now implies that the following pulling through identity holds:

\begin{equation}\label{eq:pullingthroughRRR}
\includegraphics[width=0.33\textwidth]{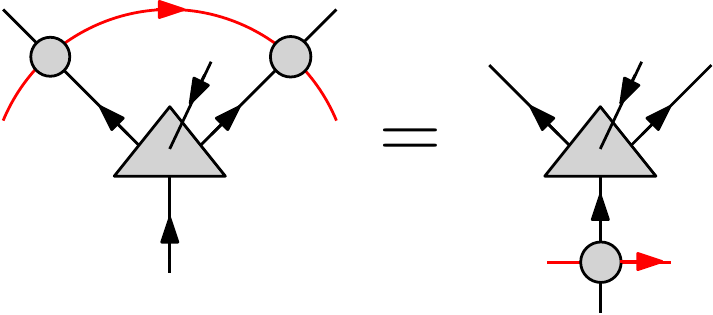}\, .
\end{equation}
Note that equation \eqref{eq:pullingthroughRRR} is only equivalent to the pentagon equation when we use the $\tilde{F}$-symbols in defining the tensor components. The underlying reason is as follows. Every index of the fixed point tensors coresponds to a fusion tensor, and the four fusion tensors from every index in a tensor together correspond to an $F$ move whose $\tilde{F}$-symbol determines the tensor component. Since the indices are defined in a super vector space, an even and an odd vector are necessarily orthogonal. However, as explained in section \ref{sec:superalgebras}, when $\epsilon_c=1$, the even and odd version of the fusion tensor $\mathsf{X}_{ab,\mu}^c$ correspond to the same fusion channel. Because of this, equation \eqref{eq:pullingthroughRRR} would only be equal to the pentagon equation up to factors of two when the tensors are defined in terms of the $F$ symbols.

Let us now define the second, left-handed, type of fMPO tensor with the internal ordering

\begin{equation}\label{eq:LH1}
\includegraphics[width=0.3\textwidth]{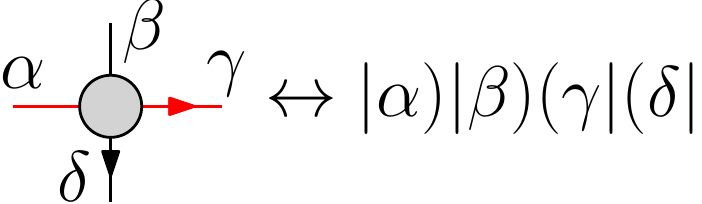}\, ,
\end{equation}
and components

\begin{equation}\label{eq:LH2}
\includegraphics[width=0.4\textwidth]{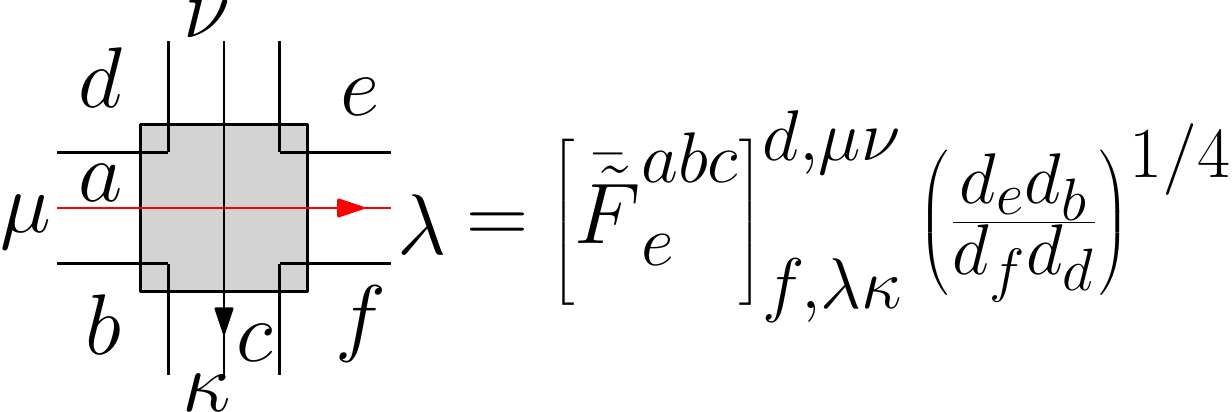}\, .
\end{equation}
We will now restrict to $\tilde{F}$-symbols that are unitary or isometric matrices, i.e. $\tilde{F}$-symbols that satisfy equations \eqref{eq:isometric1} and \eqref{eq:isometric2}, which we restate here for convenience:

\begin{equation}
\sum_{f,\lambda\kappa} [\bar{\tilde{F}}_e^{abc}]^{d',\mu'\nu'}_{f,\lambda\kappa} [\tilde{F}_e^{abc}]^{d,\mu\nu}_{f,\lambda\kappa} = 
\delta_{d',d}\begin{cases}
\delta_{\mu',\mu}\delta_{\nu',\nu'},&\epsilon_d = 0,\\
\left[\delta_{\mu',\mu} \delta_{\nu',\nu} + Y_{\mu',\mu} (M_{dc}^{e})_{\nu',\nu} \right]/2,&\epsilon_d = 1,\\
\end{cases}
\end{equation}
\begin{equation}
\sum_{d,\mu,\nu} [\tilde{F}_e^{abc}]^{d,\mu\nu}_{f,\lambda\kappa}[\bar{\tilde{F}}_e^{abc}]^{d,\mu\nu}_{f',\lambda'\kappa'}  = \delta_{f,f'}
\begin{cases}
\delta_{\kappa,\kappa'}\delta_{\lambda,\lambda'},&\epsilon_f = 0,\\
\left[\delta_{\kappa,\kappa'} \delta_{\lambda',\lambda'} + Y_{\kappa,\kappa'} (L_{af}^{e})_{\lambda,\lambda'} \right]/2,&\epsilon_f = 1.\\
\end{cases}
\end{equation}
In this case, one sees that with our definition of the left-handed fMPO tensors the following properties are satisfied

\begin{equation}\label{eq:unitary}
\includegraphics[width=0.5\textwidth]{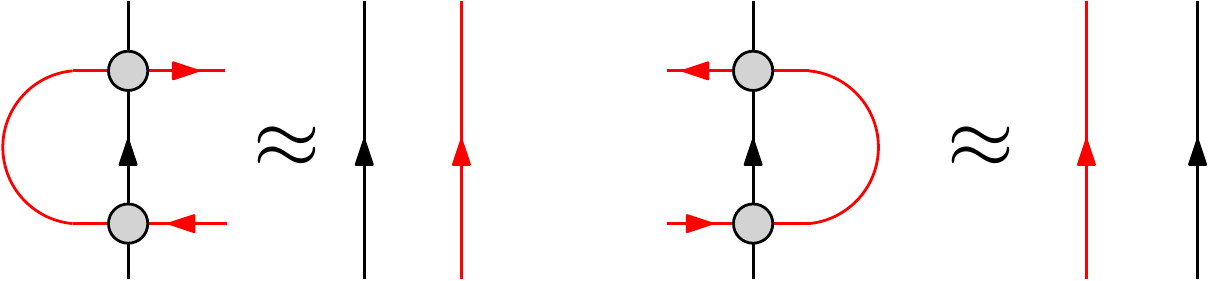}\, ,
\end{equation}
where we used approximate equality to denote that these are not strict tensor identities, but are only satisfied on the relevant subspaces. In other words, these identities should only hold when the fMPO is embedded within the fermionic PEPS. One can check that this is indeed the case for the fMPOs and fermionic PEPS just defined. As a final step, we require that the $\tilde{F}$-symbols satisfy

\begin{equation}\label{eq:pivotal2}
\left[\tilde{F}^{abc}_e\right]^{d,\mu\nu}_{f,\lambda\sigma}\frac{\sqrt{d_ed_b}}{\sqrt{d_fd_d}} = \left[\tilde{F}^{adc}_f \right]^{b,\mu\sigma}_{e,\lambda\nu}\frac{\theta^{ac,\mu}_b\theta^{ac,\lambda}_e}{\theta^{ac,\mu}_d\theta^{ac,\lambda}_f}\frac{\theta^{ac}_\sigma}{\theta^{ac}_\nu}t^{ac}_\mu s^{ac}_\lambda\, ,
\end{equation}
where $\theta\in$ U(1) and $t,s \in \{1,-1\}$. It is this condition that fixes the positive numbers $d_a$. Eq. \eqref{eq:pivotal2} is a generalization of the pivotal property for bosonic fusion categories, which together with the isometric property implies that the fMPOs also satisfy following properties:

\begin{equation}\label{eq:pivotal}
\includegraphics[width=0.5\textwidth]{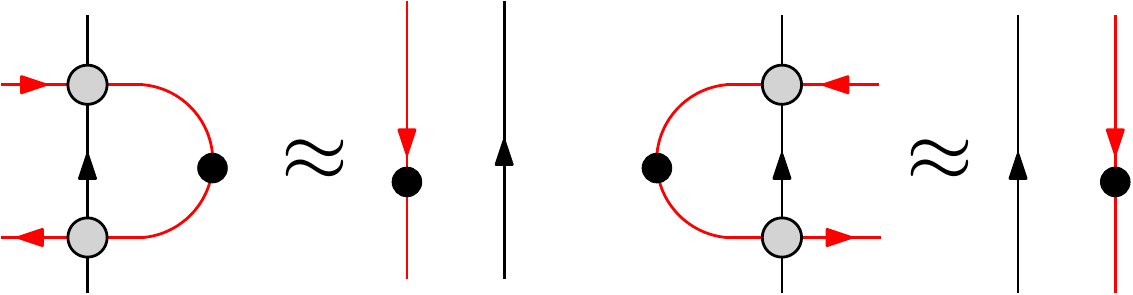}\, ,
\end{equation}
where the black dot is a graphical notation for the parity matrix $\mathsf{P} = \sum_\alpha (-1)^{|\alpha|}|\alpha)(\alpha|$. The reason for requiring unitarity and a generalization of the pivotal property is that from the pulling through identity \eqref{eq:pullingthroughRRR} we can now derive the complete set of pulling through identities for the A-sublattice:

\begin{equation}\label{eq:pullingthroughA}
\includegraphics[width=0.7\textwidth]{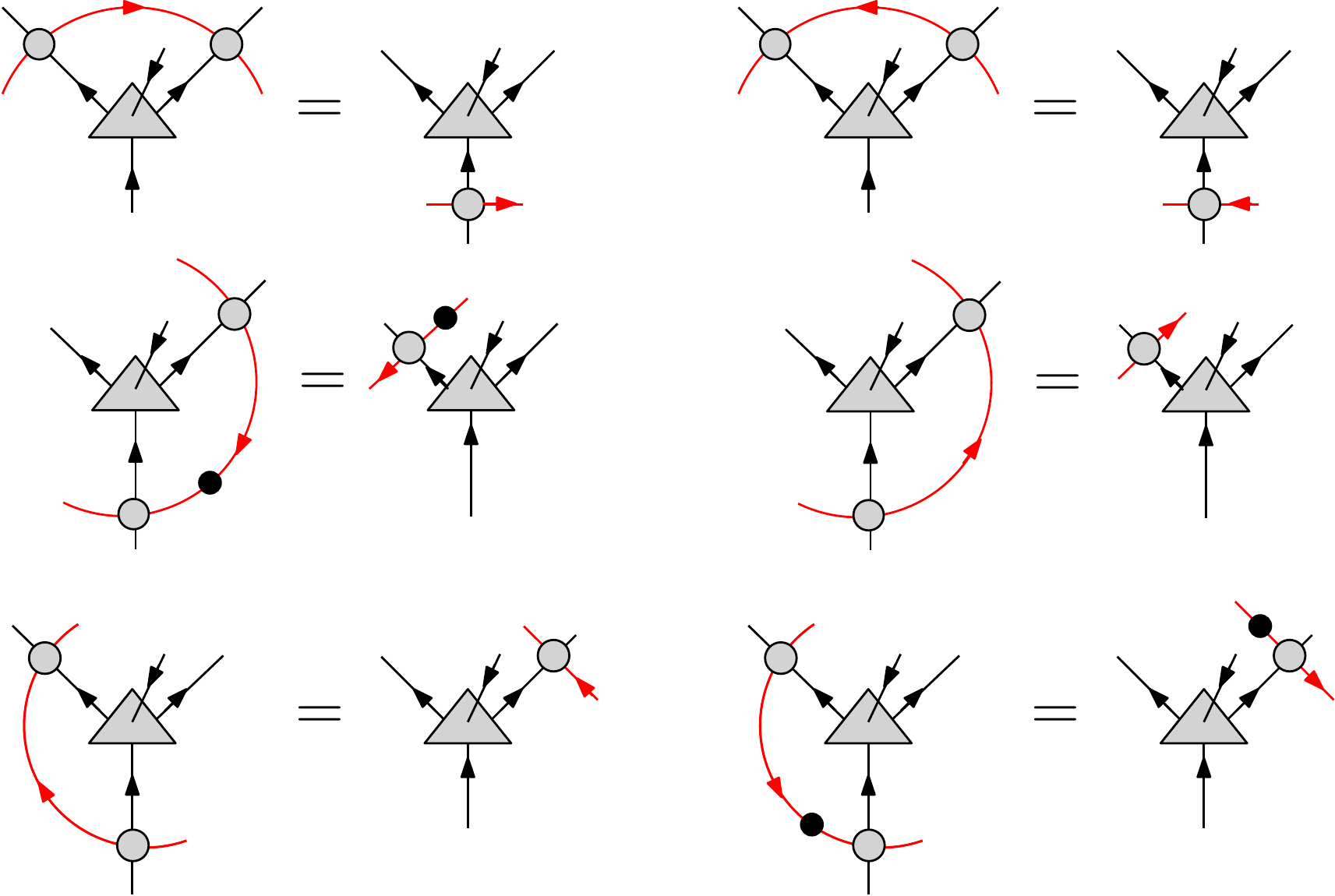}
\end{equation}
In a similar way one can derive the pulling through identities for the B-sublattice:

\begin{equation}\label{eq:pullingthroughB}
\includegraphics[width=0.7\textwidth]{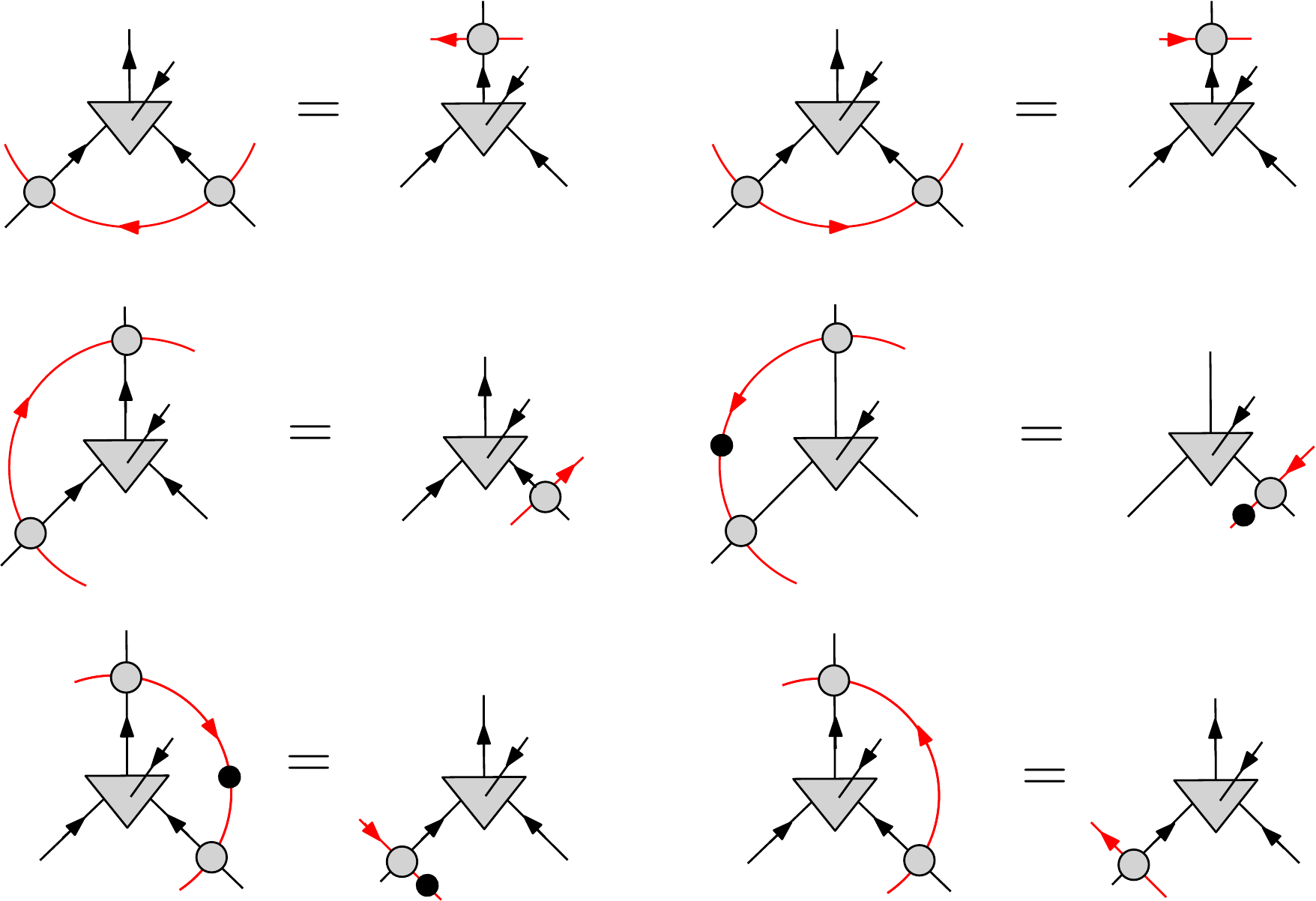}\, ,
\end{equation}
where the identity in the top left corner follows from the (complex conjugate of the) super pentagon equation, and all other identities can be derived from this one using properties \eqref{eq:unitary} and \eqref{eq:pivotal}. Note that the pulling through identities (\ref{eq:pullingthroughA}) and (\ref{eq:pullingthroughB}) imply that closed fMPOs on the virtual level of the PEPS contain parity matrices on their internal indices.  They encode the rules of how these parity matrices move or change in their total number by a multiple of two when the fMPO moves through the PEPS tensors. One can check that these rules completely determine the position of the parity matrices on every closed fMPO and imply that their number is always odd for every fMPO along a contractible cycle. This implies that our formalism survives an important consistency check. In Ref. \cite{fMPS} is was explained that an $\epsilon=1$ fMPO evaluates to zero when it is closed with an even number of parity matrices $\mathsf{P}$ inserted on its internal indices; in particular we cannot close it without inserting any parity matrix. But we just argued that the pulling through identities imply that every fMPO along a contractible cycle contains an odd number of parity matrices, thus preventing the fermionic PEPS with Majorana symmetry fMPOs from contracting to the zero vector. fMPOs along non-contractible cycles require a more detailed analysis. We will come back to this point in section \ref{sec:spinstructurestorus}.

The tensor networks we have constructed here involve a particular choice of spin structure. Apart from the spin structures related by flipping arrows around a vertex, there are still many more choices one can make. However, not all of them will be consistent with the pulling through identities \eqref{eq:pullingthroughA} and \eqref{eq:pullingthroughB}, in the sense that these local identities will not imply that the fMPOs can be moved freely through the entire tensor network. All spin structures we have found to be compatible are of the Kasteleyn type \cite{ware,Tarantino}, which means that when going around a plaquette in a particular direction the number of arrows on the edges bounding that plaquette pointing in the opposite direction is odd.

In this section we have constructed fermionic PEPS tensors on the honeycomb lattice and fMPO tensors, both right- and left-handed, such that the pulling through identities hold. The pulling through identities are a fingerprint of non-trivial topological order in PEPS, which can --for example-- be seen by defining the fermionic PEPS on a torus. In this situation, one can place fMPOs on the virtual level along non-contractible cycles. This will lead to PEPS that are locally indistuinguishable from each other, since the fMPOs can move freely on the virtual level. This results in a topological ground state degeneracy.

\section{Gu-Wen symmetry-protected phases}\label{sec:guwen}

Up to this point we have studied fMPO super algebras to construct fermionic tensor networks that have non-trivial topological order. But as explained in \cite{Chen2,Williamson} fMPO group representations $\{O^L_g|\,g\in G \}$ are also relevant for symmetry-protected topological (SPT) phases. In this section we will restrict to the case $\epsilon_g = 0, \forall g\in G$. We again work on the honeycomb lattice, and the SPT PEPS tensors on the A-sublattice are

\begin{equation}\label{eq:guwentensor}
\includegraphics[width=0.55\textwidth]{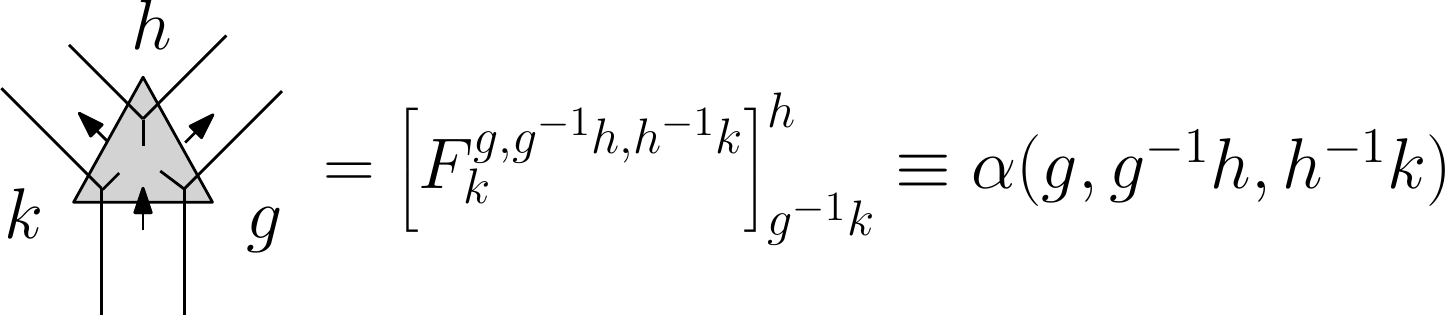}
\end{equation}
Note that this is a modified version of the PEPS tensor \eqref{eq:tensorAC} defined previously; the only difference is that we left out the middle label in the virtual indices since it is redundant in the group case and the virtual labels $g,h$ and $k$  now get copied to the physical index. The internal ordering is the same as defined in \eqref{eq:tensorA}. To completely specify this tensor we also have to specify the grading, i.e. we have to specify the parity of the basis vectors. We do this by defining a function $Z:G\times G \rightarrow \{0,1\}$. The parities of the virtual indices are then given by $Z(g,g^{-1}k)$, $Z(g,g^{-1}h)$, $Z(h,h^{-1}k)$ and the parity of the physical index is given by $Z(g^{-1}h,h^{-1}k)$. Requiring the PEPS tensor to be even implies that $Z(g,h)$ is a 2-cocycle. The tensors for the B-sublattice are obtained via a similar modification of the tensor defined in \eqref{eq:tensorB},\eqref{eq:tensorBC}. 

For fMPO group representations with $\epsilon_g \equiv 0$, the super pentagon relation can be expressed in terms of the $\alpha(g_1,g_2,g_3)$ as

\begin{equation}\label{eq:supercocycle}
\frac{\alpha(g_1,g_2,g_3)\alpha(g_1,g_2g_3,g_4)\alpha(g_2,g_3,g_4)}{\alpha(g_1g_2,g_3,g_4)\alpha(g_1,g_2,g_3g_4)} = (-1)^{Z(g_1,g_2)Z(g_3,g_4)}\, ,
\end{equation}
which is the supercocycle relation as defined previously by Gu and Wen to construct fermionic SPT phases \cite{supercohomology}. From the supercoycle relation it follows that a left-regular symmetry action on the physical indices gets intertwined to a virtual fMPO symmetry action on the virtual indices, where the fMPO is constructed from the tensors

\begin{equation}\label{eq:fmpoguwenR}
\includegraphics[width=0.33\textwidth]{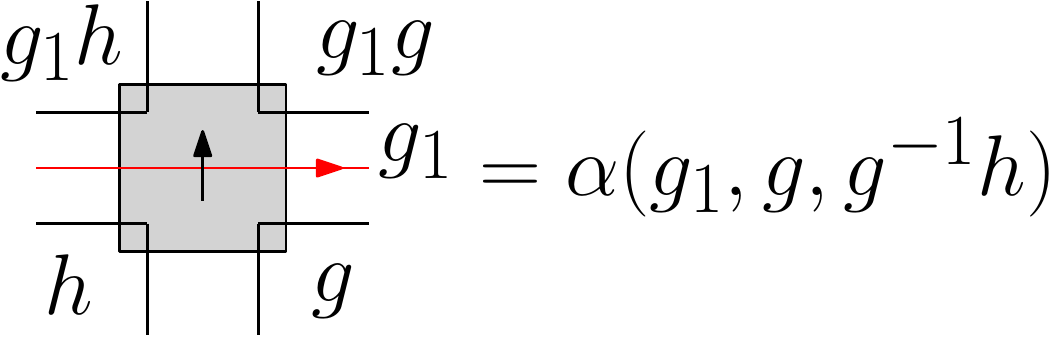}
\end{equation}
and 

\begin{equation}\label{eq:fmpoguwenL}
\includegraphics[width=0.35\textwidth]{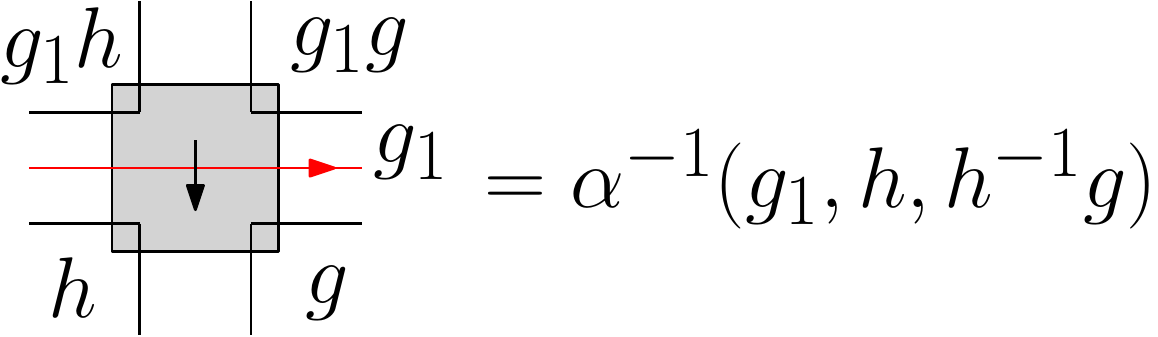}\, .
\end{equation}
The parities of the indices of the right-handed fMPO tensor are $Z(g,g^{-1}h)$, $Z(g_1,g)$, $Z(g_1g,g^{-1}h)$ and $Z(g_1,h)$. The parities of the left-handed tensor are $Z(g_1h, h^{-1}g)$, $Z(g_1,h)$, $Z(h,h^{-1}g)$ and $Z(g_1,g)$. Evenness of both tensors again follows from the fact that $Z(g,h)$ is a 2-cocycle. The internal ordering of the fMPO tensors is the same as in \eqref{eq:RH1} and \eqref{eq:LH1}

The intertwining property of the PEPS tensors \eqref{eq:guwentensor} implies that the resulting short-range entangled tensor network  has a global symmetry $G$, which contains fermion parity in its center. For more details on PEPS with a global symmetry that is realized on the virtual level by MPOs we refer to Ref. \cite{Williamson}. It was shown in Ref. \cite{Williamson} that the topologically ordered PEPS discussed in the previous section can be obtained from the SPT PEPS by gauging this global symmetry \cite{gaugingpaper}. We note that fermionic tensor networks using Grassmann variables for the gauged models were constructed in Ref.\cite{Wille}. 

The fMPOs constructed from the tensors \eqref{eq:fmpoguwenR} and \eqref{eq:fmpoguwenL} have the property that $O_{g}^\dagger = O_{g^{-1}}$. So to group elements $g_1$ satifying $g_1^2 = e$, where $e$ is the identity group element, we can associate a Frobenius-Schur indicator as defined in the general theory of fMPO super algebras in section \ref{sec:superalgebras}. Again using the supercocycle relation one finds that $Z_{g_1}$ is given by

\begin{equation}
\includegraphics[width=0.55\textwidth]{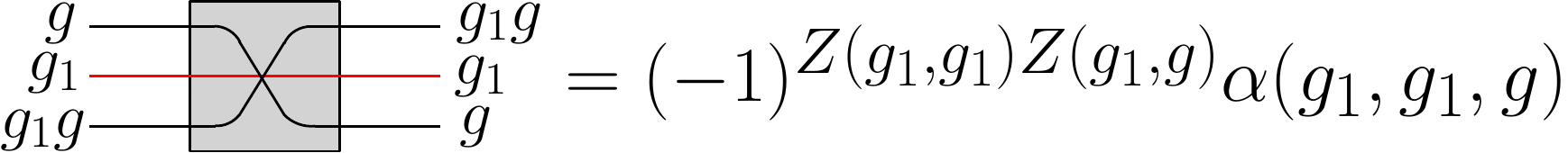}\, ,
\end{equation}
where without loss of generality we have taken representative cocycles satisfying $\alpha(e,g,h) = 1$ and $Z(e,g)=0$ \footnote{This form can always be obtained by the coboundary rescaling $\alpha(g,h,k)\rightarrow \alpha'(g,h,k) = \alpha(g,h,k)\frac{\beta(g,h)\beta(gh,k)}{\beta(g,hk)\beta(h,k)}$ with $\beta(e,g) = \alpha^{-1}(e,g,g^{-1})$.}. The parity of $Z_{g_1}$ is $Z(g_1,g_1g) + Z(g_1,g) = Z(g_1,g_1)$ mod 2 (since $Z(e,g) = 0$). If $Z(g_1,g_1) = 0$ one can verify that

\begin{equation}
Z_{g_1}\bar{Z}_{g_1} = \alpha^{-1}(g_1,g_1,g_1)\mathds{1}\, ,
\end{equation}
while if $Z(g_1,g_1) =1$ it holds that

\begin{equation}
Z_{g_1}\bar{Z}_{g_1} = \alpha^{-1}(g_1,g_1,g_1)P\, .
\end{equation}
Since the super cocycle relation implies that $\alpha(g_1,g_1,g_1)^2 = (-1)^{Z(g_1,g_1)}$, these results are indeed compatible with the general theory of the Frobenius-Schur indicator discussed in section \ref{sec:superalgebras}.

\subsection{Group structure}

We define the fusion tensors $\mathsf{X_{g_2,g_1}}$ associated to the fMPO group representation constructed from tensors \eqref{eq:fmpoguwenR} and \eqref{eq:fmpoguwenL} with components

\begin{equation}\label{eq:guwenfusiontensor1}
\includegraphics[width=0.36\textwidth]{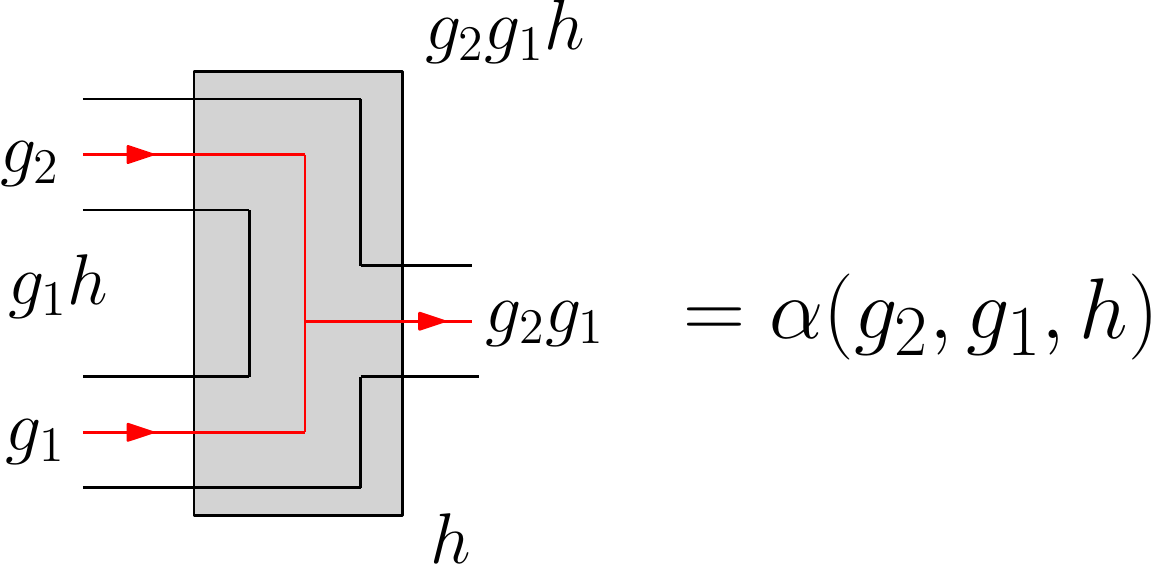}\, .
\end{equation}
in the basis 

\begin{equation}\label{eq:guwenfusiontensor2}
\includegraphics[width=0.28\textwidth]{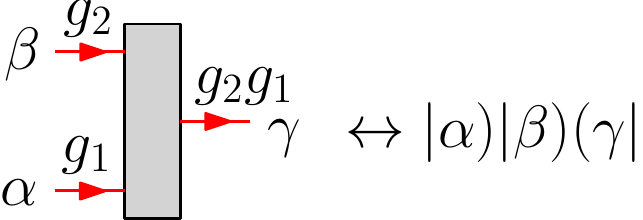}\, .
\end{equation}
Note that the parity of this fusion tensor is $Z(g_1,h) + Z(g_2,g_1h) + Z(g_2g_1,h) = Z(g_2,g_1)$ mod 2. At this point we would like to note that the parity of the internal fMPO indices has no physical value, we could as well interchange even with odd for the internal fMPO indices for any of the $O_g$. If we denote with $x(g) \in \{0,1\}$ whether or not we have interchanged even and odd for the fMPO $O_g$, then the parity of the fusion tensors changes as $Z(g_2,g_1) \rightarrow Z(g_2,g_1)+x(g_2) +x(g_1)+x(g_2g_1)$ mod 2. So we see that the only invariant information associated to the fMPO is the second cohomology class $H^2(G,\mathbb{Z})$ represented by $Z(g_2,g_1)$. One can also check that PEPS constructed from different $Z(g_2,g_1)$ in the same cohomology class are equivalent in the following way: after taking the tensor product with product states such that the local physical super vector spaces are the same, there exists a strictly on-site unitary that maps one PEPS to the other and intertwines both left regular symmetry actions. Similarly to the bosonic case,  multiplying $\mathsf{X_{g_2,g_1}}$ with the phase $\gamma(g_2,g_1)$ changes $\alpha(g_3,g_2,g_1)$ by a coboundary $\gamma(g_3,g_2)\gamma(g_3g_2,g_1)\bar{\gamma}(g_2,g_1)\bar{\gamma}(g_3,g_2g_1)$. This implies that only $\alpha(g_3,g_2,g_1)$ modulo coboundaries contains invariant information.

The super cocycle relation \eqref{eq:supercocycle} implies that the fusion tensors defined above indeed satisfy the zipper condition:

\begin{equation}
\includegraphics[width=0.35\textwidth]{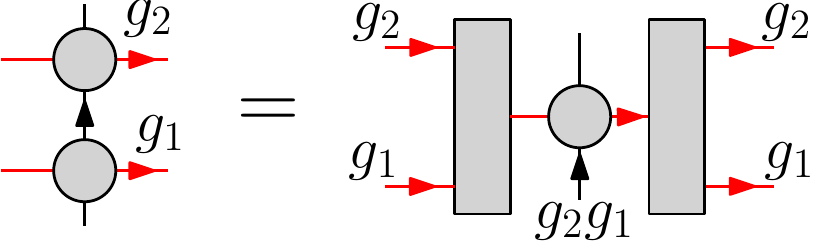}\, .
\end{equation}
Again applying the super cocycle relation shows that the $F$-move for these fusion tensors produces the super cocycle that we used to construct the PEPS:

\begin{equation}\label{eq:fmovesupercocycle}
\includegraphics[width=0.52\textwidth]{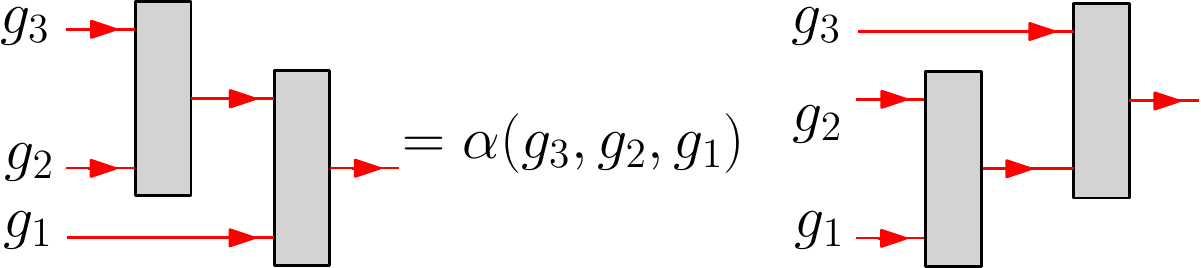}\, .
\end{equation}
This $F$-move is written down as an equation in the following way

\begin{equation}
\mathcal{C}\left(\mathsf{X_{g_3,g_2}}\otimesg \mathsf{X_{g_3g_2,g_1}} \right) =
\alpha(g_3,g_2,g_1) \mathcal{C}\left(\mathsf{X_{g_2,g_1}} \otimesg \mathsf{X_{g_3,g_2g_1}} \right)\, ,
\end{equation}
where $\mathcal{C}$ represents the proper fermionic contraction as depicted in \eqref{eq:fmovesupercocycle}.

In the previous section we showed that the Frobenius-Schur indicator associated to an $\epsilon\equiv 0$ fMPO group representation is completely fixed by the supercocycle. The invariant algebraic data associated to the fMPO representation is therefore given by $Z(g_2,g_1)$ and $\alpha(g_3,g_2,g_1)$. Since the fMPO describes all possible anomalous symmetry actions on the boundary of the two-dimensional system, this data should directly classify the SPT phase of the short-range entangled bulk. Let us now ask the question of what happens to this data when we stack different SPT phases, i.e. when we take the graded tensor product of PEPS with the same global symmetry. It is clear that the stacked PEPS has a virtual symmetry given by the graded tensor product of the original fMPO representations. The fusion tensor of the graded tensor product of two fMPOs is also just the graded tensor product of the individual fusion tensors, which we denote by $\mathsf{X_{g_2,g_1}^1}$ and $\mathsf{X_{g_2,g_1}^2}$. Using the rules of fermionic contraction with super vector spaces we can now easily obtain the supercocycle for the stacked PEPS by evaluating the $F$-move for $\mathsf{X_{g_2,g_1}^1}\otimesg \mathsf{X_{g_2,g_1}^2}$: 

\begin{eqnarray}
 & & \mathcal{C}\left[\left( \mathsf{X_{g_3,g_2}^1}\otimesg \mathsf{X_{g_3,g_2}^2}\right)\otimesg \left(\mathsf{X_{g_3g_2,g_1}^1}\otimesg \mathsf{X_{g_3g_2,g_1}^2}\right)\right] \nonumber\\
 & = &  (-1)^{Z_1(g_3g_2,g_1)Z_2(g_3,g_2)}\mathcal{C}\left[\mathsf{X_{g_3,g_2}^1}\otimesg \mathsf{X_{g_3g_2,g_1}^1}\otimesg \mathsf{X_{g_3,g_2}^2}\otimesg \mathsf{X_{g_3g_2,g_1}^2}\right] \nonumber\\
& = & (-1)^{Z_1(g_3g_2,g_1)Z_2(g_3,g_2)}\alpha_1(g_3,g_2,g_1)\alpha_2(g_3,g_2,g_1) \nonumber \\
 & & \mathcal{C}\left[\mathsf{X_{g_2,g_1}^1}\otimesg \mathsf{X_{g_3,g_2g_1}^1}\otimesg\mathsf{X_{g_2,g_1}^2}\otimesg \mathsf{X_{g_3,g_2g_1}^2}\right] \nonumber\\
& = & (-1)^{Z_1(g_3g_2,g_1)Z_2(g_3,g_2)+Z_1(g_3,g_2g_1)Z_2(g_2,g_1)}\alpha_1(g_3,g_2,g_1)\alpha_2(g_3,g_2,g_1) \nonumber \\
 & & \mathcal{C}\left[\left(\mathsf{X_{g_2,g_1}^1}\otimesg\mathsf{X_{g_2,g_1}^2}\right)\otimesg\left( \mathsf{X_{g_3,g_2g_1}^1}\otimesg\mathsf{X_{g_3,g_2g_1}^2}\right)\right] \, .
\end{eqnarray}
The parity of $\mathsf{X_{g_2,g_1}^1}\otimesg \mathsf{X_{g_2,g_1}^2}$ is of course just given by $Z_1(g_2,g_1) + Z_2(g_2,g_1)$. We therefore find that the stacked SPT PEPS is described by the following algebraic data:

\begin{eqnarray}
\tilde{\alpha}(g_3,g_2,g_1)  & = & (-1)^{Z_1(g_3g_2,g_1)Z_2(g_3,g_2)+Z_1(g_3,g_2g_1)Z_2(g_2,g_1)}\alpha_1(g_3,g_2,g_1)\alpha_2(g_3,g_2,g_1) \nonumber \\
\tilde{Z}(g_2,g_1) & = & Z_1(g_2,g_1) + Z_2(g_2,g_1)\;\;\;\text{ mod } 2\, .
\end{eqnarray}
This shows how the algebraic data changes under stacking and allows one to calculate the group structure of Gu-Wen SPT phases.

\subsection{Projective transformation of symmetry defects}

One of the characterizing physical properties of SPT phases is that symmetry defects can carry fractional quantum numbers. In this section we will discuss how the projective nature of defects in Gu-Wen phases is derived from the defining algebraic data $Z(g_2,g_2)$ and $\alpha(g_3,g_2,g_1)$.

\subsubsection{$\pi$-flux defects}

\begin{figure}
  \centering
    \includegraphics[width=0.4\textwidth]{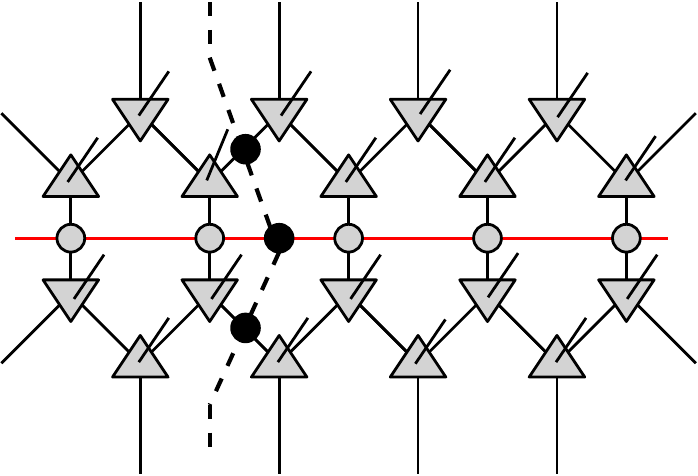}
  \caption{Section of a path on the dual lattice and the associated position of parity matrices (represented by the black dots) in the PEPS.}\label{fig:dualpath}
\end{figure}

In section \ref{sec:fixedpoint} we explained how the fixed-point PEPS obtained via the bootstrap method incorporate a lattice spin structure. Different spin structures can be obtained by choosing a closed path on the dual lattice and putting a parity matrix $\mathsf{P}$ on every virtual index that crosses this cut. It is important to note that internal fMPO indices crossing the path should also gain a parity matrix. In figure \ref{fig:dualpath} we show a part of such a path and the associated parity matrices in the PEPS. 

If we now choose an open path on the dual lattice and again insert parity matrices on links that cross the path we have created $\pi$-flux defects on the plaquettes where the path ends. Symmetry fMPOs on the virtual level of the PEPS that encircle one of these $\pi$-flux defects contain an even number of parity matrices on their internal indices. One can verify that group fMPOs $O_g$ as constructed above satisfy 

\begin{equation}
\tilde{O}_g\tilde{O}_h = (-1)^{Z(g,h)}\tilde{O}_{gh}\, ,
\end{equation}
where the tilde denotes the fact that the fMPOs contain an even number of parity matrices. This gives an explicit physical interpretation to $Z(g,h)$: it is the projective representation under which $\pi$-flux defects transform.

\begin{figure}
  \centering
    \includegraphics[width=0.35\textwidth]{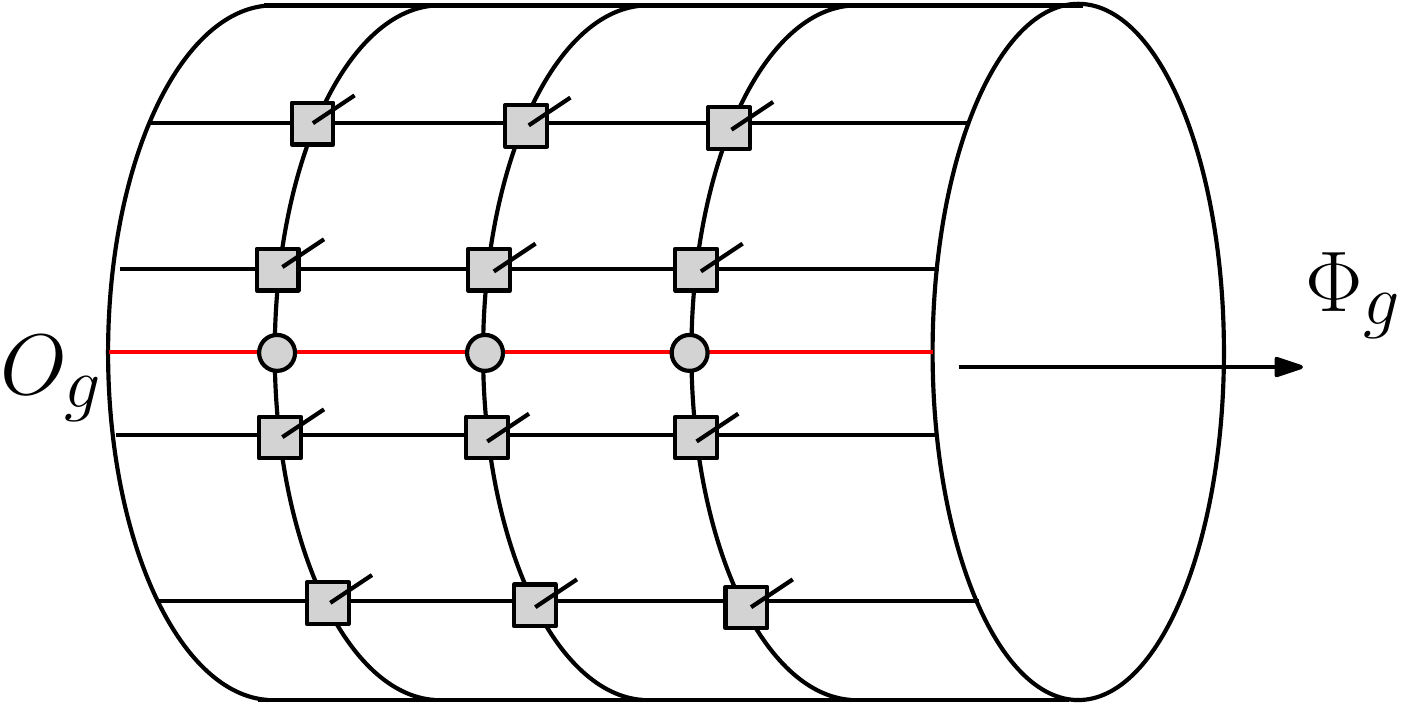}
  \caption{A PEPS on the cylinder with a flux $\Phi_g$ through the hole, or equivalently, with boundary conditions twisted by $g$ along the periodic direction. The flux (or twisted boundary conditions) is realized by placing the fMPO $O_g$ on the virtual level of the tensor network.}\label{fig:flux}
\end{figure}

A convenient way to think about symmetry defects is the following: if we put the PEPS on a cylinder and we twist the boundary conditions along the periodic direction by the group element $g$, then there are symmetry defects at both ends of the cylinder. This boils down to simply placing a fMPO $O_g$ on the virtual level going from one end of the cylinder to the other. Figure \ref{fig:flux} contains a graphical representation of this situation. There is one subtlety if we apply this reasoning to $\pi$-flux defects. Let us consider the PEPS on the cylinder with periodic boundary conditions along the periodic direction. In this case a fMPO wrapping the non-contractible cycle will contain an even number of parity matrices on its internal indices. But as explained above, in this case the fMPOs form a projective representation. We can also define the PEPS with anti-periodic boundary conditions by choosing a path on the dual lattice extending from one end of the cylinder to the other and again inserting the appropriate parity matrices. Now the fMPOs wrapping the cylinder contain an odd number of parity matrices and form a non-projective representation. This shows that the PEPS with periodic boundary conditions should be intepreted as having a $\pi$-flux through the cylinder. The cylinder with anti-periodic boundary conditions contains no flux and can in principle be `capped off' to a sphere. This is the tensor network analogue of the fact that the Neveu-Schwarz spin structure on the circle can be extended to the unique spin structure on a disc, while the Ramond spin structure does not have this property.

\subsubsection{General defects}\label{sec:generaldefects}

To study general symmetry defects we have to use a second type of fusion tensor, which in the basis

\begin{equation}
\includegraphics[width=0.3\textwidth]{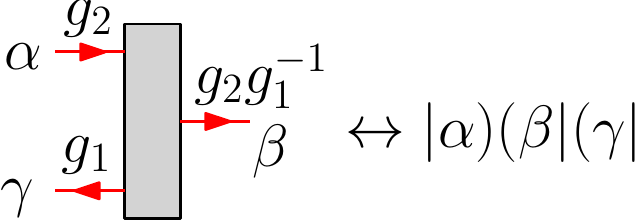}\, ,
\end{equation}
has following components:

\begin{equation}
\includegraphics[width=0.55\textwidth]{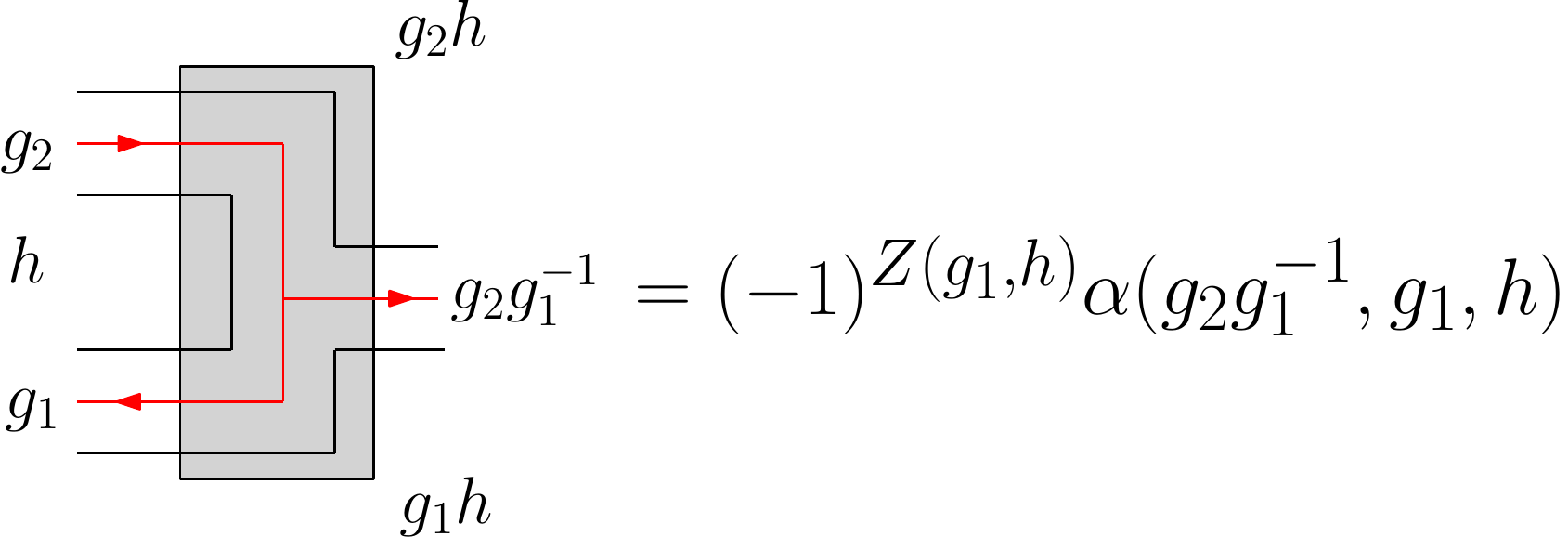}\, .
\end{equation}
This second type of fusion tensor can be obtained by reducing a right-handed and a left-handed fMPO tensor to a right-handed one. We now again consider the cylinder with boundary conditions twisted by $g$ as in figure \ref{fig:flux}. We also impose anti-periodic boundary conditions such that there is no $\pi$-flux through the cylinder. One can check that the physical symmetry action of elements in $\mathcal{Z}_g$, the center of $g$, gets intertwined to an action on the left virtual indices and the right virtual indices. The action on the left virtual indices is given by the following fMPO:

\begin{equation}
\includegraphics[width=0.13\textwidth]{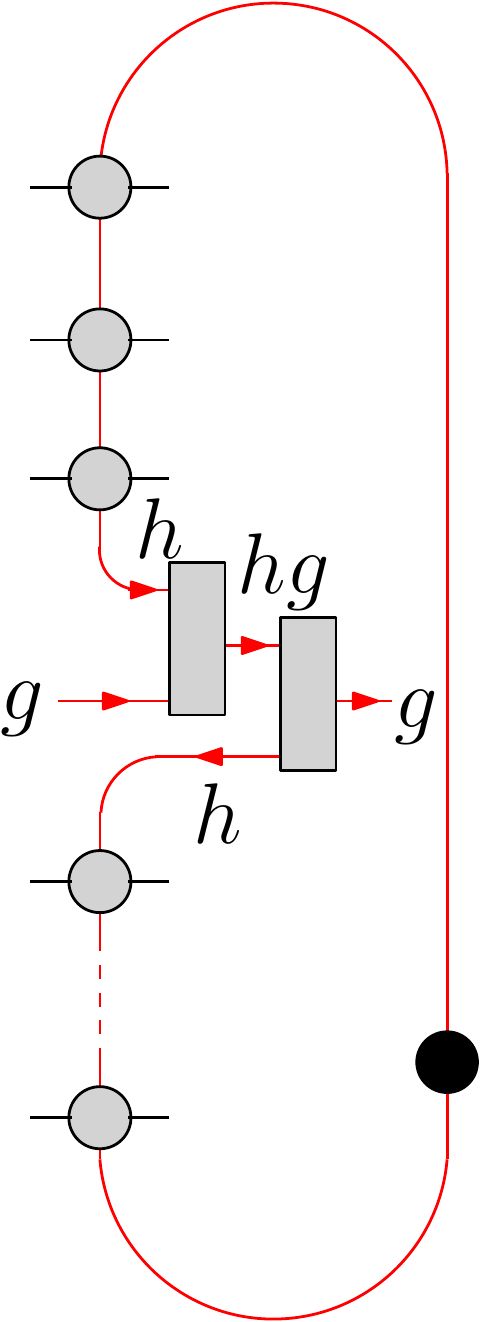}\, .
\end{equation}
A tedious, but straightforward calculation shows that this fMPO is a projective representation of $\mathcal{Z}_g$ with 2-cocycle

\begin{equation}\label{eq:2cocycle}
\omega_g(h,k) = (-1)^{Z(h,k)(Z(hk,g)+Z(g,hk))}\frac{\alpha(h,g,k)}{\alpha(h,k,g)\alpha(g,h,k)}
\end{equation}
For $h,k$ and $l$ commuting with $g$ the supercocycle relation implies that this phase indeed satisfies the 2-cocycle relation:

\begin{equation}
\omega_g(h,k)\omega_g(hk,l) = \omega_g(h,kl)\omega_g(k,l)\, .
\end{equation}
The virtual symmetry action on the right boundary indices is of course also projective, but with 2-cocycle $\bar{\omega}_g(h,k)$. Equation \eqref{eq:2cocycle} thus describes the fractionalization of symmetry defects in Gu-Wen SPT phases. It is a generalization of the slant product for bosonic SPT phases.

\subsection{Modular transformations}

Let us now consider the Gu-Wen tensor network on a torus. In this case we can twist the boundary conditions in both the $x$ and $y$ direction with group elements $h$ and $g$, provided that $[g,h] = ghg^{-1}h^{-1} = e$, by putting fMPOs along the non-contractible cycles. These fMPOs labeled by $h$ and $g$ meet in one point, where they have to be connected using fusion tensors. There are many different possibilities to connect the fMPOs in this way, but using the pivotal properties of Gu-Wen fusion tensors discussed in appendix \ref{GuWenpivotal} one can show that all these different choices only differ by a phase factor for the twisted wavefunction. In this section we find it convenient to work in the following basis for the twisted Gu-Wen states:

\begin{equation}\label{standardbasis}
\includegraphics[width=0.2\textwidth]{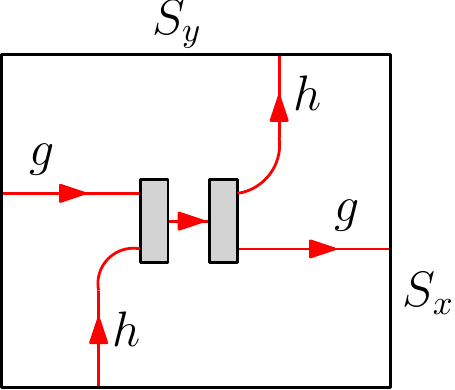}\, ,
\end{equation}
where opposite sides should be identified. This figure shows the position of the fMPOs $O_h$ and $O_g$ on the virtual level of the Gu-Wen tensor network on the torus and how they are connected using fusion tensors. $S_x$ and $S_y$ denote whether the boundary conditions are periodic or anti-periodic along the two non-contractible cycles of the torus, where $S=0\,(1)$ means anti-periodic (periodic). Note that since all PEPS and fMPO tensors are even, the parity of the twisted state \eqref{standardbasis} is determined by the parity of the fusion tensors, which gives $Z(g,h)+Z(h,g)$ mod 2.

We can now define the $\mathcal{S}$ transformation on these states as

\begin{equation}\label{eq:s}
\includegraphics[width=0.55\textwidth]{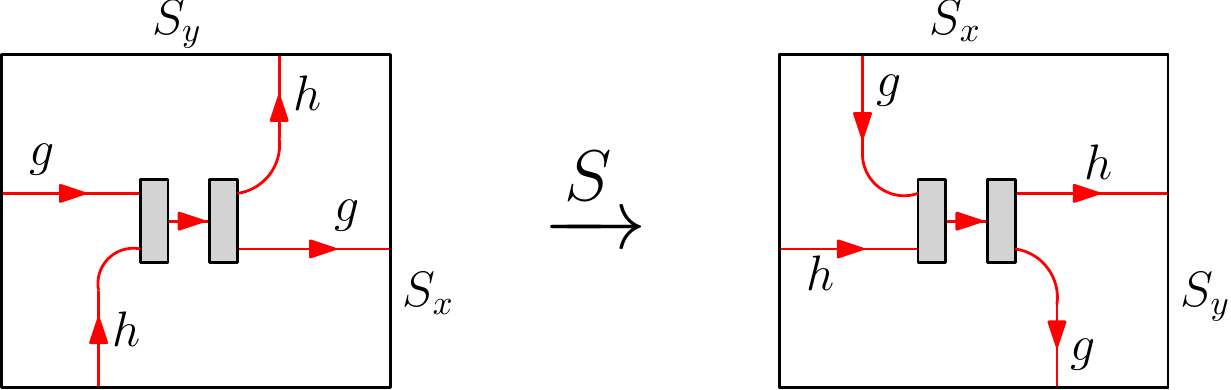}\, .
\end{equation}
Using the $F$ move \eqref{eq:fmovesupercocycle} and the pivotal properties \eqref{linebending} introduced in appendix \ref{GuWenpivotal} the twisted state after the $\mathcal{S}$ transformation can be expressed back in the standard basis \eqref{standardbasis}. If we denote the basis state \eqref{standardbasis} as $|(h,g);(S_x,S_y)\rangle$ then the $\mathcal{S}$ transformation takes following matrix form

\begin{eqnarray}
S = & \sum_{\underset{[g,h]=e}{g,h}}\sum_{S_x,S_y} (-1)^{Z(g,g^{-1})S_x + (Z(g,h)+Z(h,g))S_y}\,\frac{\alpha(g^{-1},h,g)}{\alpha(h,g^{-1},g)\alpha(g^{-1},g,h)} \nonumber \\ & |(g^{-1},h);(S_y,S_x)\rangle\langle (h,g);(S_x,S_y)|\, .
\end{eqnarray}
From the supercocycle relation it follows that the $S$ matrix satisfies $S^4 = (-1)^{Z(g,h)+Z(h,g)}\mathds{1}$, which is to be expected since $S^4$ represents a $2\pi$ rotation and $Z(g,h)+Z(h,g)$ is the fermion parity of the twisted state $|(h,g);(S_x,S_y)\rangle$.

We can now define the $\mathcal{T}$ transformation, corresponding to a Dehn twist on the twisted states:

\begin{equation}\label{eq:t}
\includegraphics[width=0.6\textwidth]{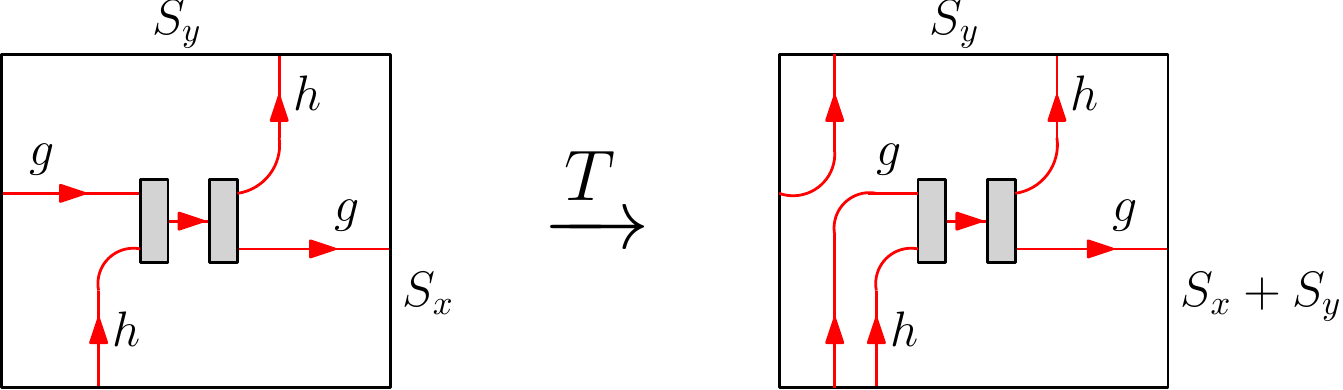}\, .
\end{equation}
The state after the $\mathcal{T}$ transformation can again be brought back into the standard basis \eqref{standardbasis} using $F$-moves and the pivotal properties of the fusion tensors. This gives following expression for the $T$ matrix:

\begin{equation}
T = \sum_{\underset{[g,h]=e}{g,h}}\sum_{S_x,S_y} (-1)^{Z(g,h)(S_y+Z(g,h)+Z(h,g))} \alpha(g,h,g)|(gh,g);(S_x+S_y,S_y)\rangle\langle (h,g);(S_x,S_y)|\, .
\end{equation}
The $S$ and $T$ matrices obviously depend on the representative cocycles $Z(g,h)$ and $\alpha(g,h,k)$. However, under a coboundary transformation 

\begin{eqnarray}
Z(g,h) & \rightarrow &  Z(g,h) + x(g)+x(h)+x(gh) \text{ with } x(g) \in \{0,1\}\nonumber \\
\alpha(g,h,k) & \rightarrow & \alpha(g,h,k)\frac{\gamma(g,h)\gamma(gh,k)}{\gamma(g,hk)\gamma(h,k)}
\end{eqnarray}
the $S$ matrix transforms as $USU^\dagger$, with $U$ a diagonal unitary matrix. The $T$ matrix does not have this property under coboundary transformations of $Z(g,h)$. This seems to imply that $T$ is not an object containing universal information about the Gu-Wen phase. However, $T^2$ does have the desired property $T^2\rightarrow UT^2U^\dagger$ under general coboundary transformations, implying that its eigenvalues are relevant invariants. This ambiguity has a physical meaning if we interpret the eigenvalues of $T$ as $e^{i2\pi h}$, where $h$ are the topological spins of the defects. Because the transparant particle in Gu-Wen phases is a fermion, the topological spins are only defined modulo $1/2$. This sign ambiguity in the eigenvalues of $T$ can be avoided by looking at $T^2$. As explained above, the coboundary transformation on $Z(g,h)$ can be interpreted as attaching a fermion to the virtual fMPO indices. Since $g$ defects are connected via fMPOs $O_g$ such a coboundary transformation indeed has the net effect of attaching fermions to the defects, changing the topological spin by $1/2$.
The ambiguity in $T$ also manifests itself in the relation $(ST)^3 = (-1)^{Z(g,h)}S^2$, which follows from the supercocycle relation. This shows that $S$ and $T$ only form a representation of $SL(2,\mathbb{Z})$ up to a minus sign which changes under coboundary transformations.

Finally, we want to point out that the $S$ and $T$ matrices for the gauged, topologically ordered PEPS can be obtained from those of the SPT phase \cite{Williamson,Barkeshli}. It was shown in Ref.~\cite{Williamson} that the $S$ and $T$ matrices for the gauged theory are obtained by applying $\mathcal{S}$ and $\mathcal{T}$ on the states 

\begin{equation}\label{eq:gaugedstate}
\frac{1}{|G|}\sum_{x\in G}U(x)^{\otimes L_xL_y}|(h,g);(S_x,S_y)\rangle\, ,
\end{equation}
where $U(x)$ is the on-site physical symmetry action and $L_xL_y$ is the size of the torus. If $k \in \mathcal{Z}_{h,g}$, the centralizer of both $h$ and $g$, then the twisted states satisfy $U(k)^{\otimes L_xL_y}|(h,g);(S_x,S_y)\rangle = \epsilon_{h,g}^{S_x,S_y}(k)|(h,g);(S_x,S_y)\rangle$. Using the results of section \ref{sec:generaldefects} it follows that the one-dimensional representation $\epsilon_{h,g}^{S_x,S_y}(k)$ of $\mathcal{Z}_{h,g}$ is given by

\begin{equation}
\epsilon_{h,g}^{S_x,S_y}(k) = (-1)^{(Z(h,k)+Z(k,h))S_y + (Z(g,k)+Z(k,g))S_x + (Z(g,h)+Z(h,g))(Z(g,k)+Z(k,g))}\frac{\omega_g(h,k)}{\omega_g(k,h)}\, .
\end{equation}
Since the states \eqref{eq:gaugedstate} are obtained by a projection on the symmetric subspace, only those states for which $\epsilon_{h,g}^{S_x,S_y}(k) = 1$ for all $k \in \mathcal{Z}_{h,g}$ are non-zero. Both $\mathcal{S}$ and $\mathcal{T}$ commute with the global symmetry action $U(x)^{\otimes L_xL_y}$. For $\mathcal{S}$, this is immediate, but for $\mathcal{T}$ this follows from the results in \cite{MPOpaper}. From the commutativity of $\mathcal{S}, \mathcal{T}$ and $U(x)^{\otimes L_xL_y}$ one can easily infer the $S$ and $T$ matrices of the gauged theory from those of the Gu-Wen SPT. However, note that the $S$ and $T$ matrices obtained in this way are not expressed in the basis that has a definite anyon flux through one of the holes of the torus. To compare the $S$ and $T$ matrices before and after gauging, note that the action of $\mathcal{S}$ and $\mathcal{T}$ on the states $\sum_{x\in G} \Gamma^\mu_{ij}(x)U(x)^{\otimes L_xL_y}|(h,g);(S_x,S_y)\rangle$, with $\Gamma^\mu(g)$ an irrep of $G$, is independent of $\mu,i$ and $j$. This shows that the $S$ and $T$ matrices of the SPT phase consist of multiple copies of the same block (up to diagonal unitary similarity transformations), and the gauging proces selects only one of these identical blocks.

\section{$\mathbb{Z}_2$ Majorana phases}\label{sec:Z2}

In this section we consider the example of a fMPO representation of $\mathbb{Z}_2$, where the non-trivial group element corresponds to an irreducible fMPO of the type $\epsilon=1$. Concretely, we start from two fMPOs $O^L_1$ and $O^L_\sigma$, satisfying

\begin{equation}
O^L_1O^L_1 = O^L_1\;\;\;\; O^L_1O^L_\sigma = O^L_\sigma O^L_1 = O^L_\sigma \;\;\;\; O^L_\sigma O^L_\sigma =O^L_1\, ,
\end{equation}
for every $L$. The matrix algebras spanned by the tensors of $O^L_1$ and $O^L_\sigma$ are of the type $\epsilon_1 = 0$ and $\epsilon_\sigma = 1$. Note that $N_{\sigma\sigma}^1 = 1$, which implies that we have defined $O^L_\sigma$ with a global factor $1/\sqrt{2}$ as explained in section \ref{sec:superalgebras}. 

Without loss of generality we can take the parity of $\mathsf{X^1_{11}}$ to be zero. With this convention we can solve the pentagon equation to get following the independent $F$-symbols:

\begin{equation}\label{eq:fsymbols1}
\begin{matrix}
\left[F_1^{111} \right]^{1,00}_{1,00} = 1 \\
\left[F_\sigma^{11\sigma} \right]^{\sigma,00}_{1,00} = \left[F_\sigma^{1\sigma 1} \right]^{\sigma,00}_{\sigma,00} = \left[F_\sigma^{\sigma 11} \right]^{1,00}_{\sigma,00} = 1 \\
\left[F_1^{1\sigma\sigma} \right]^{1,00}_{\sigma,00} = \left[F_1^{\sigma 1\sigma} \right]^{\sigma,00}_{\sigma,00} = \left[F_1^{\sigma \sigma1} \right]^{\sigma,00}_{1,00} = 1 \\
\left[F_\sigma^{\sigma\sigma\sigma} \right]^{1,00}_{1,00} = (-1)^\rho\frac{1}{\sqrt{2}}\;\;\;\;\;\; \left[F_\sigma^{\sigma\sigma\sigma} \right]^{1,00}_{1,11} = (-1)^{\eta+\rho}i \frac{1}{\sqrt{2}}\\
\left[F_\sigma^{\sigma\sigma\sigma} \right]^{1,11}_{1,00} = (-1)^{\rho+1}\frac{1}{\sqrt{2}}\;\;\;\; \left[F_\sigma^{\sigma\sigma\sigma} \right]^{1,11}_{1,11} = (-1)^{\eta+\rho}i \frac{1}{\sqrt{2}}\, ,
\end{matrix}
\end{equation}
where $\rho \in\{0,1\}$. Since $\mu$ and $\hat{\mu}$ as defined in the general formalism of fMPO super algebras are now one-dimensional (because $N_{ab}^c \in \{0,1\}$) the row and column indices of the $F$-symbols consist only of the labels $1$ and $\sigma$ and the parities of the fusion tensors. The super pentagon equation also implies that $\zeta = \chi = 0$. So we have found four different solutions of the super pentagon equation for a $\mathbb{Z}_2$ fMPO representation with $\epsilon_\sigma=1$, labeled by $\eta$ and $\rho$.  The set of $F$-symbols given above is not complete, one can obtain other ones by changing the parity of fusion tensors $\mathsf{X_{ab}^c}$ with $c=\sigma$ or $a=b=\sigma$ via suitable contractions with $\mathsf{Y}$. However, these additional $F$-symbols are completely determined by the $F$-symbols given above and the relations (\ref{eq:Ymove1}) ,(\ref{eq:Ymove2}) and (\ref{eq:Ymove3}). We note that these $F$-symbols were first presented in \cite{walker}.

\begin{figure}
  \centering
    \includegraphics[width=0.72\textwidth]{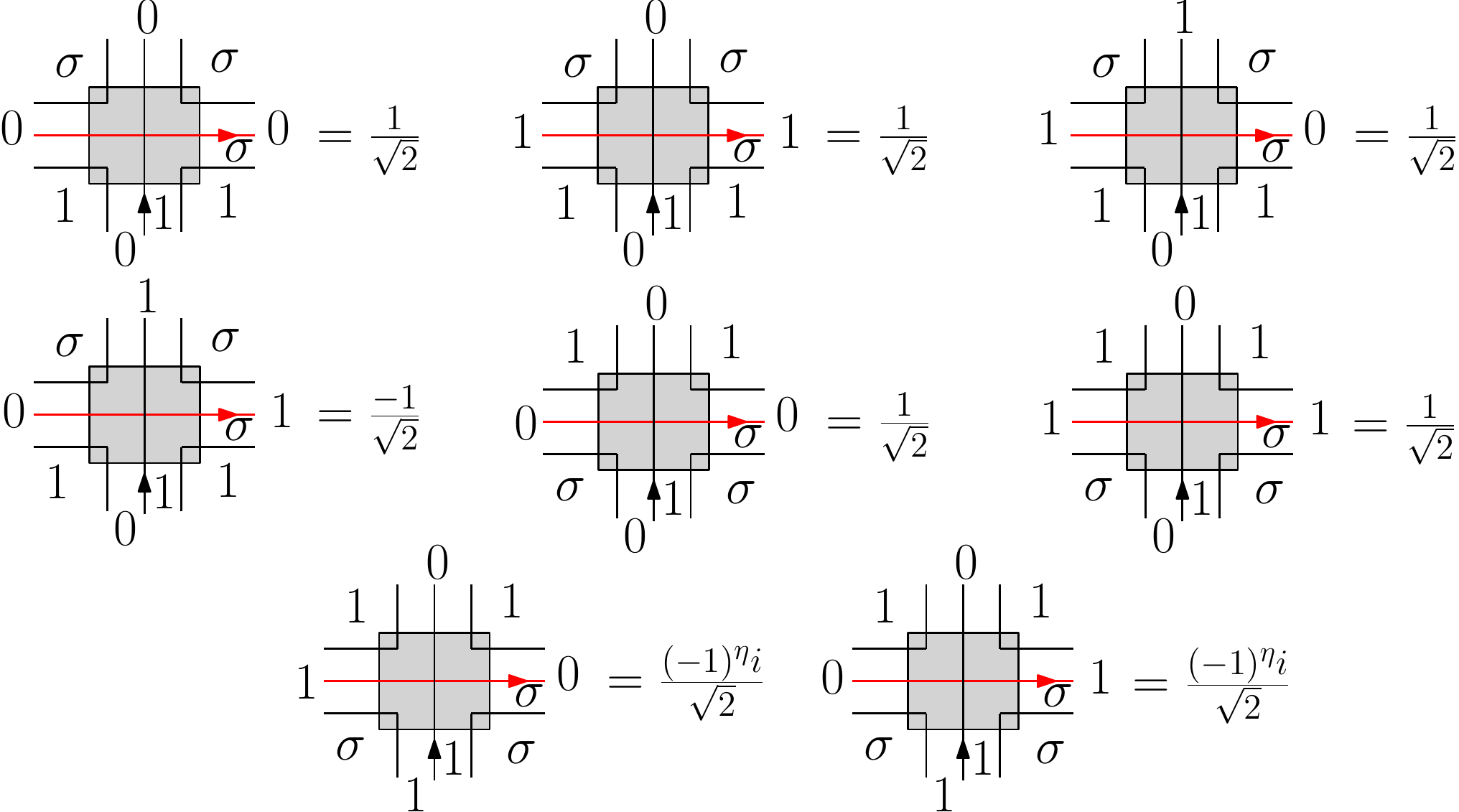}\\
    \includegraphics[width=0.72\textwidth]{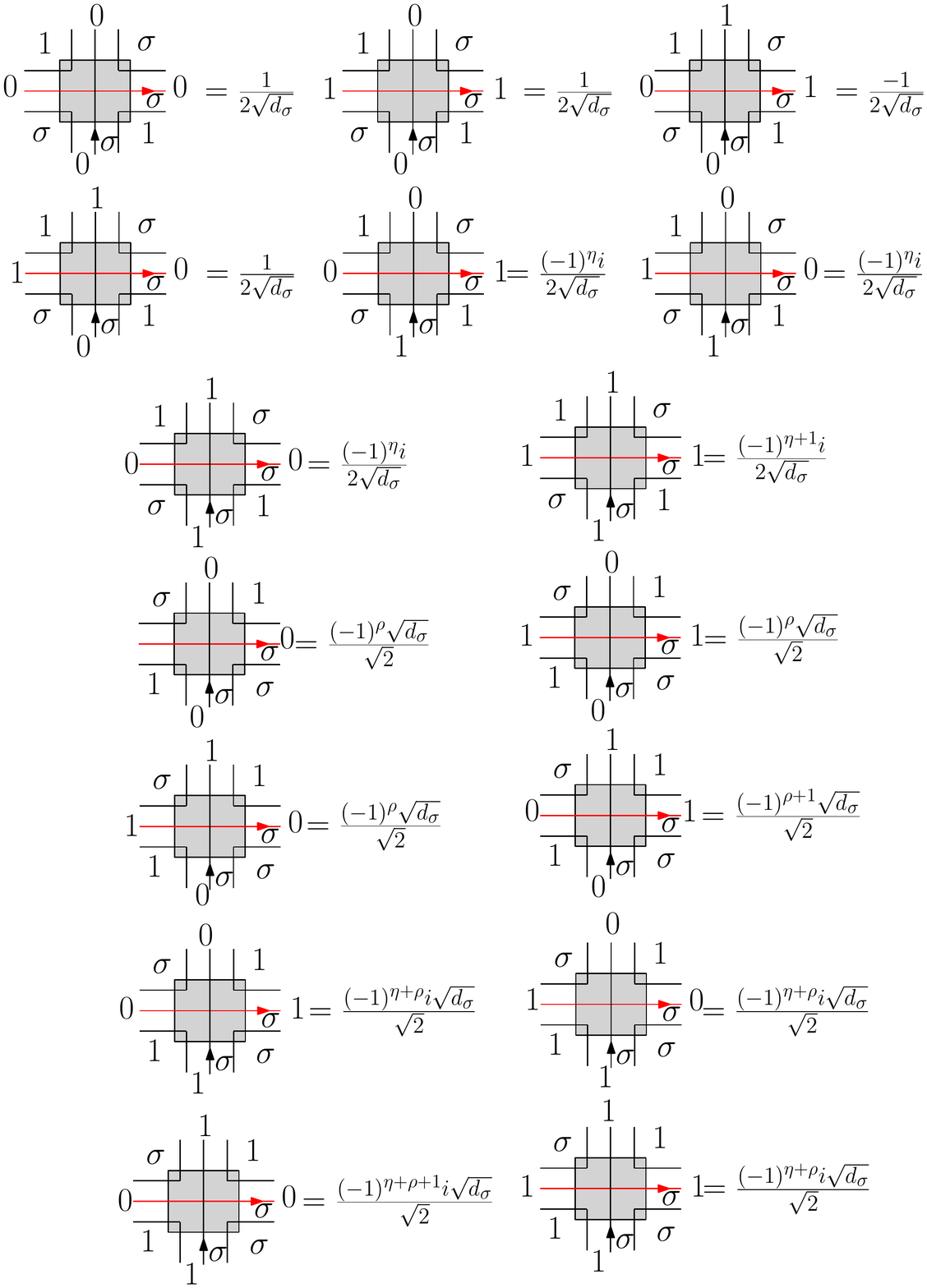}
 \caption{Non-zero tensor components of the fMPO $O_\sigma$ in the basis \eqref{eq:RH1} and with $d_\sigma = \frac{1}{\sqrt{2}}$. The outer most labels 0 and 1 denote the parity of the indices. }\label{fig:osigma}
\end{figure}

\begin{figure}
  \centering
    \includegraphics[width=0.66\textwidth]{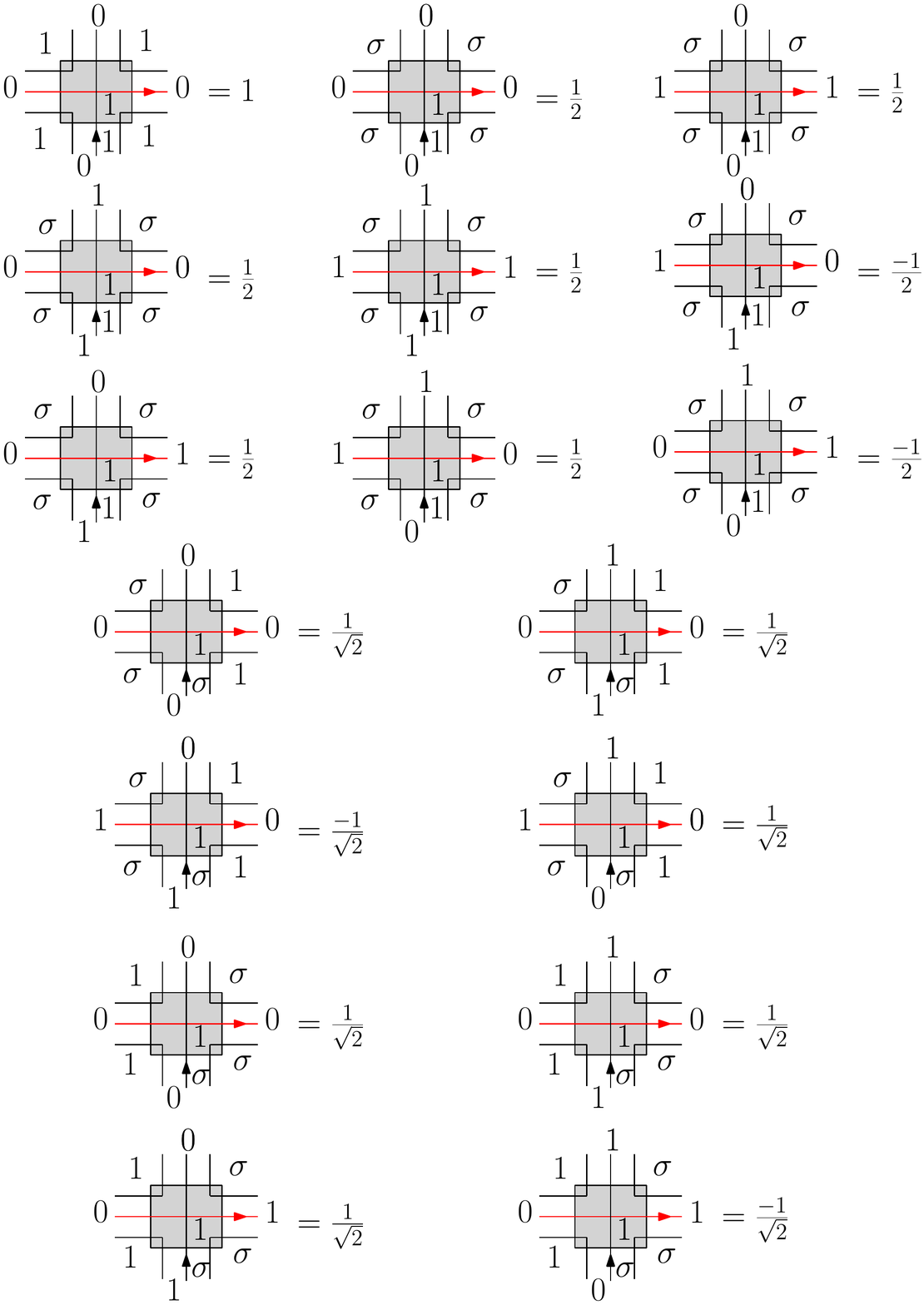}
 \caption{Non-zero tensor components of the fMPO $O_1$ in the basis \eqref{eq:RH1}. The outer most labels 0 and 1 denote the parity of the indices.}\label{fig:oone}
\end{figure}

Applying the general recipe of section \ref{sec:fmovepentagon} we find that the $F$-symbols need to be rescaled in the following way to obtain the $\tilde{F}$-symbols:

\begin{equation}
\begin{matrix}
\tilde{F}^{111}_1 = F^{111}_1 & \tilde{F}^{11\sigma}_\sigma = \frac{1}{\sqrt{2}}F^{11\sigma}_\sigma & \tilde{F}^{1\sigma 1}_\sigma = \frac{1}{2}F^{1\sigma 1}_\sigma & \tilde{F}^{1\sigma\sigma}_1 = \frac{1}{\sqrt{2}}F^{1\sigma\sigma}_1 \\
\tilde{F}^{\sigma 11}_\sigma = \frac{1}{\sqrt{2}} F^{\sigma 11}_\sigma & \tilde{F}^{\sigma\sigma 1}_1 = \frac{1}{\sqrt{2}} F^{\sigma\sigma 1}_1 & \tilde{F}^{\sigma 1\sigma}_1 = \frac{1}{2}F^{\sigma 1\sigma}_1 & \tilde{F}^{\sigma\sigma\sigma}_\sigma = F^{\sigma\sigma\sigma}_\sigma\, .
\end{matrix}
\end{equation}
One can explicitly verify that these $\tilde{F}$-symbols satisfy the isometric properties \eqref{eq:isometric1} and \eqref{eq:isometric2}. From the pivotal property \eqref{eq:pivotal2} one finds that $d_1 =1$ and $d_\sigma = 1/\sqrt{2}$. In figure \ref{fig:osigma} we explicitly give the non-zero tensor components of the fMPO $O_\sigma$, and in figure \ref{fig:oone} we give the components of $O_1$. We note that the tensor components of $O_\sigma$ are of the form $B[\sigma]^{ij} = y^{|i|+|j|}\otimes C^{ij}$ in the basis $\sum_{ij\alpha\beta} B[\sigma]^{ij}_{\alpha\beta}|\alpha)|i\rangle\langle j|(\beta|$. In Ref.\cite{fMPS} it was shown that this indeed corresponds to the normal form of $\epsilon = 1$ fMPOs.

Making use of the general expressions \eqref{eq:tensorA},\eqref{eq:tensorAC} and \eqref{eq:tensorB},\eqref{eq:tensorBC} one can construct the fixed-point PEPS corresponding to the $\{O_1,O_\sigma\}$ fMPO algebra \footnote{From private discussions we learned that this tensor network is found independently by different authors and will appear in Ref. \cite{Aasen2}}. The fixed-point PEPS construction might not be very insightful. However, some physical intuition can be gained by analyzing the PEPS tensors. Keeping in mind the fMPS expression for the Majorana chain \cite{fMPS} one can  convince oneself that the PEPS wavefunction represents a superposition of all coverings of the honeycomb lattice with closed Majorana chains. In recent work an explicit commuting projector Hamiltonian stabilizing this type of ground state wave function was constructed \cite{ware}. It was pointed out that  this system has the same topological properties as a $(p+ip)\times (p-ip)$ bilayer system where fermion parity in one of the layers is gauged. The corresponding phase of matter was first envisioned by starting from the Ising string-net and condensing the $\psi$ anyon \cite{walker}, which appears to be a general mechanism to obtain fermionic topological phases \cite{gaiotto,bhardwaj,thorngren2}.

\subsection{Spin structures and ground states on the torus}\label{sec:spinstructurestorus}

As explained in section \ref{sec:fixedpoint}, it is essential that every fMPO closed along a contractible loop has an odd number of parity matrices $\mathsf{P}$ inserted on its internal indices. In section \ref{sec:guwen} we explained that the number of parity matrices modulo two on fMPOs along non-contracible cycles is determined by the boundary conditions, or equivalently, by the spin structure. Here we will show that this leads to a non-trivial interplay between spin structure, ground state degeneracy and ground state parity for the topologically ordered fermionic PEPS contructed from the fMPO superalgebra $\{O^L_1,O^L_\sigma\}$ with $\epsilon_\sigma = 1$ via equations \eqref{eq:tensorAC} and \eqref{eq:tensorBC}.

\begin{figure}
  \centering
    \includegraphics[width=0.6\textwidth]{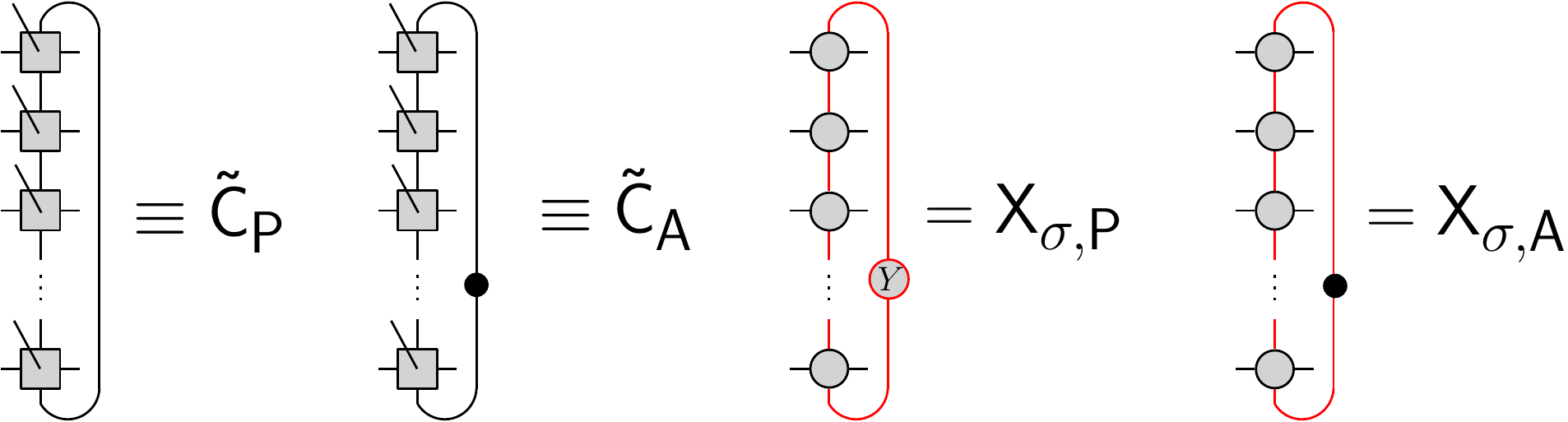}
  \caption{Definition of the tensors $\mathsf{\tilde{C}_P}$, $\mathsf{\tilde{C}_A}$,  $\mathsf{X_{\sigma,P}}$ and  $\mathsf{X_{\sigma,A}}$.} \label{fig:column}
\end{figure}

We start by showing that the fermionic PEPS on the torus with periodic boundary conditions in both directions (PP) evaluates to zero if no fMPO $O_\sigma$ is inserted on the virtual level. To see this, first construct a tensor $\mathsf{\tilde{C}_P}$ by contracting all PEPS tensors that lie in the same column, where P denotes that we use periodic boundary conditions in the direction along the column. The ordering convention for the indices of $\mathsf{\tilde{C}_P}$ is as follows: first the virtual indices corresponding to the left hand side of the column, then the physical indices and lastly the virtual indicices on the right hand side. The virtual indices are ordered such that contracting neighboring columns corresponds to matrix multplication of the components of $\mathsf{\tilde{C}_P}$. The procedure just described is of course just the fermionic version of standard reinterpretation of a PEPS on the cylinder as a matrix product state with tensors $\mathsf{\tilde{C}_P}$. We will denote the fMPO $O^{L_y}_\sigma$ going along the periodic direction, with the external indices reordered in the same way as the virtual indices of $\mathsf{\tilde{C}}$, as $\mathsf{X_{\sigma,P}}$. It is crucial to note that $\mathsf{X_{\sigma,P}}$ has odd parity while $\mathsf{X_{\sigma,A}}$ is even. This is because $\epsilon_\sigma =1$ and it was shown in Ref.\cite{fMPS} that such fMPOs have to be closed with $\mathsf{Y}$ on the internal indices under periodic boundary conditions in order to be non-zero. With anti-periodic boundary conditions $O^{L_y}_\sigma$ has to be closed without $\mathsf{Y}$ and is therefore even. Figure \ref{fig:column} gives a graphical representation of the tensors just defined. 

Now we can easily show that the fermionic PEPS in the PP sector without any fMPO is zero. Its coefficients on a torus consisting of $L_x$ columns are given by tr($P^{\otimes L_y}\tilde{C}^{i_1}\tilde{C}^{i_2}\dots\tilde{C}^{i_{L_x}}$), where $i_j$ represents the collection of all physical indices in the $j$th column and $P^{\otimes L_y}$, the tensor product of $L_y$ parity matrices, is generated as a supertrace by the fermionic contraction (see Ref.\cite{fMPS} for more detail). In appendix \ref{app:OP} we show that $\mathsf{X_{\sigma,P}}\mathsf{X_{\sigma,P}} = (-1)^\eta i \mathsf{X_{1,P}}$, which can now be used to show that

\begin{eqnarray}
\text{tr}(P^{\otimes L_y}\tilde{C}^{i_1}\tilde{C}^{i_2}\dots\tilde{C}^{i_{L_x}}) & = & (-1)^{\eta+1}i\, \text{tr}(P^{\otimes L_y}\tilde{C}^{i_1}\tilde{C}^{i_2}\dots\tilde{C}^{i_{L_x}}X_{\sigma,P}X_{\sigma,P}) \nonumber \\
 & = &(-1)^{\eta}i\, \text{tr}(P^{\otimes L_y}X_{\sigma,P}\tilde{C}^{i_1}\tilde{C}^{i_2}\dots\tilde{C}^{i_{L_x}}X_{\sigma,P}) \nonumber\\
 & = & (-1)^{\eta}i\, \text{tr}(P^{\otimes L_y}\tilde{C}^{i_1}\tilde{C}^{i_2}\dots\tilde{C}^{i_{L_x}}X_{\sigma,P}X_{\sigma,P})\nonumber\\
 & = & - \text{tr}(P^{\otimes L_y}\tilde{C}^{i_1}\tilde{C}^{i_2}\dots\tilde{C}^{i_{L_x}})\, ,
\end{eqnarray}
where the second equality follows from the fact that $\mathsf{X_{\sigma,P}}$ is odd and the third equality follows from the pulling through property. 

The non-zero states in the PP sector can be schematically represented as

\begin{equation}
\includegraphics[width=0.45\textwidth]{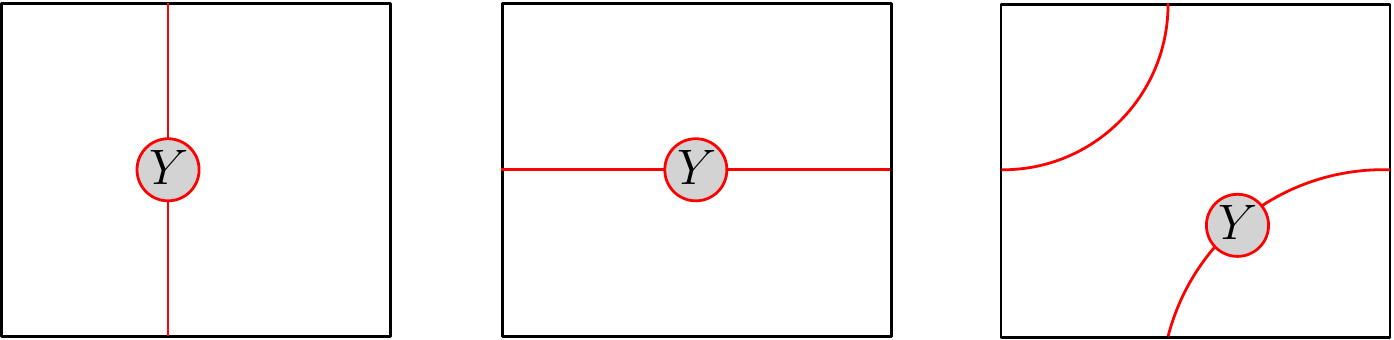}\, ,
\end{equation}
where the torus is depicted as a rectangle with opposite sides identified. The red line represents a fMPO $\mathsf{X_\sigma}$ on the virtual level of the PEPS wrapping a non-contractible cycle. The state on the left has coefficients $\text{tr}(X_{\sigma,P}\tilde{C}^{i_1}\tilde{C}^{i_2}\dots\tilde{C}^{i_{L_x}})$, where the fermionic contraction now does not generate a matrix $P^{\otimes L_y}$ because $X_{\sigma,P}\tilde{C}^{i_1}\tilde{C}^{i_2}\dots\tilde{C}^{i_{L_x}}$ has odd parity. For this reason we cannot conclude that this state is zero. Similar reasoning shows that the other two states in the PP sector may also be non-zero. Note that the three ground states in the PP sector all have odd fermion parity because of the matrix $\mathsf{Y}$ on the internal fMPO indices.

In the AP sector, with anti-periodic boundary conditions in the $x$-direction, one can show that the following state is zero:

\begin{equation}
\includegraphics[width=0.2\textwidth]{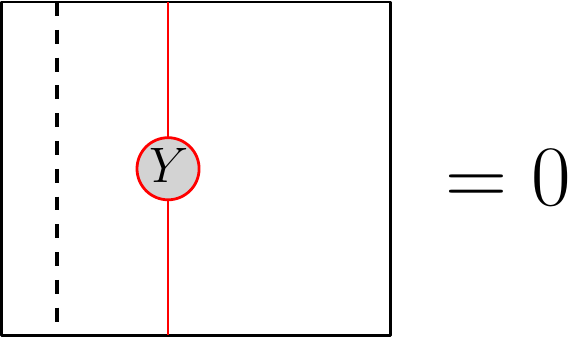}\, ,
\end{equation}
where the dashed line represents the anti-periodic boundary conditions, i.e. along that line on the dual lattice we have inserted parity matrices $P$ on the virtual indices. Note that the periodic boundary conditions in the $y$-direction imply that the fMPO $\mathsf{O^{L_y}_\sigma}$ is odd. The coefficients for this state are $\text{tr}(P^{\otimes L_y}X_{\sigma,P}\tilde{C}^{i_1}\tilde{C}^{i_2}\dots\tilde{C}^{i_{L_x}})$, where $P^{\otimes L_y}$ is now not generated by the fermionic contraction because $\mathsf{X_{\sigma,P}}$ is odd but is inserted by hand because of the anti-periodic boundary conditions. This trace expression for the coefficients can easily be seen to be zero. The three non-zero states in the AP sector are

\begin{equation}
\includegraphics[width=0.45\textwidth]{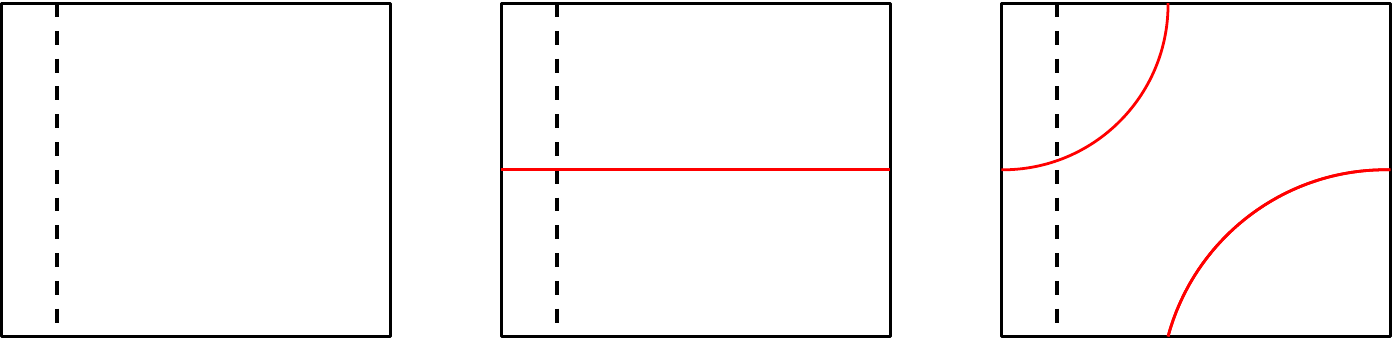}\, ,
\end{equation}
where both $\sigma$-fMPOs are even because they cross the dashed line an odd number of times. Note that the coefficients of the state in the AP sector without any fMPO are $\text{tr}(P^{\otimes L_y}\tilde{C}^{i_1}\tilde{C}^{i_2}P^{\otimes L_y}\dots\tilde{C}^{i_{L_x}})$, where one $P^{\otimes L_y}$ is generated by the supertrace of an even tensor and the second $P^{\otimes L_y}$ comes from the anti-periodic boundary conditions. Analogously, one can show that the three non-zero states in the PA sector are:

\begin{equation}
\includegraphics[width=0.45\textwidth]{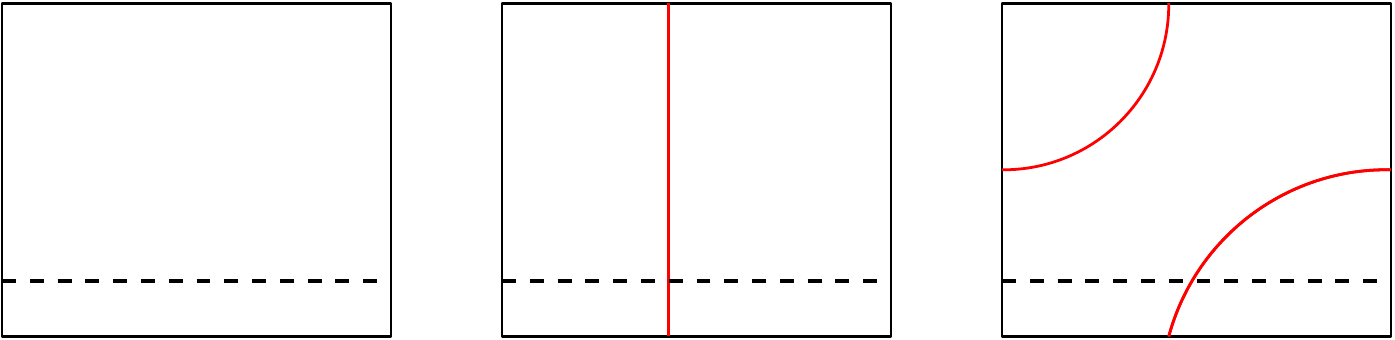}\, .
\end{equation}

In the AA sector one can show that the following state is zero:

\begin{equation}
\includegraphics[width=0.38\textwidth]{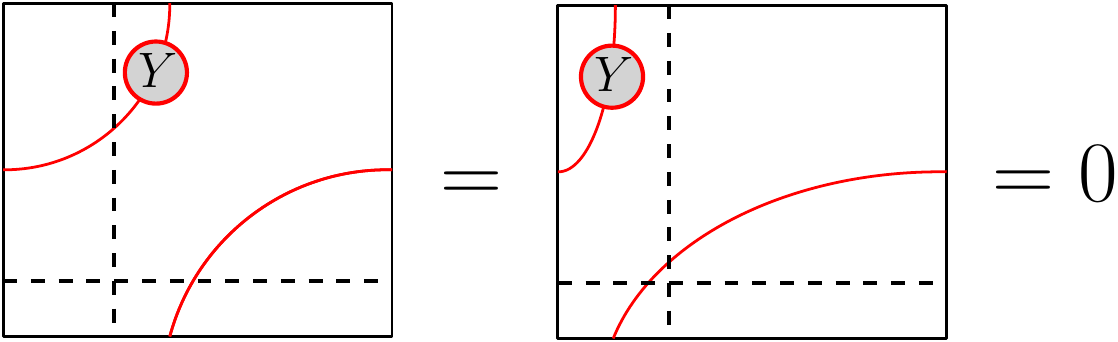}\, ,
\end{equation}
where the fMPO is odd because it crosses a dashed line an even number of times. The state above is zero because the two graphical expressions given for it differ by a minus sign, as can easily be seen by using the pulling through property and the fact that $\mathsf{Y}$ is odd. The non-zero states in the AA sector are then given by

\begin{equation}
\includegraphics[width=0.45\textwidth]{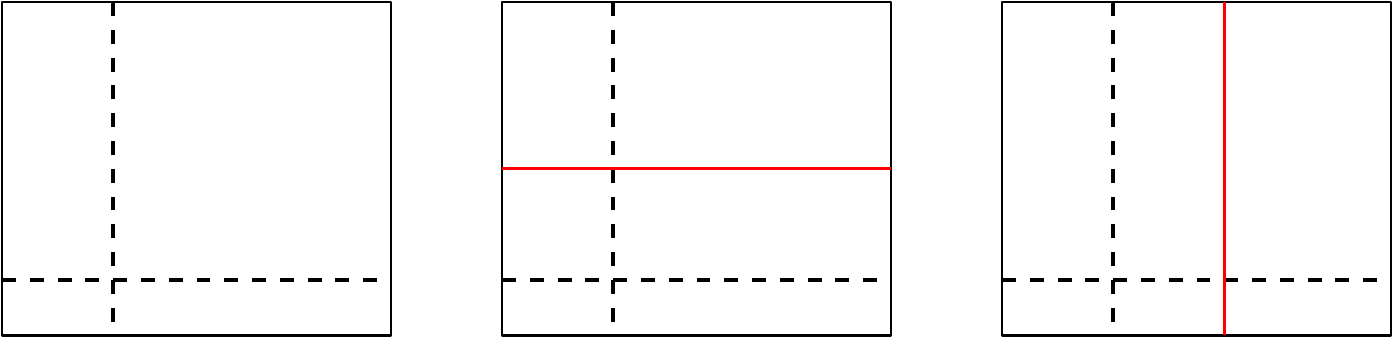}\, .
\end{equation}

So to conclude, we have found that the fermionic PEPS constructed from the fMPO algebra $\{O^L_1, O^L_\sigma\}$ with $\epsilon_\sigma = 1$ has three non-zero ground states in each spin structure sector. In the PP sector these states have odd parity, while in the AP, PA and AA sectors they have even parity. This agrees with the results of Ref.\cite{ware}, where an explict commuting projector Hamiltonian was constructed for the topological phases captured by the fermionic PEPS described in this section.

\subsection{Symmetry-protected phases}

Above we used the fMPO group representations $\{O^L_1,O^L_\sigma\}$ with $\epsilon_\sigma =1$ to construct fermionic PEPS with non-trivial topological order. Here we will discuss applications of these fMPOs for $\mathbb{Z}_2$ symmetry-protected phases. In analogy to section \ref{sec:guwen}, where we treated the case $\epsilon\equiv 0$, we construct the short-range entangled PEPS on the hexagonal lattice using the tensors

\begin{equation}\label{eq:z2spt}
\includegraphics[width=0.48\textwidth]{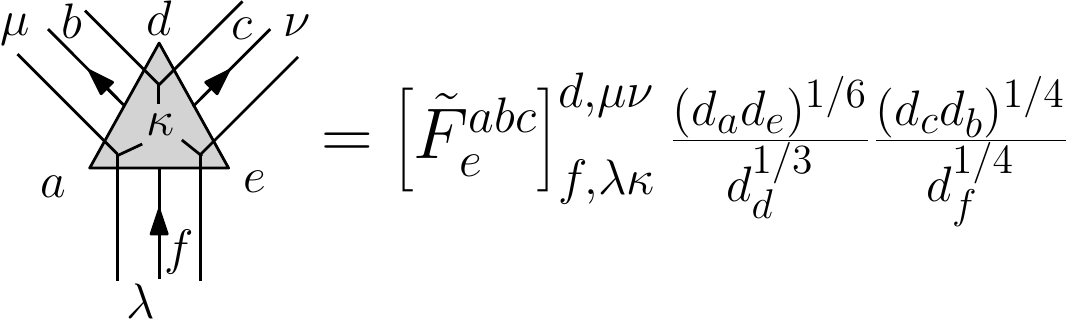}
\end{equation}
for the A-sublattice and a similar modification of \eqref{eq:tensorBC} for the B-sublattice. The resulting PEPS has a global $\mathbb{Z}_2$ symmetry, where the physical on-site symmetry action gets intertwined to a fMPO $O_\sigma$ on the virtual indices, where $O_\sigma$ is the same fMPO as before constructed from the tensor shown in figure \ref{fig:osigma}.

The PEPS obtained via the tensors \eqref{eq:z2spt} describes a wave function where Majorana chains are bound to domain walls of the plaquette variables. An explicit commuting projector Hamiltonian with this type of ground state was constructed in \cite{Tarantino}. A physical property of this SPT phase is that $\mathbb{Z}_2$ symmetry defects bind Majorana modes. In the tensor network language this can easily be seen by defining the PEPS on a cylinder with twisted boundary conditions along the non-contracible cycle. This is done by simply placing the fMPO $O_\sigma$ along the cylinder on the virtual level, going from one end of the cylinder to the other. At the two boundaries of the cylinder this results in a symmetry defect. Because $O_\sigma$ is of the type $\epsilon_\sigma = 1$, the resulting fMPS on the cylinder has a non-trivial center corresponding to $\mathsf{Y}$ acting on the internal fMPO index. One can use similar reasoning as in Ref. \cite{fMPS} to conclude that there will be Majorana modes at the ends of the cylinder. 

Other immediate concequences of the results in Ref.\cite{fMPS} involve the entanglement spectrum and the physical systems that can realize this phase. First, the PEPS on a cylinder with periodic boundary conditions has at least a two-fold degeneracy in its entanglement spectrum for cuts wrapping the non-contractible cycle. This follows from the fact that the $\mathbb{Z}_2$ symmetry action $\tilde{O}_\sigma$ on the boundary is odd and therefore anti-commutes with fermion parity. If we twist the boundary conditions with $O_\sigma$, then there will again be at least a two-fold degeneracy, both in the periodic and the anti-periodic sector. This degeneracy follows from the fact that $\epsilon_\sigma = 1$. By interpreting the PEPS on the cylinder with boundary conditions twisted by $O_\sigma$ as a MPS and applying the results of Ref.\cite{fMPS} it follows that the $\mathbb{Z}_2$ Majorana SPT phases cannot occur in systems with (unbroken) particle number conservation.

In appendix \ref{app:OP} we show that under periodic boundary conditions, the fMPOs $\tilde{O}_\sigma$ satisfy $\tilde{O}_\sigma \tilde{O}_\sigma = (-1)^\eta i \tilde{O}_1$. This gives a physical interpretation to the invariant $\eta$: it determines the projective representation of the global $\mathbb{Z}_2$ symmetry on $\pi$-flux defects. Note that since $\tilde{O}_\sigma$ is odd, this projective representation is consistent with the fact that two $\pi$-flux defects fuse to the vacuum. The combined action on two $\pi$-flux defects is given by $\tilde{O}_\sigma\otimesg \tilde{O}_\sigma$, which satisfies 

\begin{equation}
\left(\tilde{O}_\sigma\otimesg \tilde{O}_\sigma\right)\left(\tilde{O}_\sigma\otimesg \tilde{O}_\sigma\right) = -\tilde{O}_\sigma\tilde{O}_\sigma\otimesg \tilde{O}_\sigma\tilde{O}_\sigma = -(-1)^\eta i(-1)^\eta i \tilde{O}_1\otimesg \tilde{O}_1 = \tilde{O}_1\otimesg \tilde{O}_1\, .
\end{equation} 
This shows that the symmetry action on two $\pi$-flux defects is indeed non-projective.

For the fMPO $O_\sigma$ we readily determine the Frobenius-Schur indicator as defined in the general theory of fMPO super algebras. We find that $Z_\sigma$ as defined as in section \ref{sec:superalgebras} takes the following form:

\begin{equation}
\includegraphics[width=0.65\textwidth]{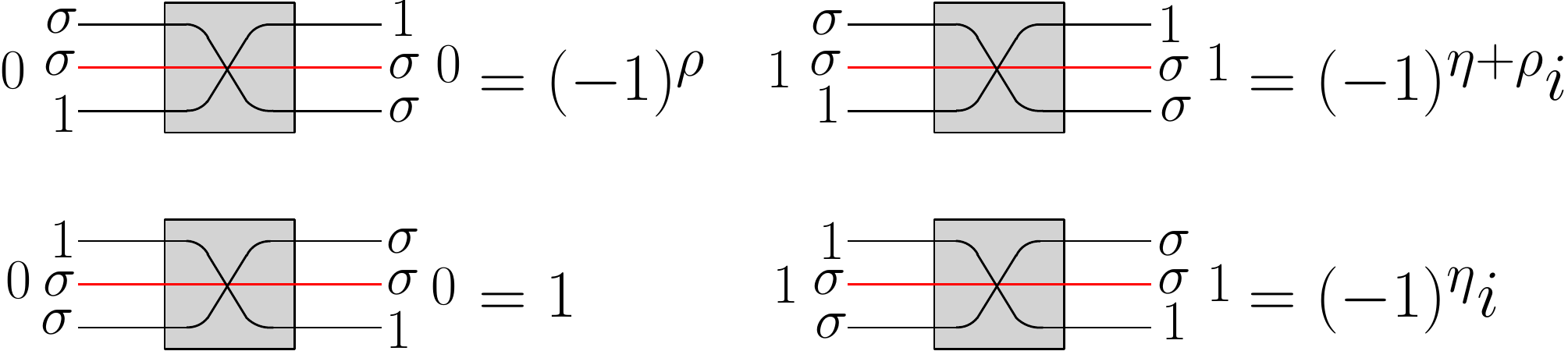}\, .
\end{equation}
Note that $Z_\sigma$ is even. The first invariant associated to the Frobenius-Schur indicator can now easily be obtained from

\begin{equation}
Z_\sigma \bar{Z}_\sigma = (-1)^\rho \mathds{1}\, ,
\end{equation}
The second invariant determining the Frobenius-Schur indicator we get from

\begin{equation}
\left(YZ_\sigma\right)\left(\bar{Y}\bar{Z}_\sigma\right) = (-1)^{\rho+\eta+1}iP\, .
\end{equation}
From this we see that the value of the Frobenius-Schur indicator uniquely determines the $F$-symbols for the $\{O_1,O_\sigma\}$ group representation with $\epsilon_\sigma =1$. The same holds for the case $\epsilon_\sigma = 0$, where the Frobenius-Schur indicator completely fixes one of the four supercohomology classes for $\mathbb{Z}_2$. So in total we have eight different fermionic SPT phases with a global $\mathbb{Z}_2$ symmetry. In section \ref{sec:superalgebras} we also mentioned that the Frobenius-Schur indicator is isomorophic to $\mathbb{Z}_8$, implying that the $\mathbb{Z}_2$ SPT phases form a $\mathbb{Z}_8$ group under stacking, agreeing with previous studies \cite{ryu,qi,yao,gulevinZ2}. Since the Frobenius-Schur indicator has the same mathematical origin as the invariants associated to time-reversal or reflection invariant fMPS, we have thus connected the classification of two-dimensional unitary $\mathbb{Z}_2$ SPT phases to the classification of one-dimensional SPT phases with time-reversal or reflection symmetry. This is the tensor network manifestation of the Smith isomorphism, which relates the cobordism groups conjectured to describe both types of SPT phases \cite{thorngren}.

\section{Discussion and outlook}

In this work we have studied the properties of fMPO super algebras. The resulting algebraic structure was used to construct explict fermionic topological PEPS models, both for phases with intrinsic and symmetry-protected topological order. The fermionic string-nets and supercohomology phases were reproduced as a special ($\epsilon =0$) subset of the general formalism.

The fixed-point fermionic PEPS models allow for a straightforward calculation of many interesting universal properties associated with the topological phases. We illustrated this for Gu-Wen SPT phases, where we determined the projective symmetry properties of defects and the modular matrices associated with symmetry-twisted states on the torus. Also for the $\mathbb{Z}_2$ Majorana phases, the PEPS construction enables us to relate the algebraic data classifying the different phases to physical properties of the system.

Starting from the tensor networks constructed here, there are many different directions to explore in future work. Perturbing the fixed-point models yields interesting PEPS to be studied numerically, which could give rise to new insights in e.g. entanglement properties and topological phase transitions. The SPT phases considered in this work only have discrete on-site unitary symmetries. However, we expect that fMPOs should also capture the phases associated with continuous, anti-unitary and/or spatial symmetries.  The global $\mathbb{Z}_2$ symmetry of fermionic PEPS corresponding to fermion parity can be gauged by applying the gauging map as introduced in \cite{gaugingpaper}. This gives an explicit realization of the connection between fermionic topological phases and bosonic topological phases with an emergent fermion \cite{walker,gaiotto,bhardwaj,thorngren}.  For the fermionic PEPS with intrinsic topological order one would like to determine the anyons and their braiding properties as was done for spin systems \cite{MPOpaper,Aasen}. We refer to  \cite{Aasen2} for details on this construction. Once the anyons and their topological properties are understood, an interesting question is how they intertwine with a possible global symmetry in the system, which leads to the study of fermionic symmetry-enriched topological phases.
\\ \\
\emph{Acknowledgements - } We would like to thank David Aasen for explaining us the results presented in \cite{walker} and for discussions about the connections to tensor networks. We ackowledge Bela Bauer, Nicolas Tarantino and Brayden Ware for many inspiring discussions about the fixed-point models constructed in \cite{ware,Tarantino}, and possible realizations as tensor networks. We also thank Matthias Bal for helpful comments on an earlier version of the manuscript. The authors thank KITP for supporting the progams `\emph{Symmetry, Topology, and Quantum Phases of Matter: From Tensor Networks to Physical Realizations}' (Sept.-Dec. 2016) and `\emph{Synthetic quantum matter}' (Sept. - Dec. 2016), where part of this work was done. This work was supported by the Austrian Science Fund (FWF) through grants ViCoM and FoQuS, and the EC through the grant QUTE. J.H. and F.V. acknowledge the support from the Research Foundation Flanders (FWO).

\appendix

\section{Fusion of fMPOs}\label{app:fusionfMPO}

In this appendix we provide further details about the fusion of fMPO tensors $\mathsf{B[a]}$ and $\mathsf{B[b]}$ into a tensor $\mathsf{B[c]}$, and study the properties of the fusion tensors and the interplay with the fMPO types $\epsilon_{a}$, $\epsilon_b$, $\epsilon_c$ in full generality.

We use the same notation and conventions as in Section~\ref{sec:superalgebras}. Furthermore, we denote the virtual space of the fMPO tensor $\mathsf{B}_a$ as the super vector space $V_a \cong \mathbb{C}^{D_a^0 \vert D_a^1}$ with $D_a = D_a^0+D_a^1$ the total bond dimension, and $D_a^0$ ($D_a^1$) the dimension of the even (odd) part. Upon multiplying $O_a$ and $O_b$, we obtain a new fMPO with tensor
\begin{equation}
\mathsf{B}_{ab} = \sum_{\alpha,\alpha',i,k,\beta,\beta'} (B_{ab}^{ik})_{(\alpha,\alpha'),(\beta,\beta')} |\alpha) |\alpha') |i\rangle \langle k | (\beta'|(\beta|
\end{equation}
with
\begin{align*}
(B_{ab}^{ik})_{(\alpha,\alpha'),(\beta,\beta')} &= \sum_j (-1)^{|\alpha'|(|i|+|j|)}(B_a^{ij})_{\alpha,\beta} (B_b^{jk})_{\alpha',\beta'}\\
&= (-1)^{|\alpha'|(|\alpha|+|\beta|)}\sum_j (B_a^{ij})_{\alpha,\beta} (B_b^{jk})_{\alpha',\beta'}.
\end{align*}
We can also write the right hand side of $O_a O_b = \sum_{c} N_{ab}^c O_c$ by taking a direct sum of $N_{ab}^c$ copies of every tensor $\mathsf{B}_c$, i.e. the tensor components would be equivalent to the matrices $\bigoplus_{c} \mathds{1}_{N_{ab}^c} \otimes B_c^{ik}$.

As the trace expression for fMPOs with antiperiodic boundary conditions is (with the choice of ordering in $\mathsf{B}_{ab}$) equivalent to that of a bosonic MPO/MPS (namely a product of matrices), we can use the fundamental theorem of MPS to show the existence of a gauge transform $X_{ab}$ that brings the matrices $B_{ab}^{ik}$ in a canonical form (block upper triangular) where the diagonal blocks can be equated with those of $\bigoplus_{c} \mathds{1}_{N_{ab}^c} \otimes B_c^{ik}$. We furthermore assume that off diagonal blocks vanish, so that we obtain a strict equality
\begin{align}
	B_{ab}^{ik} X_{ab} = X_{ab} \bigoplus_{c} \mathds{1}_{N_{ab}^c} \otimes B_c^{ik}.
\end{align}
This equation is referred to as the zipper condition in the main text. The gauge transformation $X_{ab}$ is not unique, but any other gauge transformation $\tilde{X}_{ab}$ that establishes the same relation is related to $X_{ab}$ by an element in the center, i.e.
\begin{align*}
	\tilde{X}_{ab} = X_{ab} \bigoplus_c \begin{cases} M_c \otimes \mathds{1}_c, &\epsilon_c = 0\\
 M_c \otimes \mathds{1}_c + M_c' \otimes Y_c,&\epsilon_c=1	
 \end{cases}
\end{align*}
with $M_c$ and $M_c'$ matrices acting on the $N_{ab}^c$-dimensional degeneracy space. Using $(-1)^{|i|+|j|} B^{ij}_a = P_a B^{ij}_a P_a$ with $P_a = P_a^{-1} = \mathds{1}_{D_a^0} \oplus (-\mathds{1}_{D_a^1})$ the parity matrices, we can construct a different $\tilde{X}_{ab} = (P_a \otimes P_b) X_{ab} (\bigoplus_{c} \mathds{1}_{N_{ab}^c} \otimes P_{c})$, from which we infer
\begin{align*}
	(P_a \otimes P_b) X_{ab} = X_{ab} \bigoplus_c \begin{cases} M_c \otimes P_c, &\epsilon_c = 0\\
 M_c \otimes P_c + M_c' \otimes Y_c P_c,&\epsilon_c=1	
 \end{cases}
\end{align*}
Applying this relation twice leads to $M_c^2=\mathds{1}_c$ if $\epsilon_c=0$, and to $M_c^2+M_c^{\prime 2}=\mathds{1}_c$ and $[M_c,M_c']=0$ if $\epsilon_c=1$. In the first case $\epsilon_c=0$, $M_c$ is seen to have eigenvalues $\pm 1$ and thus to act as a parity matrix in the degeneracy space $V_{ab}^c$. By an appropriate basis transform in this degeneracy space, it takes the standard form $(M_c)_{\mu,\nu} = (-1)^{|\mu|}\delta_{\mu,\nu}$ thus providing a definition of $|\mu|$. This clearly shows that $V_{ab}^c$ is itself a $\mathbb{Z}_2$ graded vector space. For $\epsilon_c=1$, a basis transform in the degeneracy space can be used to simultaneously diagonalize $M_c$ and $M_c'$ into $(M_c)_{\mu,\nu} = \cos(\theta_\mu) \delta_{\mu,\nu}$ and $(M'_c)_{\mu,\nu} = \sin(\theta_\mu) \delta_{\mu,\nu}$. However, a further transformation with $\bigoplus_{\mu} \cos(\theta_\mu/2) \mathds{1}_c + \sin(\theta_\mu/2) Y_c$ results in $M_c = \mathds{1}$, $M_c'=0$. 

Using this choice of basis, we now select the columns of $X_{ab}$ and the rows of $X_{ab}^{-1}$ corresponding to a single block $c$, which we denote as $X_{ab,\mu}^{c}$ and $X_{ab,\mu}^{c+}$ respectively. From these, we can build fermionic (splitting and) fusion tensors
\begin{align}
\mathsf{X}_{ab,\mu}^{c} &= \sum_{\alpha,\beta,\gamma} (\mathsf{X}_{ab,\mu}^{c})_{(\alpha,\beta),\gamma} |\alpha)|\beta) (\gamma|\\
\mathsf{X}_{ab,\mu}^{c+} &= \sum_{\alpha,\beta,\gamma} (\mathsf{X}_{ab,\mu}^{c+})_{\gamma,(\alpha,\beta)} |\gamma)(\beta|(\alpha|
\end{align}
that satisfy the properties discussed in Section~\ref{sec:superalgebras}. Furthermore, when $\epsilon_c = 0$, the parity of the tensor $\mathsf{X}_{ab,\mu}^{c}$ is given by $|\mu|$. When $\epsilon_c=1$, we have ensured that the parity of $\mathsf{X}_{ab,\mu}^{c}$ is even, but there exists an equivalent odd choice $\mathcal{C}(\mathsf{X}_{ab,\mu}^{c} \otimesg \mathsf{Y}_c)$. Ultimately, this is a consequence of the fact that, at the level of the matrices, the Majarona type fMPOs have a further decomposition into a block diagonal form with two blocks, but which is protected by the $\mathbb{Z}_2$ grading (i.e.\ the fermion parity). For simplicity of notation below, we also denote the parity of the fusion tensor $\mathsf{X}_{ab,\mu}^{c}$ as $|\mu|$ for the case $\epsilon_c=1$, and of course have $|\mu|=0$ since we restrict to even fusion tensors in that case. 

Before moving on to the fusion of three fMPOs and the $F$-move, let us also discuss the influence of $\epsilon_a$ and $\epsilon_b$. Note that there is a priori no relation beween $\epsilon_a$, $\epsilon_b$ and $\epsilon_c$ that we can deduce from the local fusion property of the fMPO tensors. As a global object, fMPOs with periodic boundary conditions have a total fermion parity that is equal to the fMPO type $\epsilon$, and the latter therefore seems to follow the $\mathbb{Z}_2$ group structure of the former. This is however a global consequence of the properties we discuss below, and does not manifest itself when working with anti-periodic boundary conditions as arise on contractible loops in our topological fermionic PEPS.

If $\epsilon_a = 1$, we can define $\mathcal{C}(\mathsf{Y}_a\otimesg \mathsf{X}_{ab,\mu}^{c})$ as an equivalent tensor, but with opposite parity of $\mathsf{X}_{ab,\mu}^{c}$. Considering the case $\epsilon_c=0$, this implies the relation
\begin{equation}
\mathcal{C}(\mathsf{Y}_a\otimesg \mathsf{X}_{ab,\mu}^{c}) = \sum_{\nu} (M_a)_{\nu,\mu} \mathsf{X}_{ab,\mu}^{c}
\end{equation}
where $M_a$ is nonzero only if $|\mu| \neq |\nu|$. Applying this relation twice leads to $M_a^2 = -\mathds{1}$, e.g.\ $M_a$ acts as a $Y$ matrix in the degeneracy space. This requires the degeneracy space $V_{ab}^c$ to be even-dimensional with equal dimensions of even and odd parity. We can choose a suitable basis such that $M_a$ takes a standard form and replace the labeling $\mu$ to $(\hat{\mu},0)$ and $(\hat{\mu},1)$ defined by
\begin{equation}
\mathcal{C}(\mathsf{Y}_a\otimesg \mathsf{X}_{ab,(\hat{\mu},0)}^{c}) = \mathsf{X}_{ab,(\hat{\mu},1)}^{c},\quad \mathcal{C}(\mathsf{Y}_a\otimesg \mathsf{X}_{ab,(\hat{\mu},1)}^{c}) = -\mathsf{X}_{ab,(\hat{\mu},0)}^{c}.\label{eq:defmu01}
\end{equation}
An equivalent result holds when $\epsilon_b=1$ (still assuming $\epsilon_c=0$). However, if both $\epsilon_a=\epsilon_b=1$, more care is required. As both $\mathsf{Y}_a$ and $\mathsf{Y}_b$ are odd tensors, their order of contraction matters (at the level of the matrices, contracting with $\mathsf{Y}_a$ and $\mathsf{Y}_b$ amounts to left multiplication of $X_{ab,\mu}^c$ with $Y_a \otimes \mathds{1}_b$ and $P_a \otimes Y_b$ respectively). Hence, while the general relation with a generic $M_a$ and $M_b$ remains valid, we furthermore obtain $\{M_a,M_b\}=0$ and only one of the two matrices $M_a$ and $M_b$ can be brought into standard form. Choosing Eq.~\eqref{eq:defmu01} to be still valid, we obtain for the contraction with $\mathsf{Y}_b$ the relation
\begin{equation}
\mathcal{C}(\mathsf{Y}_b\otimesg \mathsf{X}_{ab,(\hat{\mu},0)}^{c}) = \sum_{\hat{\nu}} (\hat{M}_b)_{\hat{\nu},\hat{\mu}} \mathsf{X}_{ab,(\hat{\nu},1)}^{c} = \sum_{\hat{\nu}} (\hat{M}_b)_{\hat{\nu},\hat{\mu}}\mathcal{C}(\mathsf{Y}_a\otimesg \mathsf{X}_{ab,(\hat{\nu},0)}^{c}).
\end{equation}
with $\hat{M}_b^2=-1$ resulting from applying this relation twice. $\hat{M}_b$ thus has eigenvalues $+\mathrm{i}$ or $-\mathrm{i}$ and can be be diagonalized by a further basis transformation in the $\hat{\mu}$ space. Working in this basis, we have thus obtained
\begin{equation}
\mathcal{C}(\mathsf{Y}_b\otimesg \mathsf{X}_{ab,\mu}^{c}) = (-1)^{\eta^{c}_{ab,\hat{\mu}}} \mathrm{i} \mathcal{C}(\mathsf{Y}_a\otimesg \mathsf{X}_{ab,\mu}^{c}).
\end{equation}

If $\epsilon_a=\epsilon_c=1$, the contraction of $\mathsf{Y}_a$ and $\mathsf{X}_{ab,\mu}^{c}$ yields an odd tensor, so that we have the relation
\begin{equation}
\mathcal{C}(\mathsf{Y}_a\otimesg \mathsf{X}_{ab,\mu}^{c}) = \sum_{\nu} (L_a)_{\nu,\mu} \mathcal{C}(\mathsf{X}_{ab,\mu}^{c}\otimesg \mathsf{Y}_c)
\end{equation}
and applying this relation twice learns that $L_a^2=1$. A proper choice of basis diagonalizes $L_a$ and results in
\begin{equation}
\mathcal{C}(\mathsf{Y}_a\otimesg \mathsf{X}_{ab,\mu}^{c}) = (-1)^{\zeta^c_{ab,\mu}} \mathcal{C}(\mathsf{X}_{ab,\mu}^{c} \otimesg \mathsf{Y}_c).
\end{equation}
Similarly, if $\epsilon_b=\epsilon_c=1$ we can choose a basis where
\begin{equation}
\mathcal{C}(\mathsf{Y}_b\otimesg \mathsf{X}_{ab,\mu}^{c}) = (-1)^{\xi^c_{ab,\mu}} \mathcal{C}(\mathsf{X}_{ab,\mu}^{c} \otimesg \mathsf{Y}_c).
\end{equation}
However, if $\epsilon_a=\epsilon_b=\epsilon_c=1$, we again obtain $\{L_a,L_b\}=0$ and both matrices cannot be diagonalized simultaneously. This relation requires the degeneracy space to be even dimensional and $L_a$ and $L_b$ to have equally many $+1$ and $-1$ eigenvalues; e.g.\ the simplest representation could be $L_a = Z$ and $L_b = X$.

\section{Fixed-point fMPO representation}\label{app:fMPO}

In this appendix we show that the fixed-point fMPOs constructed from the tensors \eqref{eq:RH1}, \eqref{eq:RH2} form an explicit representation of the fMPO algebra whose $\tilde{F}$-symbols were used to define the tensor components.

We define the fusion tensor $\mathsf{\tilde{X}}_{ab,\mu}^c$ with internal ordering

\begin{equation}
\includegraphics[width=0.3\textwidth]{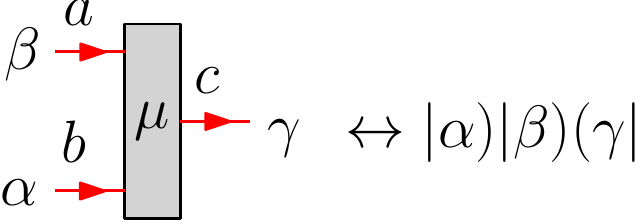}
\end{equation}
and components

\begin{equation}
\includegraphics[width=0.33\textwidth]{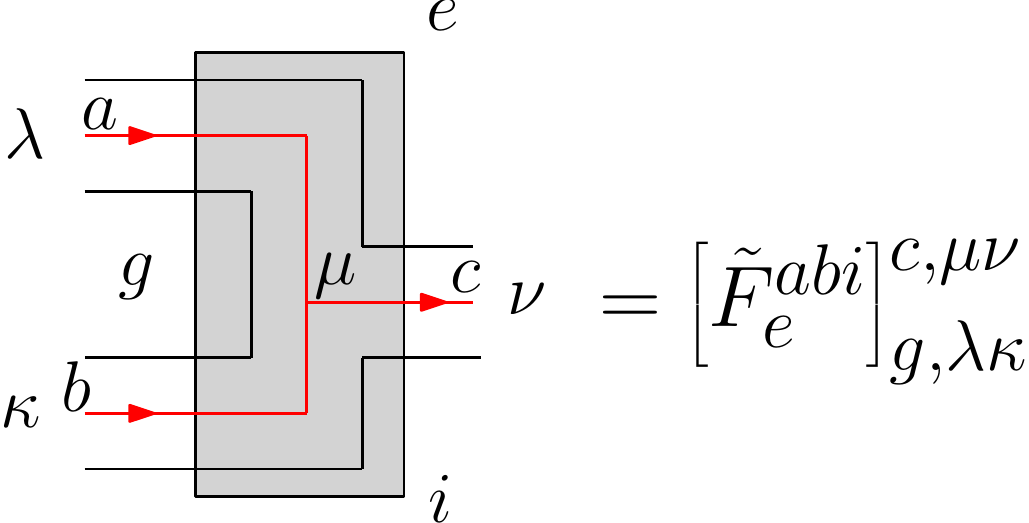}\, .
\end{equation}
One can now check that following tensor identity is equivalent to the super pentagon equation \eqref{eq:pentagon}:

\begin{equation}\label{eq:fusionstringnet3}
\includegraphics[width=0.3\textwidth]{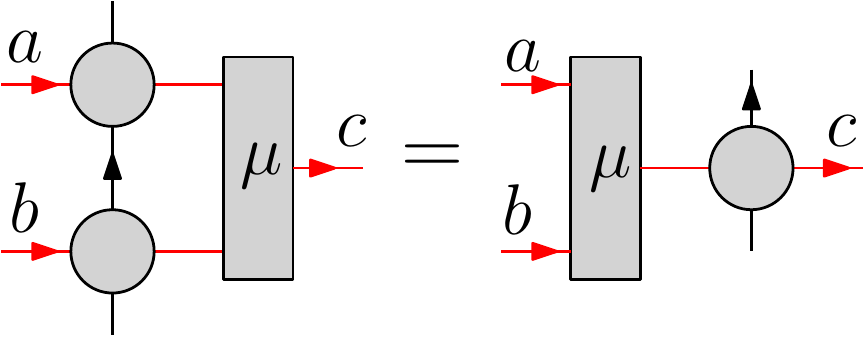}\, .
\end{equation}
Combining this relation with the isometric property of the $\tilde{F}$-symbols implies that identities \eqref{eq:reduction2} and \eqref{eq:orthogonality} hold, from which it follows that the fMPOs $O_a$ constructed from tensors \eqref{eq:RH1}, \eqref{eq:RH2} indeed satisfy the correct multiplication properties $O_aO_b = \sum_c N_{ab}^c O_c$.  Note that also the stronger property \eqref{eq:zipper} follows from \eqref{eq:fusionstringnet3} and unitarity. Taking the explicit expressions for the fusion tensors $\mathsf{\tilde{X}}_{ab,\mu}^c$ it is straightforward to check that the $F$-move indeed produces the same $\tilde{F}$ symbols as those defining all tensor components.

In this appendix we only considered right-handed fMPO tensors. However, similar to the bosonic case \cite{MPOpaper}, all fMPOs consisting of an arbitrary number of right-handed and left-handed tensors form a representation of the fMPO algebra $O_aO_b = \sum_c N_{ab}^c O_c$ with the correct $\tilde{F}$-symbols.

\section{Pivotal properties of Gu-Wen fusion tensors}\label{GuWenpivotal}

To study the pivotal properties of Gu-Wen fusion tensors we first introduce two new tensors. The first tensor has in the basis

\begin{equation}
\includegraphics[width=0.25\textwidth]{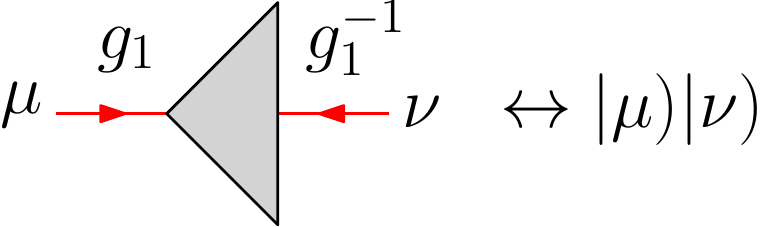}\, ,
\end{equation}
coefficients which take following form:

\begin{equation}
\includegraphics[width=0.47\textwidth]{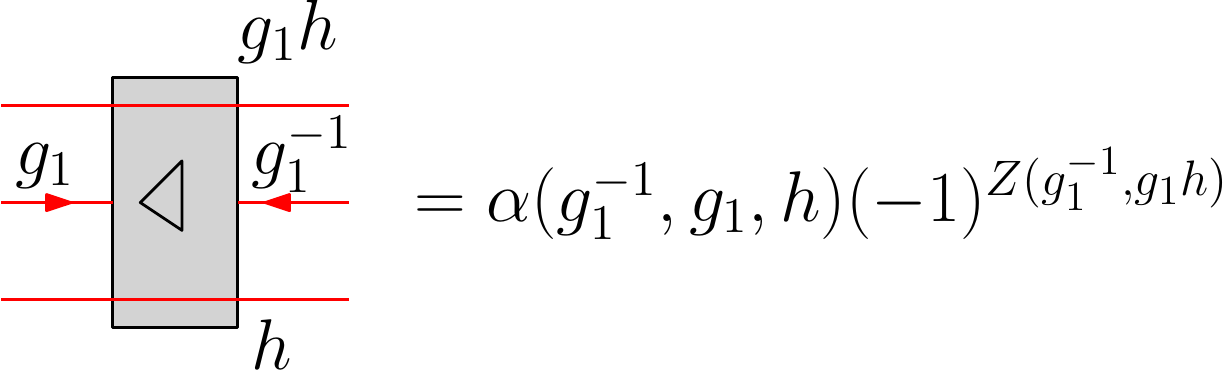}\, .
\end{equation}
The parity of its indices is given by $Z(g_1,h)$ and $Z(g_1^{-1},g_1h)$, implying that the total parity of this tensor is $Z(g_1^{-1},g_1)$ (using that $Z(e,g) = 0)$. The second tensor is defined in the basis

\begin{equation}
\includegraphics[width=0.25\textwidth]{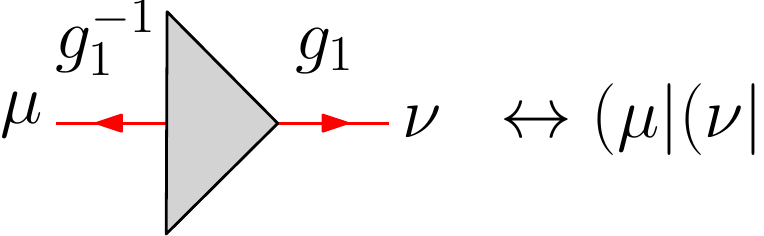}
\end{equation}
and has coefficients given by

\begin{equation}
\includegraphics[width=0.35\textwidth]{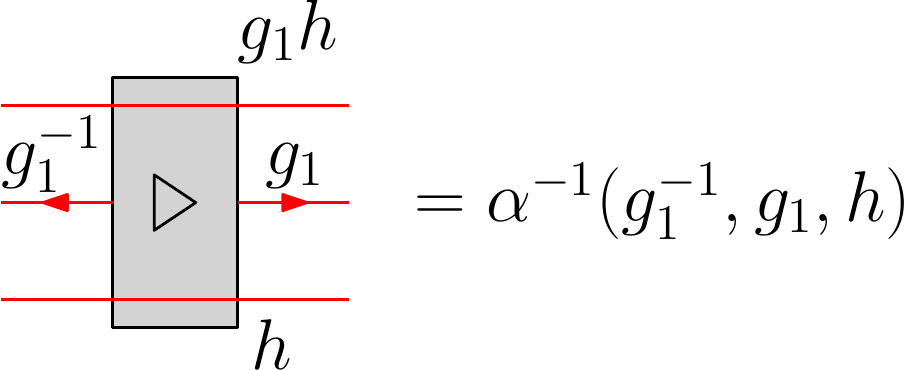}\, .
\end{equation}
The parities of the indices are again $Z(g_1,h)$ and $Z(g_1^{-1},g_1h)$, such that the total parity is $Z(g_1^{-1},g_1)$, similar to the previous tensor. One can verify that these tensors satisfy following relations

\begin{equation}
\includegraphics[width=0.75\textwidth]{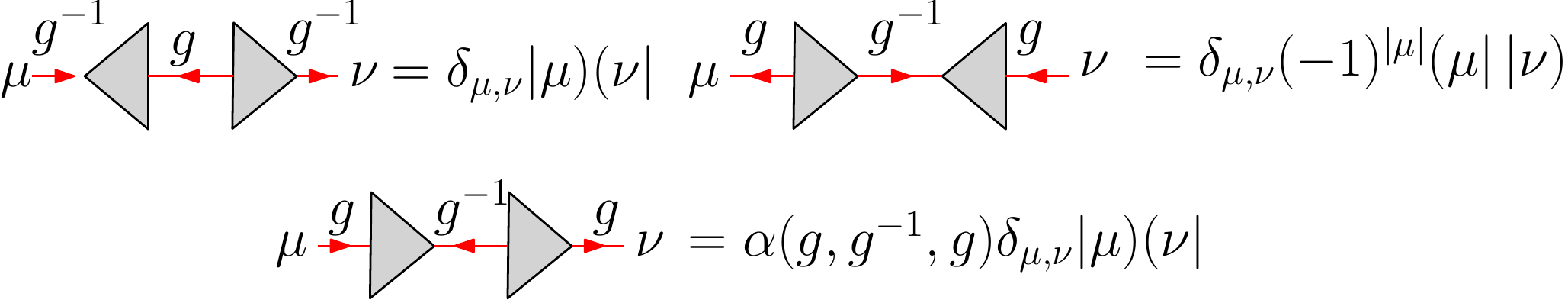}\, ,
\end{equation}
where we, again without loss of generality, work with representative cocycles satisfying $\alpha(e,g,h)=1$. Note that these tensors are very similar to the matrices $Z_g$ as defined at the beginning of section \ref{sec:guwen}. For details about the precise connection in the bosonic case we refer to \cite{MPOpaper,Williamson}. The reason for introducing these new tensors is that now we have following important tensor identity, relating right- and left-handed fMPO tensors:

\begin{equation}\label{eq:leftright}
\includegraphics[width=0.32\textwidth]{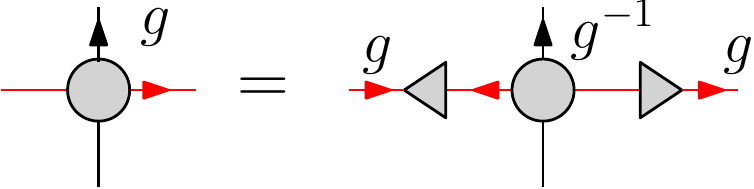}\, .
\end{equation}
From \eqref{eq:leftright} one can show that the fusion tensors should satisfy following relations:

\begin{equation} \label{linebending}
\includegraphics[width=0.55\textwidth]{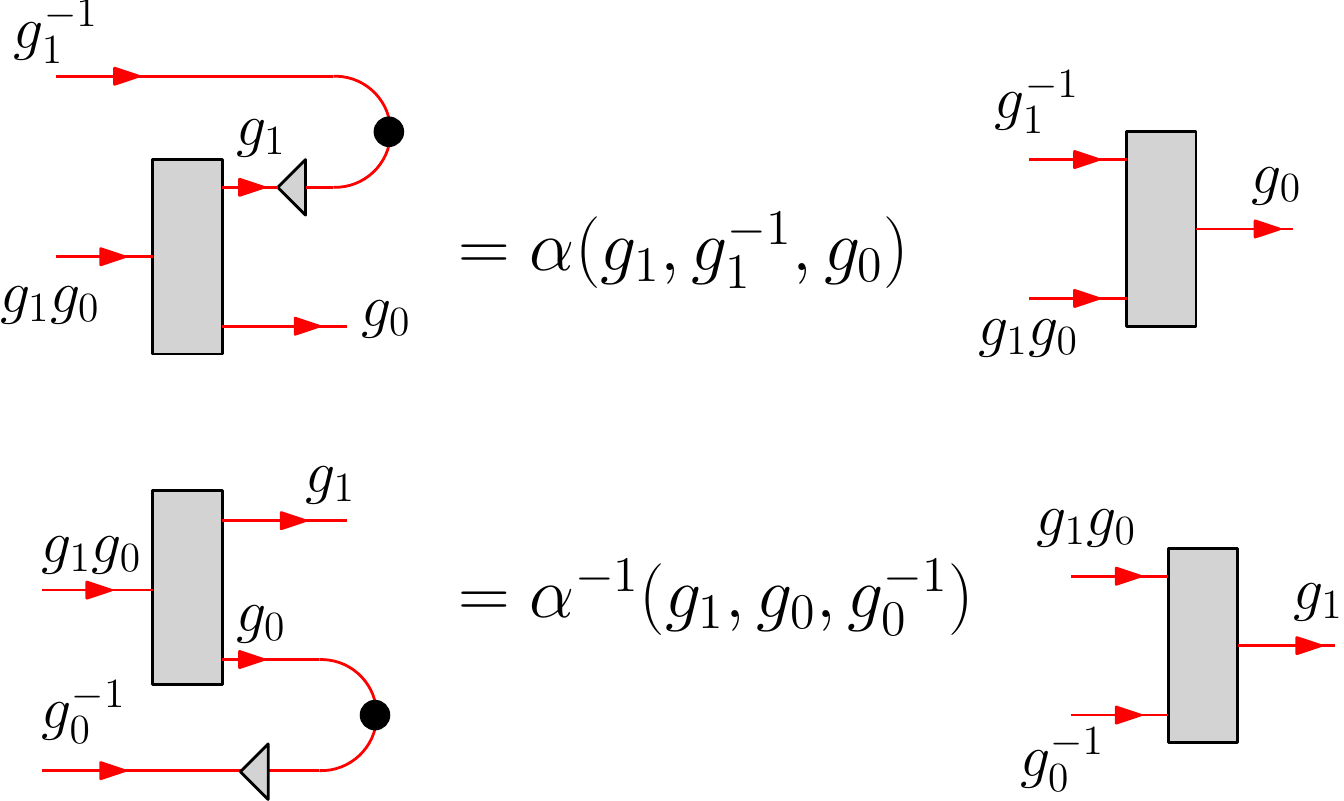}\, .
\end{equation}
Of course this can also be verified directly by taking the explicit expression \eqref{eq:guwenfusiontensor1},\eqref{eq:guwenfusiontensor2} for $\mathsf{X}_{g,h}$. These expressions are of great value since they allow for a graphical calculation of many interesting properties.

\section{$\{\tilde{O}^L_1,\tilde{O}^L_\sigma \}$ $\mathbb{Z}_2$ representation with periodic boundary conditions}\label{app:OP}

In this appendix we derive the projective group action of $\tilde{O}_\sigma$, which is an fMPO constructed from the same tensor as $O_\sigma$, but with an even number of parity matrices on the internal indices. For concreteness, let us take $\tilde{O}_\sigma$ to be 

\begin{equation}
\includegraphics[width=0.45\textwidth]{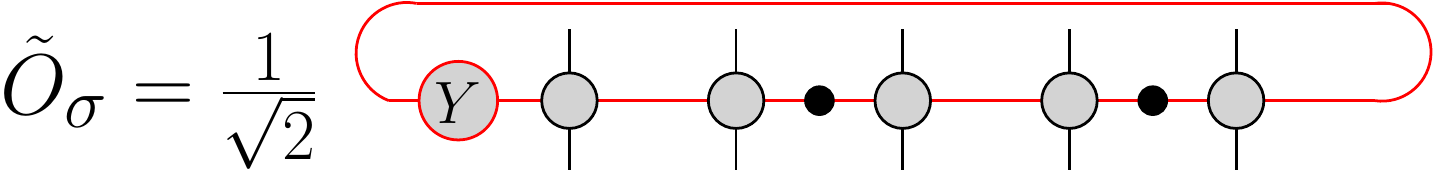}\, .
\end{equation}
Of course, the length $L$ of $\tilde{O}_\sigma$, which we took to be five here, and the specific even number of parity matrices and their positions on the internal fMPO indices is just an arbitrary choice and the result of this appendix does not depend on these choices. For example, as already explained in the main text, regardless of the length and specific even number of parity matrices, we always have to insert the odd matrix $\mathsf{Y}$ on the internal index for $\tilde{O}_\sigma$ to be non-zero.

The product of two $\tilde{O}_\sigma$ fMPOs can be represented as 

\begin{equation}
\includegraphics[width=0.5\textwidth]{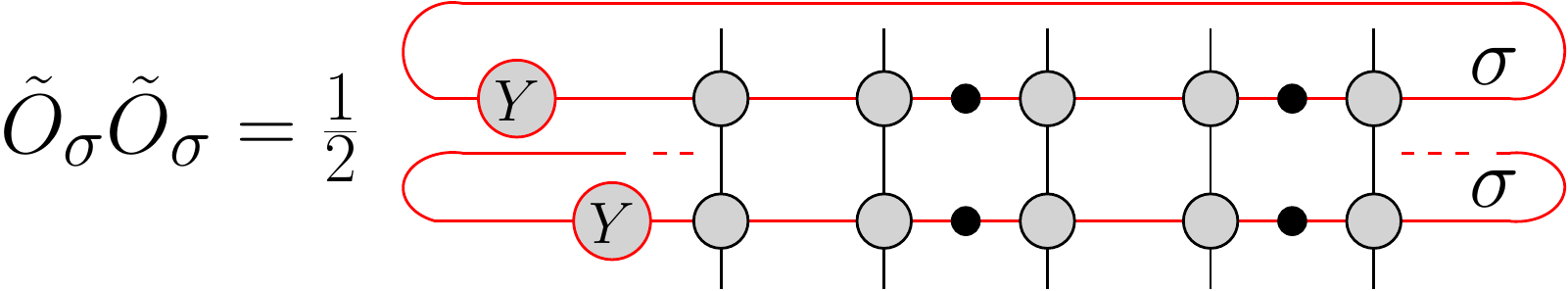}\, ,
\end{equation}
where the order of the $\mathsf{Y}$ matrices is determined by the order of multiplication of the fMPOs. Using properties \eqref{eq:zipper} and \eqref{eq:orthogonality} we obtain
\begin{equation}
\includegraphics[width=0.55\textwidth]{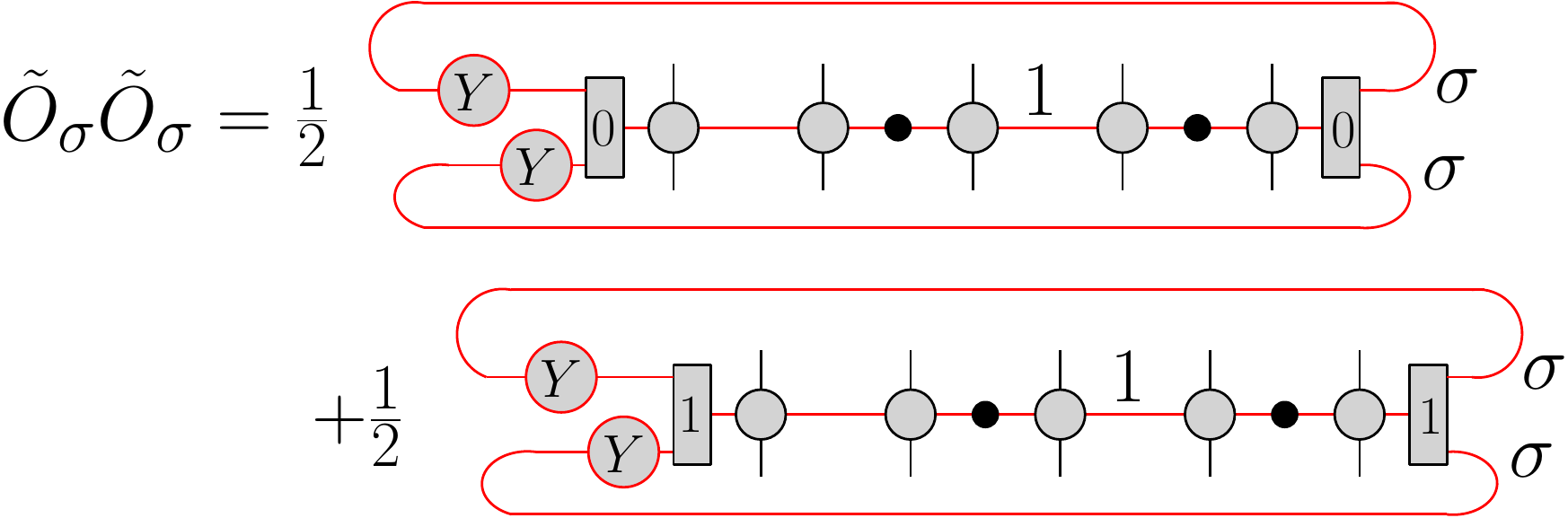}\, ,
\end{equation}
where we explicitely denote the parity of the fusion tensors. A few simple steps now lead to the desired result:

\begin{equation}
\includegraphics[width=0.55\textwidth]{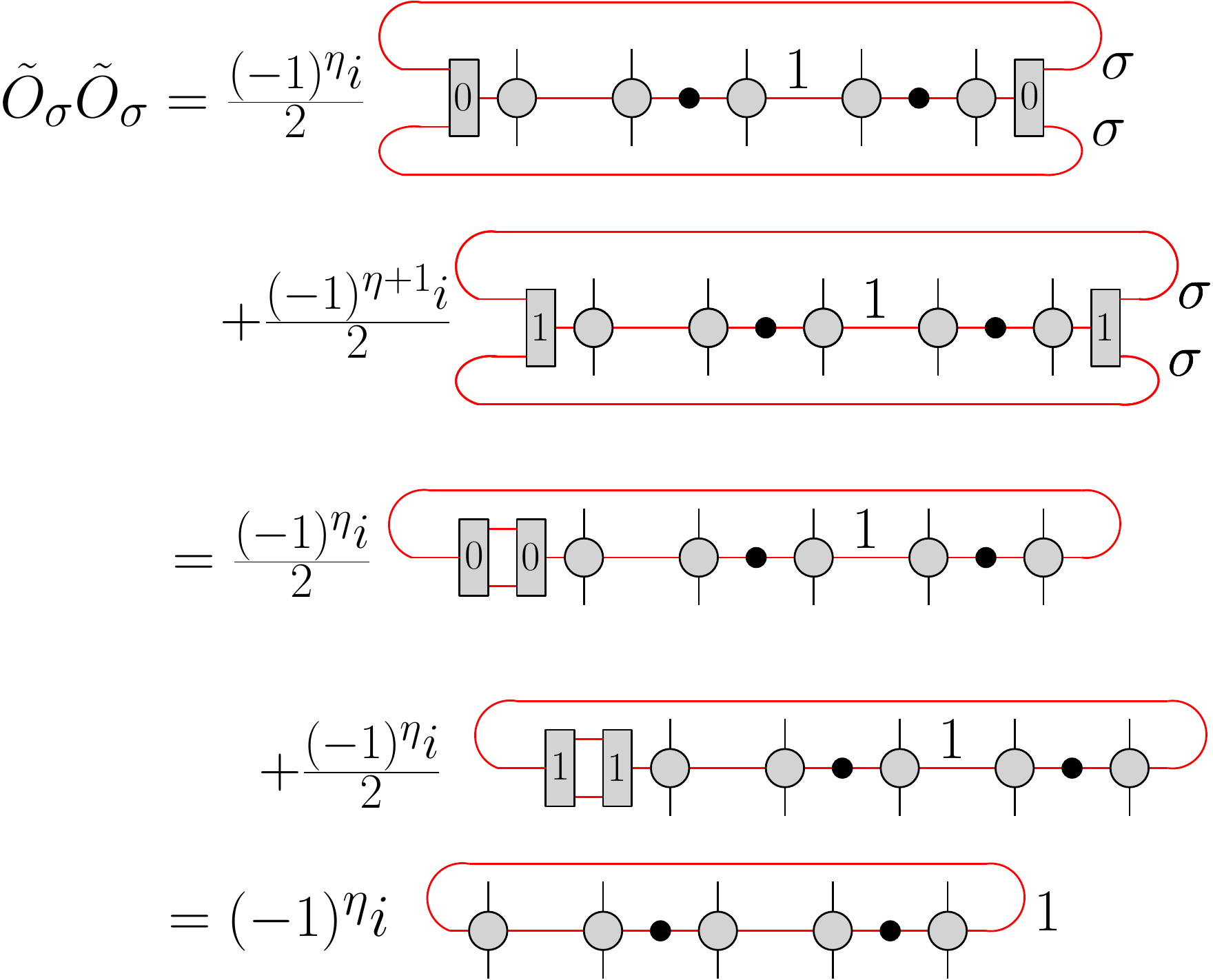}\, .
\end{equation}
In the first line we used \eqref{eq:Ymove1}, in the second line we get the additional minus sign because the fusion tensor is odd and in the last line we again used \eqref{eq:orthogonality}.

\bibliographystyle{utphys}
\bibliography{FermionicPEPS}

\end{document}